\documentclass{ws-ijbc}
\newcommand{\cn}{\mbox{cn}}
\newcommand{\mat}[1]{{\cal #1}}
\catchline{}{}{}{}{} % Publisher's Area please ignore

\begin{document}

\markboth{T. Bountis, G. Chechin and V. Sakhnenko}{Discrete Symmetry and Stability in Hamiltonian Dynamics}

\title{DISCRETE SYMMETRY AND STABILITY IN HAMILTONIAN DYNAMICS}

\author{Tassos Bountis}

\address{Department of Mathematics, University of Patras\\ Patras
26500, Greece\\
bountis@math.upatras.gr}

\author{George Chechin}
\address{Department of Physics, Southern Federal University\\
Zorge 5, Rostov--on--Don, 344090, Russia\\
chechin@phys.rsu.ru}

\author{Vladimir Sakhnenko}
\address{Department of Physics, Southern Federal University\\
Zorge 5, Rostov--on--Don, 344090, Russia\\
sakh@ip.rsu.ru} \maketitle

\begin{history}
\received{(to be inserted by publisher)}
\end{history}

\begin{abstract}

In the present tutorial we address a problem with a long history,
which remains of great interest to date due to its many important
applications: It concerns the existence and stability of periodic
and quasiperiodic orbits in $N$ - degree of freedom Hamiltonian
systems and their connection with discrete symmetries. Of primary
importance in our study is what we call nonlinear normal modes
(NNMs), i.e periodic solutions which represent continuations of
the system's linear normal modes in the nonlinear regime. We
examine questions concerning the existence of such solutions and
discuss different methods for constructing them and studying their
stability under fixed and periodic boundary conditions.

In the periodic case, we find it particularly useful to approach
the problem through the discrete symmetries of many models,
employing group theoretical concepts to identify a special type of
NNMs which we call one--dimensional ``bushes''. We then describe
how to use linear combinations of $s\geq 2$ such NNMs to construct
$s$--dimensional bushes of quasiperiodic orbits, for a wide
variety of Hamiltonian systems including particle chains, a square
molecule and octahedral crystals in 1,2 and 3 dimensions. Next, we
exploit the symmetries of the linearized equations of motion about
these bushes to demonstrate how they may be simplified to study
the destabilization of these orbits, as a result of their
interaction with NNMs not belonging to the same bush. Applying
this theory to the famous Fermi Pasta Ulam (FPU) chain, we review
a number of interesting results concerning the stability of NNMs
and higher--dimensional bushes, which have appeared in the recent
literature.

We then turn to a newly developed approach to the analytical and
numerical construction of quasiperiodic orbits, which does not
depend on the symmetries or boundary conditions of our system.
Using this approach, we demonstrate that the well--known
``paradox'' of FPU recurrences may in fact be explained in terms
of the \textit{exponential localization} of the energies $E_q$ of
NNM's being excited at the low part of the frequency spectrum,
i.e. $q=1,2,3,...$. These results indicate that it is the
stability of these low--dimensional compact manifolds called
$q$--tori, that is related to the persistence or FPU recurrences
at low energies. Finally, we discuss a novel approach to the
stability of orbits of conservative systems, expressed by a
spectrum of indices called GALI$_k$, $k=2,...,2N$, by means of
which one can determine accurately and efficiently the
destabilization of $q$ tori, leading, after very long times, to
the breakdown of recurrences and, ultimately, to the equipartition
of energy, at high enough values of the total energy $E$.
\end{abstract}


\section{Introduction}

The dynamics of Hamiltonian systems, or, more generally,
conservative mechanical systems (preserving phase space volume) is
of great importance for the understanding of many problems arising
in classical mechanics, astronomy, solid state physics, plasma
physics and nonlinear optics. The simplest states of such systems
are their equilibrium points, where all variables are fixed for
all time, and their periodic orbits, where all particles oscillate
and return to their starting values after a time interval $T$. The
first question concerning these states is their
\textit{existence}. Once this has been established, what one needs
to know is their \textit{stability}, i.e. their behavior under
small perturbations of the initial conditions and parameter
values. Two great contemporary scientists were the first to deal
with these issues in a systematic and comprehensive way: The
Russian mathematician Alexander Mikhailovich Lyapunov (1857 -
1918) and the French mathematical physicist Henri Poincar\'{e}
(1854 - 1917). Lyapunov devoted a great deal of his efforts to
`local' stability analysis, obtaining specific conditions for the
behavior of solutions of systems of ordinary differential
equations (ODEs) in the vicinity of equilibrium points
\cite{Lyapunov}. Poincar\'{e} was more concerned with `global'
properties of the dynamics, like non - integrability and the
occurrence of irregular (or chaotic) solutions, wandering over
large domains of the available state space \cite{Poincare}.
Remarkably, both of them worked on periodic solutions of $N$ -
degree - of - freedom Hamiltonian systems and were fascinated by
the problem of the stability of the solar system.

In the present paper, our purpose is to review certain recent
results concerning the dynamics of Hamiltonian systems of $N$
degrees of freedom, whose equations of motion have the form
\begin{equation}\label{eq:Hamiltonian_ODEs}
\frac{{dq}_{k}}{dt}=\frac{\partial H}{{\partial
p}_{k}},\,\,\,\frac{{dp}_{k}}{dt}=-\frac{\partial H}{{\partial
q}_{k}}, \,\,k=1,2,...N
\end{equation}
In many cases the Hamiltonian is expanded in power series in the
variables of positions and momenta, as a sum of homogeneous
polynomials $H_{m}$ of degree $m\geq2$
\begin{equation}\label{eq:Hamiltonian_expansion}
H=H_2+H_3+...+H_m=E~~,~~H_m=H_m(q_{1},...,q_{N},p_{1},...,p_{N})
\end{equation}
($E$ is the total constant energy), so that the origin,
$q_{k}=p_{k}=0,\,\,\, k=1,2,..,N$ is an equilibrium point of the
system. However, there also many examples, e.g. possessing
discrete symmetries, for which no such decomposition is necessary
as one can work directly with the original Hamiltonian.

Assume now that the linear equations resulting from
(\ref{eq:Hamiltonian_ODEs}) and (\ref{eq:Hamiltonian_expansion})
with $H_{m} = 0$, for all $m > 2$, yield a matrix, whose
eigenvalues all occur in conjugate imaginary pairs, $ \pm
i\lambda_{k}$ and provide the frequencies of the so--called normal
mode oscillations of the linearized system. According to Lyapunov,
if \textit{none} of the ratios of these eigenvalues, $\lambda_{j}
/\lambda_{k}$, is an integer, for any $j, k = 1,2.,N$, the linear
normal modes continue to exist as periodic solutions of the
nonlinear system (\ref{eq:Hamiltonian_ODEs}), when higher order
terms $H_3, H_4,...$ etc. are taken into account in
(\ref{eq:Hamiltonian_expansion}). These solutions have frequencies
close to those of the linear modes and are examples of what we
call \textit{nonlinear normal modes} (NNMs), where all variables
$(q_{k},p_{k})$ oscillate with the same frequency $(\lambda_{j} =)
\omega_{j} = 2\pi/T_{j}$, returning to the same values after a
single maximum (and minimum) in their time evolution over one
period $T_j$.

What is the importance of these periodic solutions? Once we have
determined that they exist, what can we say about their stability
properties under small perturbations? How do these change when we
vary the total energy $E$ in (\ref{eq:Hamiltonian_expansion})?
Does their loss of stability affect only their immediate vicinity
or can they also influence the dynamics of the system as a whole?
These are the questions we shall try to answer in this paper.

Let us consider, as a specific example, a mechanical system in one
dimension described by the the $N$-particle Hamiltonian
\begin{equation}\label{fpuham}
H= {1\over 2}\sum_{k=1}^N p_k^2 + {  1   \over
2}\sum_{k=0}^N(x_{k+1}-x_k)^2 + {\alpha \over 3}
\sum_{k=0}^N(x_{k+1}-x_k)^3+{\beta \over 4}
\sum_{k=0}^N(x_{k+1}-x_k)^4=E
\end{equation}
called the FPU chain in honor of Fermi, Pasta and Ulam, who were
the first to study it numerically in the early 1950's
\cite{Fermi1955}. They discovered certain very interesting and
surprising phenomena concerning its dynamics, which we will
discuss later in this review. For the time being, let us note that
when this chain is studied under \textit{fixed} boundary
conditions (f.b.c.), i.e. $x_0=x_{N+1}=0$, there are many $N$'s
for which the linear normal mode frequencies satisfy the
incommensurability condition of Lyapunov's theorem and hence they
may be rigorously continued to the nonlinear regime of $\alpha\neq
0$ and/or $\beta\neq 0$.

By contrast, when \textit{periodic} boundary conditions (p.b.c.)
are imposed, i.e. $x_j=x_{j+N}, p_j=p_{N+j}, j=1,2,...,N$, the
linear mode spectrum becomes degenerate for \textit{all} $N$ and
Lyapunov's theorem cannot be invoked. What do we do in that case?
How do we study the existence and stability of such nonlinear
normal modes, or NNMs? As we explain at length in sections II and
III, this is an important case where the identification and
analysis of the system's \textit{discrete symmetries} turns out to
be of great relevance to the understanding of the dynamics of the
problem.

Thus, we start by demonstrating in section II precisely how one
can use the powerful techniques of group theory to establish the
existence of families of such NNMs for a variety of mechanical
systems, including particle chains in one dimension (under
p.b.c.), as well as certain 2--dimensional and 3--dimensional
structures. Then, in section III, we show how one can combine such
periodic solutions to form ``bushes'' of quasiperiodic orbits and
exploit the symmetries of the system to simplify the variational
equations about these orbits and study the motion in their
vicinity. Ultimately, of course, one would like to be able to
obtain \emph{invariant manifolds} on which the dynamics of the
system is as simple as possible. This is not easy to do in
general, but if the equations of motion possess discrete symmetry
groups, one can single out some of these manifolds using regular
group theoretical methods developed in \cite{DAI, DAI2,
Chechin-Sakhnenko}. These methods provide, in fact, the
mathematical basis of the theory of \emph{bushes} of NNMs, which
plays an important role in many of the results discussed in this
review.

\begin{figure}[h]
\centering
\psfig{file=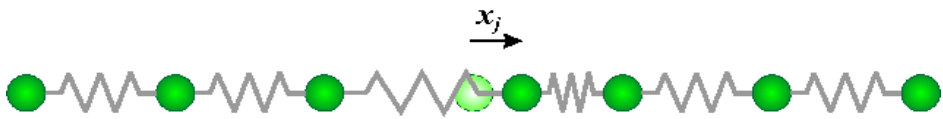,width=0.7\textwidth}
\caption{FPU chain.}\label{f1}
\end{figure}

Before discussing this theory, however, let us illustrate its main
ingredients on a seemingly simple example of an $N$ degree of
freedom Hamiltonian system. In particular, let us consider the
famous Fermi-Pasta-Ulam $\beta$-chain (FPU-$\beta$), i.e. Eq.
(\ref{fpuham}) with $\alpha =0$, representing a one--dimensional
lattice of unit masses coupled to each other by identical
nonlinear springs (see Fig.~\ref{f1}).
The ordinary differential equations (ODEs) describing longitudinal
vibrations of the FPU-$\beta$ chain can be written in the form
\begin{equation}\label{eq1}
    \ddot{x}_i=f(x_{i+1}-x_i)-f(x_i-x_{i-1}), \ \ \ \ \ \ \ i=1..N,
\end{equation}
where $x_i(t)$ is the displacement of the $i$th particle from its
equilibrium state at time $t$, while the force $f(\Delta x)$
depends on the spring deformation $\Delta x$ as
\begin{equation}\label{eq2}
    f(\Delta x)=\Delta x+\beta(\Delta x)^3
\end{equation}
For simplicity, we assume that $\beta>0$, since in that case the
potential is positive definite everywhere and particles cannot
escape to infinity. If we also impose p.b.c. we must require:
\begin{equation}\label{eq3}
    x_{N+1}(t)\equiv x_1(t), \ \ \ x_0(t)\equiv x_N(t).
\end{equation}

Thus, we shall study possible dynamical regimes of this chain by
attempting to solve (\ref{eq1}) for the ``configuration" vector
\begin{equation}\label{eq4}
    \vec{X}(t)=\{x_1(t), x_2(t), ..., x_N(t)\}
\end{equation}
whose components are the individual particle displacements.

As is well--known, it is pointless to try to obtain this vector as
a general solution of (\ref{eq1}). We will, therefore, concentrate
on studying special solutions represented by the NNMs mentioned
above. In particular, let us begin with the following simple
periodic solution
\begin{equation}\label{eq5}
    \vec{X}(t)=\{x(t), -x(t), x(t), -x(t), ..., x(t), -x(t)\}
\end{equation}
which is easily seen to exist in the FPU chain with an even number
of particles ($N \ mod \ 2 = 0$). This solution is fully
determined by \emph{only one} arbitrary function $x(t)$ and is
called "$\pi$-mode" or boundary zone mode.

From Eq.~(\ref{eq5}), it is evident that every two neighboring
particles possess equal in value but opposite in sign
displacements at any instant $t$. To investigate the dynamical
regime of this mode in the FRU-$\beta$ chain, let us substitute
$\vec{X}(t)$ from Eq.~(\ref{eq5}) in Eq.~(\ref{eq1}), taking into
account Eqs.~(\ref{eq2}) and (\ref{eq3}). As a result, all $N$
equations (\ref{eq1}) become \emph{identical} and can be written
in the form:
\begin{equation}\label{eq6}
    \ddot{x}+\omega^2 x+\gamma\Bigl(\frac{\beta}{N}\Bigr)x^3=0,
\end{equation}
where $\omega^2=4$, $\gamma=16$ and the $x_i$ have been scaled by
a factor $1/\surd N$.

This is the well-known Duffing's equation. In the case $\beta>0$
it describes periodic vibrations with arbitrary fixed amplitude
$A$ determined by initial conditions which have the form
\begin{equation}\label{eq7}
    x(0)=A, \ \ \ \dot{x}(0)=0.
\end{equation}

As is well--known, the analytical solution to Eqs. (\ref{eq6}),
(\ref{eq7}) can be expessed via the Jacobi elliptic cosine
\cite{AbramSteg}:
\begin{equation}\label{eq8}
    x(t)=A\cn(\Omega t, k^2),
\end{equation}

where
\begin{equation}\label{eq9}
    \Omega^2=\omega^2/(1-2k^2),
\end{equation}
while the modulus $k$ of this elliptic function is determined by
the relations
\begin{equation}\label{eq10}
    2k^2=\frac{b A^2}{\omega^2+b A^2}, \ \
    b=\gamma\Bigl(\frac{\beta}{N}\Bigr)  .
\end{equation}

Thus, the so--called $\pi$-mode defined by Eq.~(\ref{eq5})
describes an \emph{exact} dynamical regime of the nonlinear
system. It represents an example of the type of nonlinear normal
modes introduced by Rosenberg \cite{Rozenberg}. From
Eqs.~(\ref{eq8}, \ref{eq10}) it can be seen that in the
low-amplitude limit ($A\rightarrow 0)$ the above NNM is an exact
continuation of the \emph{linear normal mode} ($k\rightarrow0$, \
$\Omega\rightarrow\omega$), as the elliptic cosine tends to the
ordinary cosine.

NNMs exist in Hamiltonian systems with rather specific
interparticle interactions, as, for example, in cases where the
potential energy is a \emph{homogeneous} function of all its
arguments. In many cases, however, the existence of NNMs may be
established by certain symmetry related arguments, whence we refer
to such dynamical objects as \emph{symmetry-determined} NNMs.

The $\pi$-mode (\ref{eq5}) clearly represents a dynamical state of
the form:
\begin{equation}\label{eq13a}
    \vec{X}(t)=x(t)\{1, -1,| 1, -1,| ..., |1, -1\}.
\end{equation}
Hence, the following question naturally arises: Are there any other
such exact NNMs in the FPU-$\beta$ chain, under p.b.c.? Some
examples of such modes are already known in the literature
\cite{Ruffo, Rink, Shinohara, Shinohara2003, Yoshimura, LeoLeo,
Greeks, Kosevich}, under a terminology that differs among authors.
Below we list these exact states in detail:
\begin{equation}\label{eq13b}
    \vec{X}(t)=x(t)\{1, 0, -1,| 1, 0, -1,|, ..., |1, 0, -1\}, \ \ \ \
    \omega^2=3, \ \ \gamma=\frac{27}{2} \ \ (N \ mod \ 3 = 0).
\end{equation}
\begin{equation}\label{eq13c}
    \vec{X}(t)=x(t)\{1, -2, 1,| 1, -2, 1,|, ..., |1, -2, 1\}, \ \ \ \
    \omega^2=3, \ \ \gamma=\frac{27}{2} \ \ (N \ mod \ 3 = 0).
\end{equation}
\begin{equation}\label{eq13d}
    \vec{X}(t)=x(t)\{0, 1, 0, -1,|0,  1, 0, -1,|, ..., |0, 1, 0, -1\}, \ \ \ \
    \omega^2=2, \ \ \gamma=4 \ \ (N \ mod \ 4 = 0).
\end{equation}
\begin{equation}\label{eq13e}
    \vec{X}(t)=x(t)\{1, 1, -1, -1| 1, 1, -1, -1|, ..., |1, 1, -1, -1\},
\end{equation}

\centerline{$\omega^2=2, \ \ \gamma=8 \ \ (N \ mod \ 4 = 0)$.}

\begin{equation}\label{eq13f}
    \vec{X}(t)=x(t)\{0, 1, 1, 0, -1, -1| 0, 1, 1, 0, -1, -1|, ..., |0, 1, 1, 0, -1, -1\}, \ \ \ \
\end{equation}

\centerline{$\omega^2=1, \ \ \gamma=\frac{3}{2} \ \ (N \ mod \ 6 =
0)$.}

Let us comment on certain properties of the above NNMs for the
FPU-$\beta$ chain, having an appropriate number of particles in
every case:

P1) \emph{Displacement patterns} of NNMs (\ref{eq13a}) -
(\ref{eq13f}) are defined in terms of the vibrational state
primitive cells (divided by vertical lines), whose number of
elements $m$ is called the multiplicity number.

For (\ref{eq13a}) - (\ref{eq13f}) these numbers have the values:
$m=2$ for Eq.~(\ref{eq13a}), $m=3$ for Eqs.~(\ref{eq13b}) and
(\ref{eq13c}), $m=4$ for Eqs.~(\ref{eq13d}) and (\ref{eq13e}) and
$m=6$, for Eq.~(\ref{eq13f}). Note that, due to p.b.c., the
FPU-$\beta$ chain in its \emph{equilibrium} state is invariant
under translation by the interparticle distance $a$, while in its
\emph{vibrational} states it is invariant under a translation by
$ma$. Moreover, NNMs (\ref{eq13a})~-- (\ref{eq13f}) do not differ
only by translational symmetry ($ma$), but also by some additional
symmetry transformations which are discussed later in this paper.

P2) Each of the NNMs (\ref{eq13a})~-- (\ref{eq13f}) depends on
\emph{only one} function $x(t)$ and, therefore, describes a single
\emph{one-dimensional} dynamical regime. This function satisfies
Duffing's equation (\ref{eq6}), with $\omega^2$ and $\gamma$ that
differ according to the displacement pattern of each NNM.

P3) Every NNM (\ref{eq13b})~-- (\ref{eq13f}) possesses some
"dynamical domain" of its own. For example, consider the mode
(\ref{eq13b}): It is easy to check that cyclic permutations of
each primitive cell in its displacement pattern produces
\emph{other} modes: $x(t)\{1, 0, -1,| 1, 0, -1,| ..., |1, 0, -1\}
\ \rightarrow x(t)\{0, -1, 1,| 0, -1, 1,| ..., |0, -1, 1,\} \
\rightarrow x(t)\{-1, 1, 0,| -1, 1, 0,| ..., |-1, 1, 0,\}$, which
differ from each other only by the position of their
\emph{stationary} particles.

As a consequence, all dynamical properties of these NNMs turn out
to be \emph{equivalent}. In particular, as we shall show in detail
in section III, they possess the same stability regions. Because
of this reason, we only need to study one representative of the
set of equivalent NNMs. Such sets are called "dynamical domains",
borrowing the term from the theory of phase transitions in
crystals.

Many aspects of existence and stability of NNMs have been discussed
in the literature (see, for example, \cite{FPU1, FPU2, Rink, Ruffo,
Shinohara, Shinohara2003, Yoshimura, LeoLeo, Greeks, Bud, Sand,
Kosevich}. What is important at this point is to pose certain
fundamental questions concerning such NNMs, which we shall proceed
to answer in this review to the best of our ability:

Q1) Is the list (\ref{eq13a})~-- (\ref{eq13f}) of NNMs for the
FPU-$\beta$ chain complete?

Indeed, at first sight, it seems that many other NNMs can exist,
for example, modes whose multiplicity number $m$ is different from
those listed above.

Q2) What kind of NNMs exist in nonlinear chains with different
interactions than those of the FPU-$\beta$ chain?

In fact, in most papers (see \cite{Ruffo, Shinohara, Shinohara2003,
Yoshimura, Kosevich}) the NNMs listed as (\ref{eq13a})~--
(\ref{eq13f}) above have been discussed by analyzing dynamical
equations which are only connected with the FPU-$\beta$
interparticle interaction.

Q3) Do there exist NNMs for Hamiltonian systems which are more
complicated than the monoatomic chains? For example, one can pose
this question for diatomic nonlinear chains (with particles having
alternating masses), 2--dimensional (2D) lattices, or 3D crystal
structures.

Q4) Finally, there is a more subtle issue: How can one construct
exact \emph{multidimensional} dynamical regimes in nonlinear
$N$-particle Hamiltonian systems and study their stability by
locating the unstable manifolds of the corresponding quasiperiodic
orbits?

Note that the NNMs discussed so far represent exact
\emph{one-dimensional} regimes, because they are fully determined
by only one time-dependent function. As we shall discover in
subsequent sections, there exist very interesting dynamical
regimes of regular motion depending on more than one frequency and
characterized by families of quasiperiodic functions, forming
$s$--dimensional tori, with $s\geq2$.

Such exact regimes of nonlinear mechanical systems, identified by
discrete symmetry, are called \emph{bushes} of NNMs. For example,
the following two-dimensional bush exists in the FPU-$\beta$ chain
with $N \ mod \ 6 = 0$:
\begin{equation}\label{eqq20}
    \vec{X}(t)=\{0, x_1(t), x_2(t), 0, -x_2(t), -x_1(t)| ... |0, x_1(t), x_2(t), 0, -x_2(t), -x_1(t)|\}
\end{equation}
This bush, with multiplicity number $m=6$, is determined by two
functions, $x_1(t)$ and $x_2(t)$, satisfying a system of two
nonlinear autonomous ODEs.

In contrast to \emph{one-dimensional} bushes,
\emph{multidimensional} bushes also exist, representing exact
\emph{quasiperiodic} dynamical regimes. In the framework of the
bush theory, the problem of finding exact nonlinear regimes in a
physical system with discrete symmetry can be solved
\emph{without} any information about interparticle interactions.
However, the explicit form of the bush dynamical equations
essentially depends on the interparticle interactions in the
considered system.

Actually, the concept of bushes of NNMs and the group theoretical
methods used for their construction constitute an extension of the
theory of complete condensates developed in the papers
\cite{ActaCryst, StatSol, ComputMath} studying phase transitions
in solids. Note that the concept of bushes of NNMs does not apply
only to vibrations of mechanical systems, since one may consider
bushes of \emph{any other physical nature}, for example, bushes of
spin modes.

Thus, in sections II, III of this review, we attempt to show that
the theory of bushes gives definite and quite general answers to
the above questions Q1--Q3 in nonlinear systems with discrete
symmetries. We then proceed to answer question Q4, using the
recent analytical and numerical results described in section III.

More specifically, we demonstrate that it is possible to use
Poincar\'{e}-Linstedt series expansions to construct exact
quasiperiodic solutions with $s\geq 2$ frequencies, which belong
to the lower part of the linear normal mode spectrum, with
$q=1,2,3,...$. We are thus able to establish the important
property that the energies $E_q$ excited by these
low--dimensional, so--called $q$--tori, are exponentially
localized in $q$--space \cite{qtoriCBE}. The relevance of this
fact becomes apparent when we study the well-known phenomenon of
FPU recurrences, which are remarkably persistent in FPU chains
(with fixed or periodic boundary conditions), at low values of the
total energy $E$.

We also study the stability of these $q$--tori as accurately as
possible, using methods that do not rely on Floquet theory and are
quite different than those adopted in section III. In particular,
we make use of the novel method of the GALI$_k$ indicators,
$k=2,...,2N$, which distinguish rapidly and efficiently between
chaotic and quasiperiodic orbits \cite{SBAPhysD, SBAEPJ}. As
predicted theoretically and verified extensively by numerical
evidence, in the chaotic case all the GALIs decay exponentially,
while in the quasiperiodic the first $s-1$ GALIs are nearly
constant (indicating the dimension $s$ of the associated torus),
while all others go to zero by power laws. Thus, by marking the
total energy $E$ at which the GALI$_k$, with the highest $k$,
start to decrease exponentially, it is possible to determine the
destabilization threshold at which the $q$--tori are destroyed,
implying the breakdown of FPU recurrences and the eventual
equipartition of energy among all modes.

\section{Bushes of NNMs in configuration space}

The theory of bushes of nonlinear normal modes was originally
developed in \cite{DAI, DAI2, Chechin-Sakhnenko}, while its
detailed description can be found in \cite{Columbus2007}.
Low-dimensional bushes in mechanical systems with various kinds of
symmetry and structures were studied in \cite{DAI, DAI2,
Chechin-Sakhnenko, Chechin-Stokes, Gnezdilov, FPU1}. In
particular, all possible symmetry-determined NNMs, representing
one-dimensional bushes, for all $N$-particle mechanical systems
with any of 230 \emph{space groups} were found in
\cite{Chechin-Stokes}.

The important problem of bush \emph{stability} was discussed in
\cite{Gnezdilov, FPU1, FPU2}, the first two of these papers being
devoted to the vibrational bushes in Fermi-Pasta-Ulam chains. Some
theorems about bush structures and the \emph{normal forms} of
their dynamical equations were presented in
\cite{Chechin-Sakhnenko}. The general group theoretical method for
the simplification of the bush stability analysis was presented in
\cite{Chechin-Zhukov}.

Note that dynamical objects equivalent to the bushes of NNMs have
also been discussed for \emph{monoatomic chains} by several other
authors \cite{Rink, Ruffo}. Let us emphasize, however, that the
group theoretical methods developed in \cite{DAI, DAI2,
Chechin-Sakhnenko, Chechin-Zhukov} can be applied equally well not
only to monoatomic chains (as illustrated in \cite{FPU1, FPU2,
Chechin-Zhukov}), but also to \emph{any} other physical system
with discrete symmetry groups (see \cite{Columbus2007}).

In this section, we shall demonstrate the power of bush theory
methods using three simple mechanical models: A nonlinear chain of
particles moving in one-dimension under p.b.c., a "square
molecule" whose particles move in a two--dimensional plane and an
octahedral structure of particles moving in three dimensions. For
these nonlinear dynamical systems, we will consider different
\emph{vibrational} regimes and thus deal with bushes of
\emph{vibrational} modes.

Now, in order to properly define bushes of NNMs, one must first
determine the \emph{parent} symmetry group of the equations
describing the vibrations of a given mechanical system.

\subsection{Parent symmetry group}

A parent symmetry group consists of all transformations which
leave the system of dynamical equations \emph{invariant}. In
general, such transformations can include space and time variables
as well as parameters of the system. Let us consider, for example,
the dynamical equations (\ref{eq1}) for the FPU-$\beta$ chain with
p.b.c. (\ref{eq3}) and an even number of particles. It is easy to
guess some space transformations which do not change
Eqs.~(\ref{eq1}, \ref{eq3}) by considering the \emph{equilibrium
state} of our nonlinear chain. Firstly, this chain is invariant
under the action of the operator $\hat a$ which shifts it by the
lattice spacing $a$. This operator generates the translational
group
\begin{equation}
T=\{\hat e,\hat a,\hat a^2,\dotsc,\hat a^{N-1}\},\quad\hat
a^N=\hat e, \label{eq23}
\end{equation}
where $\hat e$ is the identity element and $N$ is the order of the
cyclic group~$T$. The operator $\hat a$ induces the cyclic
permutation of all particles of the chain and, therefore, acts on
the configuration vector $\vec X(t)$ as follows:
\begin{equation}
\hat{a}\vec X(t)\equiv
\hat{a}\{x_1(t),x_2(t),\dotsc,x_{N-1}(t),x_N(t)\}=
\{x_N(t),x_1(t),x_2(t),\dotsc,x_{N-1}(t)\}.
\end{equation}

Secondly, the symmetry group of the monoatomic chain contains the
\textit{inversion} $\hat\i$, with respect to the center of the
chain, which acts on the vector $\vec X(t)$ in the following way:
\begin{eqnarray}\label{e30}
\hat\i\vec X(t)\equiv
\hat\i\{x_1(t),x_2(t),\dotsc,x_{N-1}(t),x_N(t)\}=\nonumber\\
\{-x_N(t),-x_{N-1}(t),\dotsc,-x_2(t),-x_1(t)\}.
\end{eqnarray}

The complete set of symmetry transformations includes also all
products $\hat a^k\hat\i$ of the pure translations $\hat a^k$
($k=1,2,\dotsc,N-1$) with the inversion $\hat\i$ and forms the
so-called dihedral group $D$ which can be written as a direct sum
of the two \textit{cosets} $T$ and $T\cdot\hat\i$:
\begin{equation}
D=T\oplus T\cdot\hat\i. \label{eq500}
\end{equation}

This is a non-Abelian group induced by two \textit{generators}
($\hat a$ and $\hat\i$) with the following \textit{generating
relations}
\begin{equation}
\hat a^N=\hat e,\quad\hat\i^2=\hat e,\quad\hat\i\hat a=\hat
a^{-1}\hat\i. \label{eq170a}
\end{equation}

When applied to Eqs.~(\ref{eq1}), operators $\hat{a}$ and
$\hat{i}$ induce the following changes of variables:
\begin{equation}\label{eq32}
\begin{array}{l}
    \hat{a}: \ x_1(t)  \rightarrow  x_2(t),  x_2(t)  \rightarrow  x_3(t), ..., x_{N-1}(t)  \rightarrow  x_N(t), x_N(t)  \rightarrow  x_1(t);\\
    \hat{i}: \ x_1(t)  \leftrightarrow  -x_N(t),  x_2(t)  \leftrightarrow  -x_{N-1}(t), x_3(t)  \leftrightarrow  -x_{N-2}(t), ... .\\
\end{array}
\end{equation}

It is straightforward to check that upon acting on (\ref{eq1})
with transformation (\ref{eq32}) the system is transformed to an
\emph{equivalent} form. Moreover, since Eqs.~(\ref{eq1}) are
invariant under the actions of $\hat{a}$ and $\hat{i}$, they are
also invariant with respect to all products of these two operators
and, therefore, the dihedral group $D$ is indeed a symmetry group
of equations (\ref{eq1}) for a monoatomic chain with
\emph{arbitrary} interparticle interactions. As a consequence, the
dihedral group $D$ can be considered as a parent symmetry group
for \emph{all} monoatomic nonlinear chains as, for example, the
FPU-$\alpha$ chain, whose interparticle interactions are
characterized by the force
\begin{equation}\label{eq35}
    f(\Delta x)=\Delta x+\alpha(\Delta x)^2.
\end{equation}

Of course, since in the case of the FPU-$\beta$ chain the force
$f(\Delta x)=\Delta x+\beta(\Delta x)^3$ is an \emph{odd} function
of its argument $\Delta x$, the FPU-$\beta$ chain possesses a
\emph{higher} symmetry group than the FPU-$\alpha$ chain.

Indeed, let us introduce the operator $\hat{u}$ which changes the
signs of all atomic displacements without their transposition:
\begin{equation}\label{eq36}
    \hat{u}\vec{X}\equiv \hat{u}\{x_1(t), x_2(t), ..., x_{N-1}(t), x_N(t)\}=\{-x_1(t), -x_2(t), ..., -x_{N-1}(t), -x_N(t)\}.
\end{equation}
It can easily be checked that the operator $\hat{u}$ generates a
transformation of all the variables $x_i(t), \ i=1..N$ in
Eqs.~(\ref{eq1})--(\ref{eq3}), which leads to an equivalent form
of these equations. Therefore, the operator $\hat{u}$ and all its
products with elements of the dihedral group $D$ belong to the
full symmetry group of the FPU-$\beta$ chain.

Clearly, the operator $\hat{u}$ \emph{commutes} with all the
elements of the dihedral group $D$ and we can consider the group
\begin{equation}\label{e32}
    G=D\oplus D\cdot\hat{u}
\end{equation}
as the parent symmetry group of the FPU-$\beta$ chain. The group
$G$ contains \emph{twice as many} elements as the dihedral group
$D$ and, therefore, possesses a greater number of subgroups.

On the other hand, every subgroup of the parent group generates a
certain bush of NNMs. Therefore, there exists a greater number of
bushes for the FPU-$\beta$ chain (and for any other chain with
\emph{odd} force of interparticle interactions) compared with
those of the FPU-$\alpha$ chain (and for all arbitrary nonlinear
monoatomic chains, as well).

Finally, let us note that it is sufficient for our purposes to
define any symmetry group by listing only its \emph{generators}
which we denote by square brackets, for example, we write
    $T[\hat{a}], \ D[\hat{a}, \hat{i}], \ G[\hat{a}, \hat{i}, \hat{u}]$.

\subsection{Subgroups of the parent group and bushes of NNMs}

Let us consider now a specific configuration vector
$\vec{X}^{(j)}(t)$ [see Eq.~(\ref{eq4})], which determines a
displacement pattern at time $t$, and let us act on it
successively by the operators $\hat{g}$ that correspond to all the
elements of a parent group $G_0$. The full set $G_j$ of elements
of the group $G_0$, under which $\vec{X}^{(j)}(t)$ turns out to be
\emph{invariant}, generates a certain \emph{subgroup}  of the
group $G_0$ ($G_j\subset G_0$) [Note that this subgroup can be
\emph{trivial}, i.e. it can consist of only one symmetry element
$\hat{e}$. In this case, a chosen configuration vector
$\vec{X}^{(j)}(t)$ turns out to be of general form, since there
are no connections between the displacements of different
particles of the chain]. We then call $\vec{X}^{(j)}(t)$ invariant
under the action of the subgroup $G_j$ of the parent group $G_0$
and use it to determine the bush of nonlinear normal modes
corresponding to the subgroup $G_j$ of the group $G_0$.

Therefore, in the framework of the above approach, one must find
all the subgroups of the parent group $G_0$ to obtain all the
bushes of NNMs of different types,. This can be done by standard
group theoretical methods. In \cite{FPU2} a simple
crystallographic technique was developed for singling out all the
subgroups of the parent group of any monoatomic chain, following
the approach of a more general method \cite{ActaCryst, StatSol,
ComputMath}. We shall not discuss here how to find subgroups of
the parent groups; rather, we will demonstrate how one can obtain
bushes of NNMs if the subgroups are already known.

Let us consider the subgroups $G_j$ of the dihedral group $D$.
Each group $G_j$ contains its own \emph{translational subgroup}
$T_j\subset T$, where $T$ is the above discussed full
translational group~(\ref{eq23}). If $N$ is divisible by 4 (for
example, we consider below the case $N=12$) there exists a
subgroup $T_4=[\hat a^4]$ of the group $T=[\hat a]$. Note that, in
square brackets, we write down only the generators of the
considered group, while the complete set of group elements is
written in curly brackets, as, for example, in Eq.~(\ref{eq23}).

If a vibrational state of the chain possesses the symmetry group
$T_4=[\hat a^4]\equiv\{\hat e,\hat a^4,\hat a^8,\dotsc,\hat
a^{N-4}\}$, the displacements of the atoms, which are at a
distance $4a$ from each other in the equilibrium state, turn out
to be \emph{equal}, since the operator $\hat{a}^4$ leaves the
vector $\vec{X}(t)$ invariant. For example, for the case $N=12$,
the operator $\hat{a}^4$ permutes the coordinates of
$\vec{X}=(x_1,x_2,...,x_{12})$ taken in quadruplets $(x_{i},
x_{i+1}, x_{i+2}, x_{i+3}), i=1,5,9$, while from equation
$\hat{a}^4\vec{X}(t)=\vec{X}(t)$ one deduces $x_i=x_{i+4},
i=1,2,3,4$. Thus, the vector $\vec{X}(t)$ contains 3 times the
quadruplets $x_1,x_2,x_3,x_4$, where $x_i(t)$ ($i=1,2,3,4$) are
arbitrary functions of time and can be written as follows
\begin{equation}
\vec X(t)=\{~x_1(t),x_2(t),x_3(t),x_4(t)~|~x_1(t),x_2(t),x_3(t),x_4(t)~|~x_1(t),%
x_2(t),x_3(t),x_4(t)~\}. \label{eq11fpu2}
\end{equation}
In other words, the complete set of atomic displacements can be
divided into $N/4$ (in our case, $N/4=3$) \emph{identical}
subsets, which are called ``extended primitive cells'' (EPC). In
the bush (\ref{eq11fpu2}), the EPC contains four atoms, and the
vibrational state of the whole chain is described by three such
EPC. Thus, the EPC for the vibrational state with the symmetry
group $T_4=[\hat a^4]$ has size equal to $4a$, which is four times
larger than the primitive cell of the chain in the equilibrium
state.

It is essential that some symmetry elements of the dihedral group
$D$ disappear as a result of the symmetry reduction $D=[\hat
a,\hat\i]\rightarrow T_4=[\hat a^4]$. There are four other
subgroups of the dihedral group $D$, corresponding to the same
translational subgroup $T_4=[\hat a^4]$:
\begin{equation}
[\hat a^4,\hat\i],\quad [\hat a^4,\hat a\hat\i],\quad [\hat
a^4,\hat a^2\hat\i],\quad [\hat a^4,\hat a^3\hat\i].
\label{eq50fpu2}
\end{equation}
Each of these subgroups possesses \textit{two generators}, namely
$\hat a^4$ and an inversion element $\hat a^k\hat\i$
($k=0,1,2,3$). Note that the $\hat a^k\hat\i$ differ from each
other by the \textit{position} of the center of inversion.

Subgroups $[\hat a^4,\hat a^k\hat\i]$ with $k>3$ are equivalent to
those listed in~(\ref{eq50fpu2}), since the second generator $\hat
a^k\hat\i$ can be multiplied from the left by $\hat a^{-4}$,
representing the inverse element with respect to the generator
$\hat a^4$. Thus, there exist only five subgroups of the dihedral
group (with $N\mod 4=0$) constructed on the basis of the
translational group $T_4=[\hat a^4]$, namely, this group and the
four groups from the list~(\ref{eq50fpu2}).

Now, let us examine the bushes corresponding to the
subgroups~(\ref{eq50fpu2}). The subgroup $[\hat a^4,\hat\i]$
consists of the following six elements:
\begin{equation}
\hat e,\hat a^4,\hat a^8,\hat\i,\hat a^4\hat\i,\hat
a^8\hat\i\equiv\hat\i\hat a^4. \label{eq51a}
\end{equation}

The invariance of $\vec{X}(t)$ with respect to this group can be
written as follows:
\begin{equation}\label{eq40}
    \hat{a}^4\vec{X}(t)=\vec{X}(t), \ \ \
    \hat{i}\vec{X}(t)=\vec{X}(t),
\end{equation}
while the invariance of the vector $\vec{X}(t)$ under the action
of the group generators $[\hat a^4$ and $\hat\i]$ guarantees its
invariance under all elements of this group.

As explained above, the equation $\hat{a}^4\vec{X}(t)=\vec{X}(t)$
is satisfied by the vector $\vec{X}(t)$, see (\ref{eq11fpu2}),
while $\hat{i}\vec{X}(t)=\vec{X}(t)$ also holds, from which we
obtain the following relations for each of the three EPS:\
$x_1(t)=-x_4(t)$, \ $x_2(t)=-x_3(t)$.

Therefore, for $N=12$, the invariant vector $\vec{X}(t)$ of the
group [$\hat{a}^4, \hat{i}$] can be written in the form
\begin{equation}\label{eeq45}
    \vec{X}(t)=\{x_1(t), x_2(t), -x_2(t), -x_1(t)\bigl| x_1(t), x_2(t), -x_2(t), -x_1(t)\bigl| x_1(t), x_2(t), -x_2(t),
    -x_1(t)\},
\end{equation}
where $x_1(t)$ and $x_2(t)$ are arbitrary functions of time.

Thus, the subgroup [$\hat{a}^4, \hat{i}$] of the dihedral group
$D$ generates a \emph{two-dimensional} bush of NNMs. The explicit
form of the differential equations governing the two variables
$x_1(t)$ and $x_2(t)$ can now be obtained by substitution of the
ansatz (\ref{eeq45}) into the FPU-$\beta$ dynamical equations
(\ref{eq1}-\ref{eq3}). We shall, hereafter, denote the bush
(\ref{eeq45}) in the form
\begin{equation}\label{eq46a}
    B[\hat{a}^4, \hat{i}]=\bigl|x_1, x_2, -x_2, -x_1\bigl|,
\end{equation}
showing the atomic displacements \emph{in only one EPS} and
omitting the argument $t$ of variables $x_1(t)$, $x_2(t)$.

Proceeding in a similar manner, we obtain bushes of NNMs for three
other groups from the list (\ref{eq50fpu2}):
\begin{equation}\label{eq46b}
    B[\hat{a}^4, \hat{a}\hat{i}]=\bigl|0, x, 0, -x\bigl|,
\end{equation}
\begin{equation}\label{eq46c}
    B[\hat{a}^4, \hat{a}^2\hat{i}]=\bigl|x_1, -x_1, x_2, -x_2\bigl|,
\end{equation}
\begin{equation}\label{eq46d}
    B[\hat{a}^4, \hat{a}^3\hat{i}]=\bigl|x, 0, -x, 0\bigl|,
\end{equation}

Let us comment on these results: The bushes $B[\hat{a}^4,
\hat{a}\hat{i}]$ and $B[\hat{a}^4, \hat{a}^3\hat{i}]$ turn out to
be one-dimensional in the sense that they represent simple
periodic solutions corresponding to NNMs. These modes are, in
fact, \emph{equivalent} since they differ from each other only by
the numbering of the particles on the chain. In group theoretical
terms, this equivalence is a consequence of the fact that
$G_1=[\hat{a}^4, \hat{a}\hat{i}]$ and $G_2=[\hat{a}^4,
\hat{a}^3\hat{i}]$ prove to be conjugate \footnote{Two subgroups
$G_1$ and $G_2$ of the same group $G$ are called conjugate to each
other, if there exists at least one element $g_0\in G$ which
converts $G_1$ into $G_2$ via the transformation $G_2=g_0^{-1} G_1
g_0$. [note that in the case of bushes (\ref{eq46b}) and
(\ref{eq46d}), \ $g_0=\hat{a}$].} subgroups in the parent group
$D=[\hat{a}, \hat{i}]$. Both two-dimensional bushes (\ref{eq46a})
and (\ref{eq46c}) are also equivalent to each other for the same
reason.

All displacement patterns (\ref{eq46a}-\ref{eq46d}) are particular
cases of the general pattern of the bush $B[\hat{a}^4]=\bigl|x_1,
x_2, x_3, x_4\bigl|$. One may, therefore, ask what will happen
during time evolution of \emph{other} particular cases of
$B[\hat{a}^4]$. For example, if one chooses for the FPU
$\beta$-chain with $N\mod\ 4=0$ an initial one-dimensional pattern
of the form $ \vec{X}(0)=\{3x_0, -x_0, -x_0\bigl|3x_0, -x_0,
-x_0\bigl| ... \}$ or, for $N\mod\ 5=0$, an initial state of the
type $\vec{X}(0)=\{-x_0, -x_0, 4x_0, -x_0, -x_0\bigl|-x_0, -x_0,
4x_0, -x_0, -x_0\bigl| ... \}$ one easily finds that there are
\emph{no subgroups} of the parent group $D$ which produce these
displacement patterns. As a consequence, when solving the
equations of motion of the FPU-$\beta$ chain in these cases to
obtain $\vec{X}(t)$, the above structures will \emph{not be
conserved} in time.

More generally, we conclude that for \emph{sufficiently large} EPS
there are not enough symmetry elements to give rise to NNMs, since
the bushes of the corresponding displacement patterns are
multidimensional. For this reason, there exists only a very
specific number of bushes for any fixed dimension beyond the NNMs!

\subsection{Bushes of NNMs for arbitrary interparticle interactions}

As was proved in \cite{FPU2} for any nonlinear monoatomic chain
with \emph{arbitrary} interparticle interactions, there exist
\emph{only three} symmetry-determined NNMs:
\begin{equation}\label{eeq60}
\begin{array}{l}
B[\hat{a}^2, \hat{i}]=\Bigl|x, -x\Bigl|,
(\pi\mbox{-mode}),  \ \ \ \ \ B[\hat{a}^3, \hat{i}]=\Bigl|x, 0, -x\Bigl|,  \ \ \ \ \ B[\hat{a}^4, \hat{a}\hat{i}]=\Bigl|0, x, 0, -x\Bigl|.\\
\end{array}
\end{equation}
(each NNM corresponds to a certain subgroup of the \emph{dihedral}
group $D$). These modes can easily be excited in the FPU-$\alpha$
and FPU-$\beta$ chains, if we solve numerically the corresponding
dynamical equations with appropriate initial conditions.

On the other hand, it is already known that the FPU-$\beta$ chain
possesses a \emph{higher} symmetry group $G$, since changing signs
of all the displacements by the operator $\hat{u}$ [see
Eq.~(\ref{eq36})] leaves the FPU-$\beta$ dynamical equations
invariant. Therefore, all subgroups of the group $D$ also turn out
to be subgroups of the group $G$, while $G$ possesses some
\emph{additional} subgroups whose elements can contain the
operator $\hat{u}$.

As a consequence, there exist more bushes of NNMs for the
FPU-$\beta$ chain (and for any other chain with \emph{odd} force
of interparticle interactions) than for the FPU-$\alpha$ chain (or
any other chain whose interparticle interactions is not described
by odd function $f(\Delta x)$). These additional bushes were
presented in \cite{FPU2} in the following form:
\begin{equation}\label{eq61}
\begin{array}{l}
B[\hat{a}^3, \hat{i}\hat{u}]=\Bigl|x, -2x, x\Bigl|,\ \ \
B[\hat{a}^2\hat{u}, \hat{i}]=\Bigl|x, x, -x, -x\Bigl|,\ \ \
B[\hat{a}^3\hat{u}, \hat{a}\hat{i}]=\Bigl|0, x, x, 0, -x,
-x\Bigl|.\ \
\end{array}
\end{equation}
and were also obtained in \cite{Rink} by a different method.

\subsection{Bushes of NNMs for a square molecule}

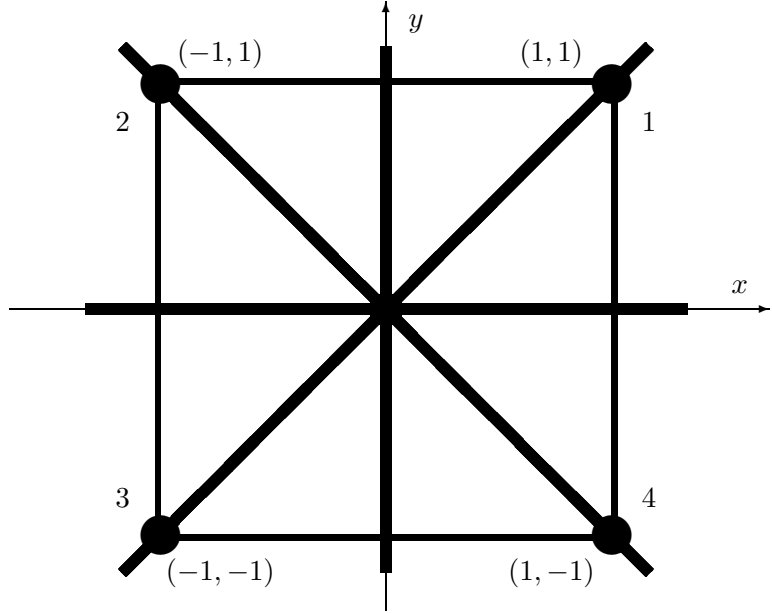
\begin{figure}[h]
\centering \unitlength 1mm \linethickness{0.4pt}
\begin{picture}(100,80)
\put(0,40){\line(1,0){100}} \put(100,40){\vector(1,0){1}}
\put(50,0){\line(0,1){80}} \put(50,80){\vector(0,1){1}}
\put(20,70){\circle*{5.2}} \put(80,70){\circle*{5.2}}
\put(20,10){\circle*{5.2}} \put(80,10){\circle*{5.2}}
\linethickness{2pt} \put(20,10){\framebox(60,60)[cc]{}}
\linethickness{4pt} \put(50,5){\line(0,1){70}}
\put(10,40){\line(1,0){80}}
\multiput(14.5,5.5)(0.01,-0.01){100}{\line(1,1){70}}
\multiput(14.5,74.5)(0.01,0.01){100}{\line(1,-1){70}}
\put(48,38){\rule{4\unitlength}{4\unitlength}}
\put(85,65){\makebox(0,0)[cc]{$1$}}
\put(15,65){\makebox(0,0)[cc]{$2$}}
\put(15,15){\makebox(0,0)[cc]{$3$}}
\put(85,15){\makebox(0,0)[cc]{$4$}}
\put(97,43){\makebox(0,0)[cc]{$x$}}
\put(54,78){\makebox(0,0)[cc]{$y$}}
\put(28,74){\makebox(0,0)[cc]{$(-1,1)$}}
\put(72,74){\makebox(0,0)[cc]{$(1,1)$}}
\put(28,5){\makebox(0,0)[cc]{$(-1,-1)$}}
\put(72,5){\makebox(0,0)[cc]{$(1,-1)$}}
\end{picture}
\caption{Mechanical model of a square molecule.\label{Fig1}}
\end{figure}

In our model, a square molecule is represented by a  mechanical
system whose equilibrium state is depicted in Fig.~\ref{Fig1}.
Four atoms of this molecule are shown as filled circles at the
vertices of the square. The number of every atom and its $(x,y)$
coordinates are as given in the figure.

Let us suppose that atoms can oscillate about their equilibrium
positions only in the $(x,y)$ plane and, therefore, eight degrees
of freedom are needed to describe this mechanical system.
Furthermore, we will not consider any specific type of
interparticle interactions and treat bushes of NNMs as purely
geometrical objects. The equilibrium configurations of our
molecule possess the symmetry group denoted by $C_{4v}$ in the
Schoenflies notation, while, at equilibrium, the molecule depicted
in Fig.~\ref{Fig1} is invariant under the action of the following
transformations:

\begin{itemize}
\item
Rotations through the angles $0^\circ$, $90^\circ$, $180^\circ$,
$270^\circ$ about the $z$ axis orthogonal to the plane of
Fig.~\ref{Fig1} and passing through the center of the square. We
will denote these rotations by $g_1$, $g_2$, $g_3$, $g_4$,
respectively.

\item
Reflections in four mirror planes orthogonal to the plane of
Fig.~\ref{Fig1} and passing through the $z$ axis. (We depict these
mirror planes by bold lines). Two of them are ``coordinate''
planes ($g_5$, $g_7$) and the other two are ``diagonal'' planes
($g_6$, $g_8$).
\end{itemize}

Analytically, the above mentioned symmetry elements can be defined
as follows:
\begin{equation}
\label{eq20}
\begin{array}{ll}
g_1(x,y)=(x,y),\ \ g_2(x,y)=(-y,x), \ \ g_3(x,y)=(-x,-y), \ \
g_4(x,y)=(y,-x)\\ g_5(x,y)=(-x,y), \ \ g_6(x,y)=(-y,-x), \ \
g_7(x,y)=(x,-y), \ \ g_8(x,y)=(y,x)

\end{array}
\end{equation}

Thus, the symmetry group of the square molecule ($C_{4v}$)
contains 8 elements $g_1,\dots,g_8$, determined by
Eq.~(\ref{eq20}). This group is non-Abelian since, for example,
$g_2\cdot g_8=g_5$, while $g_8\cdot g_2=g_7$. According to
Lagrange's theorem, the order of any subgroup is a
\textit{divisor} of the order of the full group. Therefore, for
the case of the group $G=C_{4v}$ with the order $m=\|G\|=8$, there
exist only subgroups $G_j$ with order equal to $m=1,2,4,8$ given
by the following list:
\begin{equation}\label{eq77_c}
\begin{array}{l}
m=1: \ G_1=\{g_1\}=C_1\\
m=2: \ G_2=\{g_1,g_3\}=C_2\\
\phantom{ll=11:} \ G_3=\{g_1,g_5\} \mbox{ and } G'_3=\{g_1,g_7\}=C_s^c\\
\phantom{ll=11:} \ G_4=\{g_1,g_6\} \mbox{ and } G'_4=\{g_1,g_8\}=C_s^d\\
m=4: \ G_5=\{g_1,g_2,g_3,g_4\}=C_4\\
\phantom{ll=11:} \ G_6=\{g_1,g_3,g_5,g_7\}=C_{2v}^c\\
\phantom{ll=11:} \ G_7=\{g_1,g_3,g_6,g_8\}=C_{2v}^d\\
m=8: \ G_8=\{g_1,g_2,g_3,g_4,g_5,g_6,g_7,g_8\}=C_{4v}
\end{array}
\end{equation}

Note that throughout this paper, we use Schoenflies notation for
the point symmetry groups and distinguish between coordinate and
diagonal settings of some subgroups using the superscripts ``c''
and ``d'', respectively.

Let us suppose that the equilibrium state of our molecule is
\textit{stable} under arbitrary infinitesimal displacements of the
atoms in the $xy$-plane. Moreover, we will also assume that this
state is \textit{isolated} in the sense that within a finite
radius neighborhood around it there are no other equilibrium
states.

Next, we consider the planar vibrations, i.e.\ vibrations of the
molecule in the plane of its equilibrium configuration. Let us
excite a vibrational regime of our molecule by displacing the
atoms from their equilibrium positions in a specific manner. As a
result of such displacements, the \textit{initial configuration}
of the molecule will have a well defined symmetry which is
described by one of the \textit{subgroups} of the group $G_{4v}$
listed in (\ref{eq77_c}).

\begin{figure}
\centering
\begin{tabular}{|rlc|rlc|}
  \hline
  \vspace*{-2.5cm}& &\psfig{file=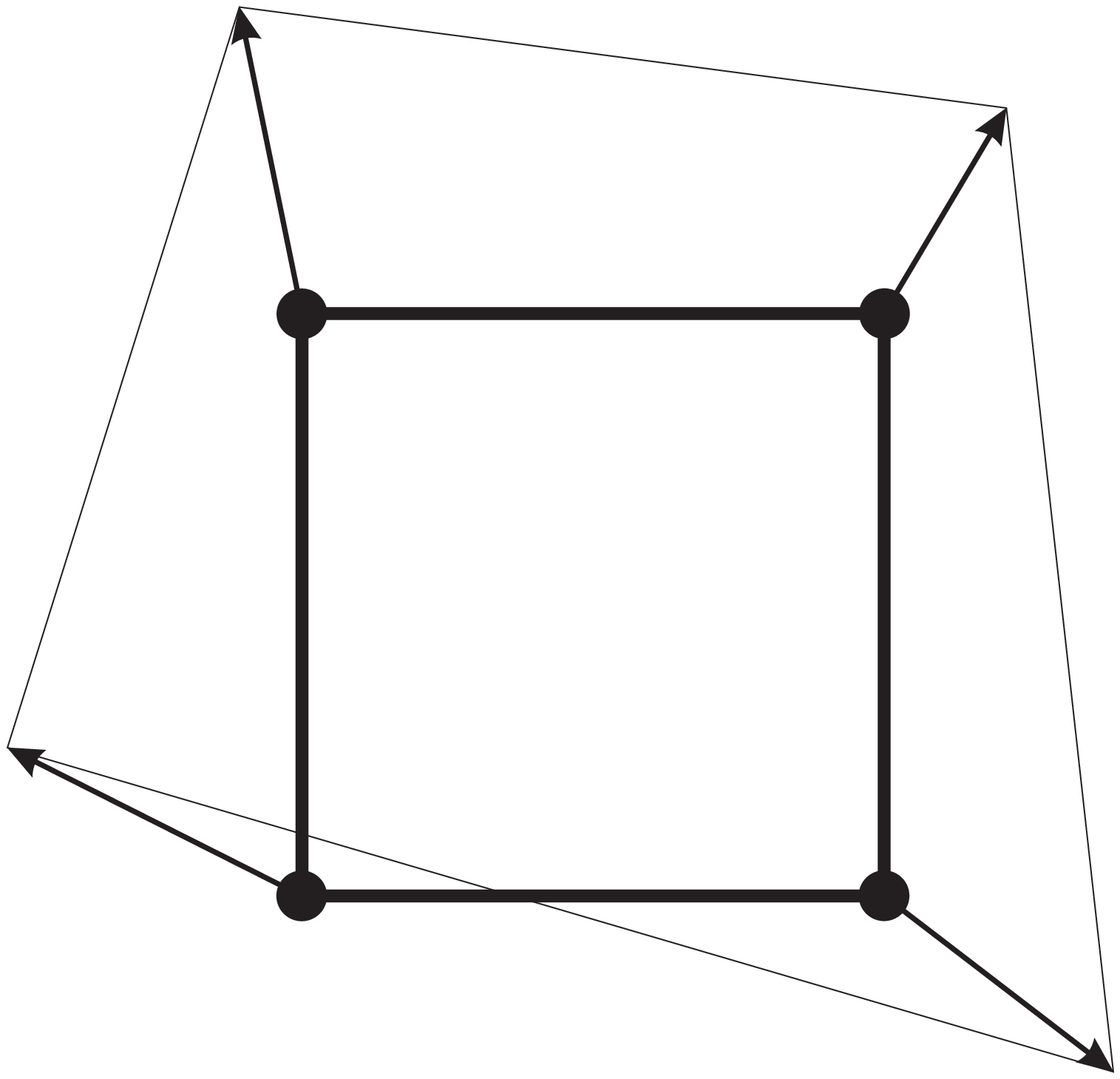,width=0.3\textwidth} &
                  & &\psfig{file=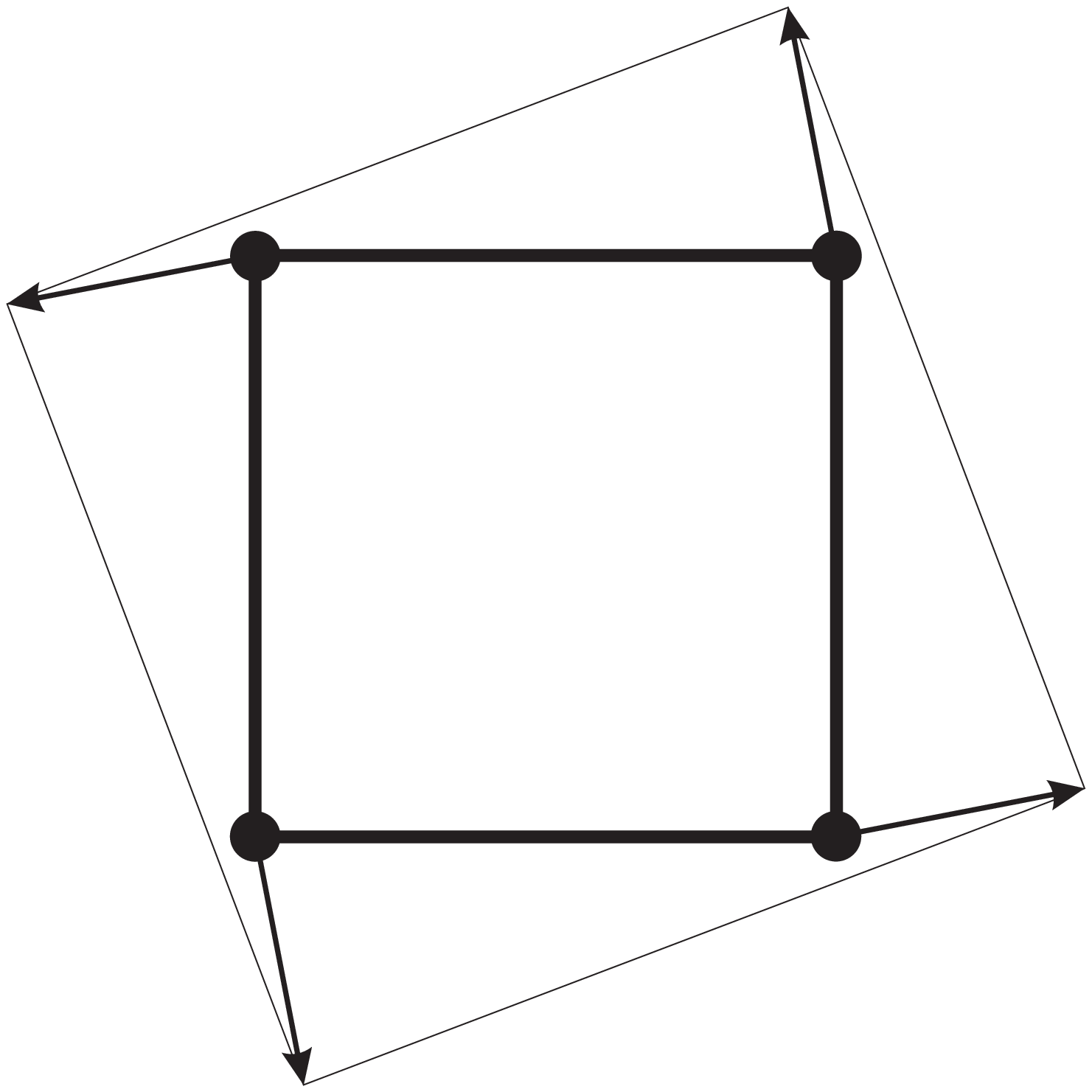,width=0.3\textwidth} \\
  \vspace*{1.8cm}1) & $C_1$ &  & 5) & $C_4$ & \\
  \hline
  \vspace*{-2.5cm}& &\psfig{file=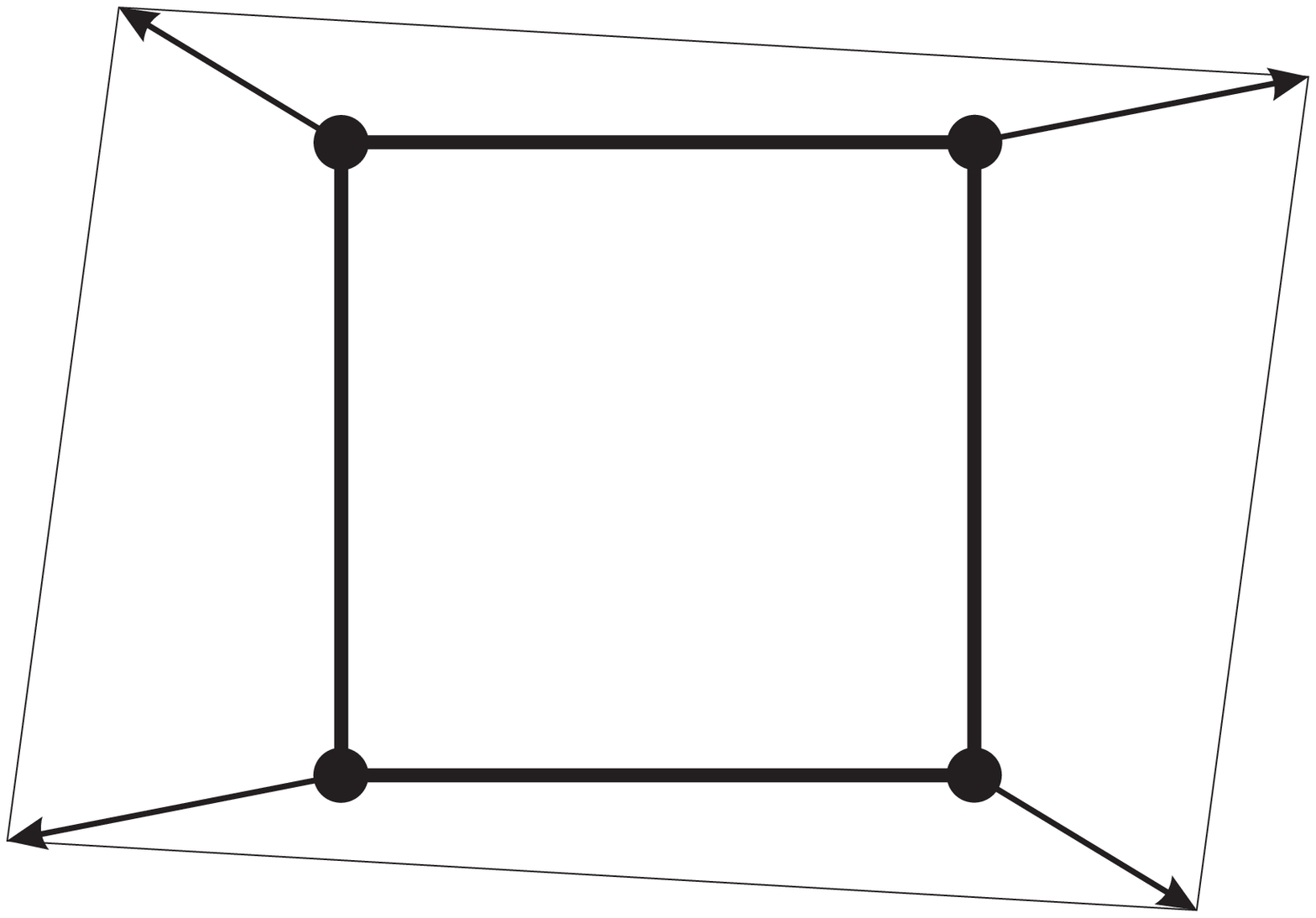,width=0.3\textwidth} &
                  & &\psfig{file=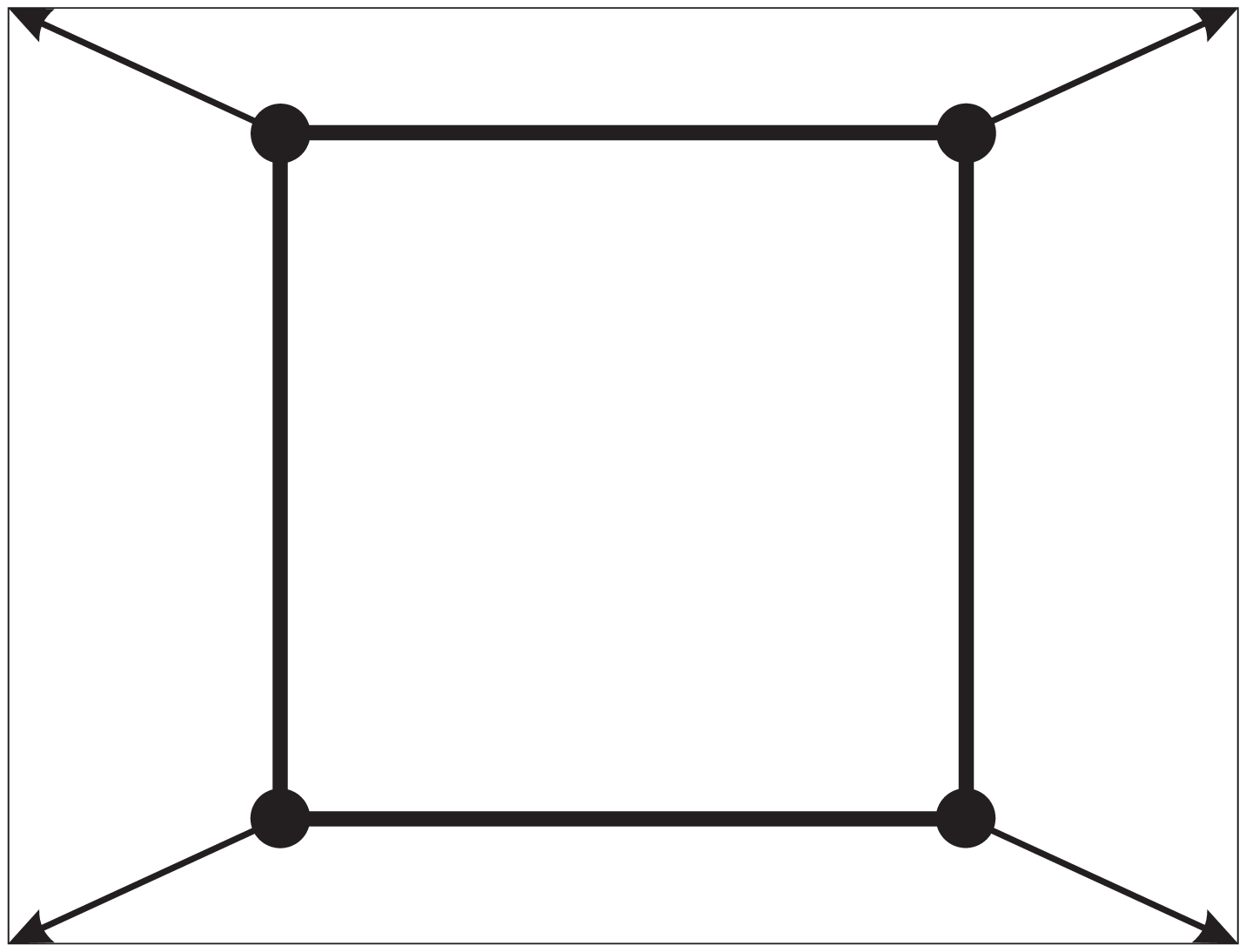,width=0.3\textwidth} \\
  \vspace*{1.8cm}2) & $C_2$ &  & 6) & $C_{2v}^c$ & \\
  \hline
  \vspace*{-2.5cm}& &\psfig{file=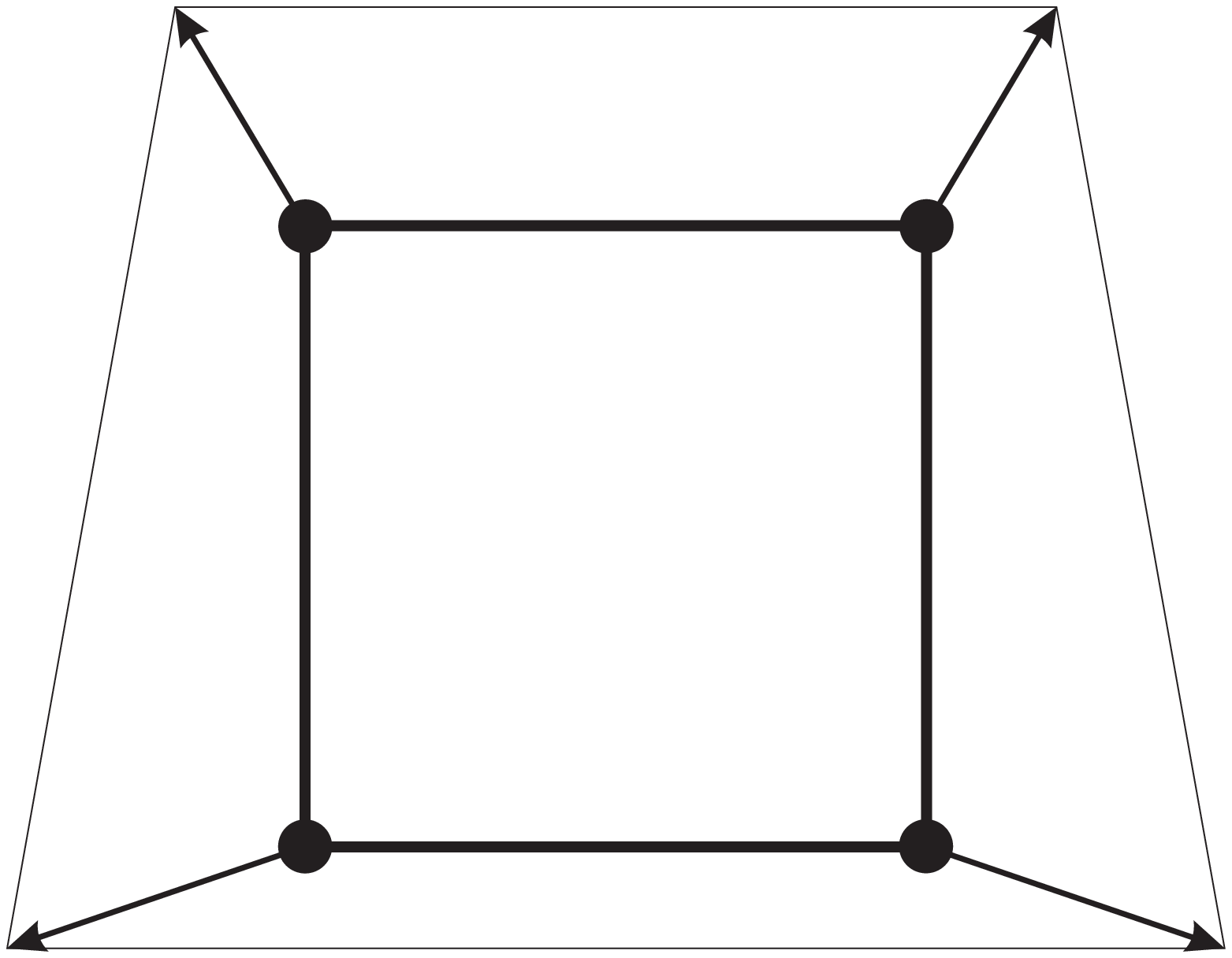,width=0.3\textwidth} &
                  & &\psfig{file=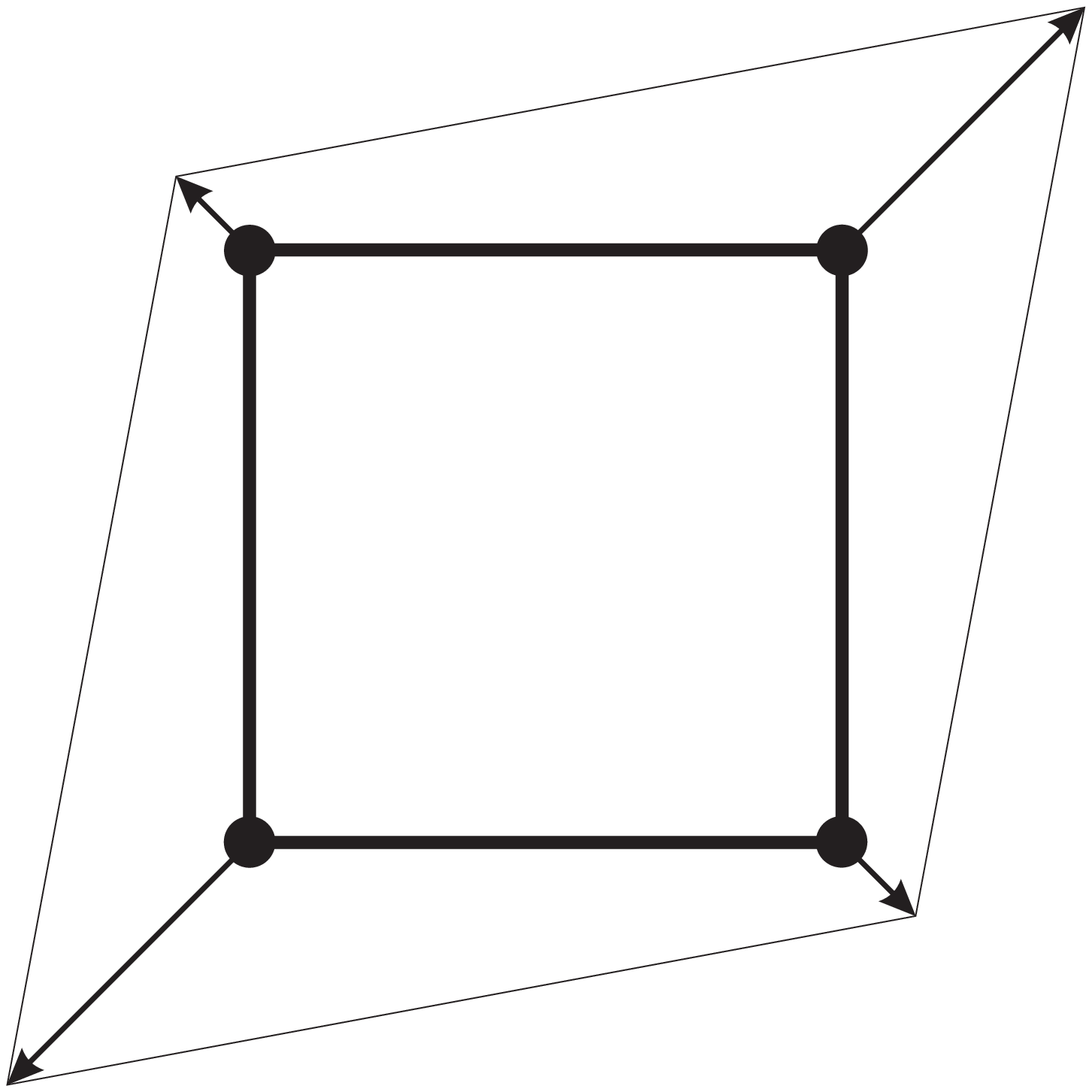,width=0.3\textwidth} \\
  \vspace*{1.8cm}3) & $C_s^c$ &  & 7) & $C_{2v}^d$ & \\
  \hline
  \vspace*{-2.5cm}& &\psfig{file=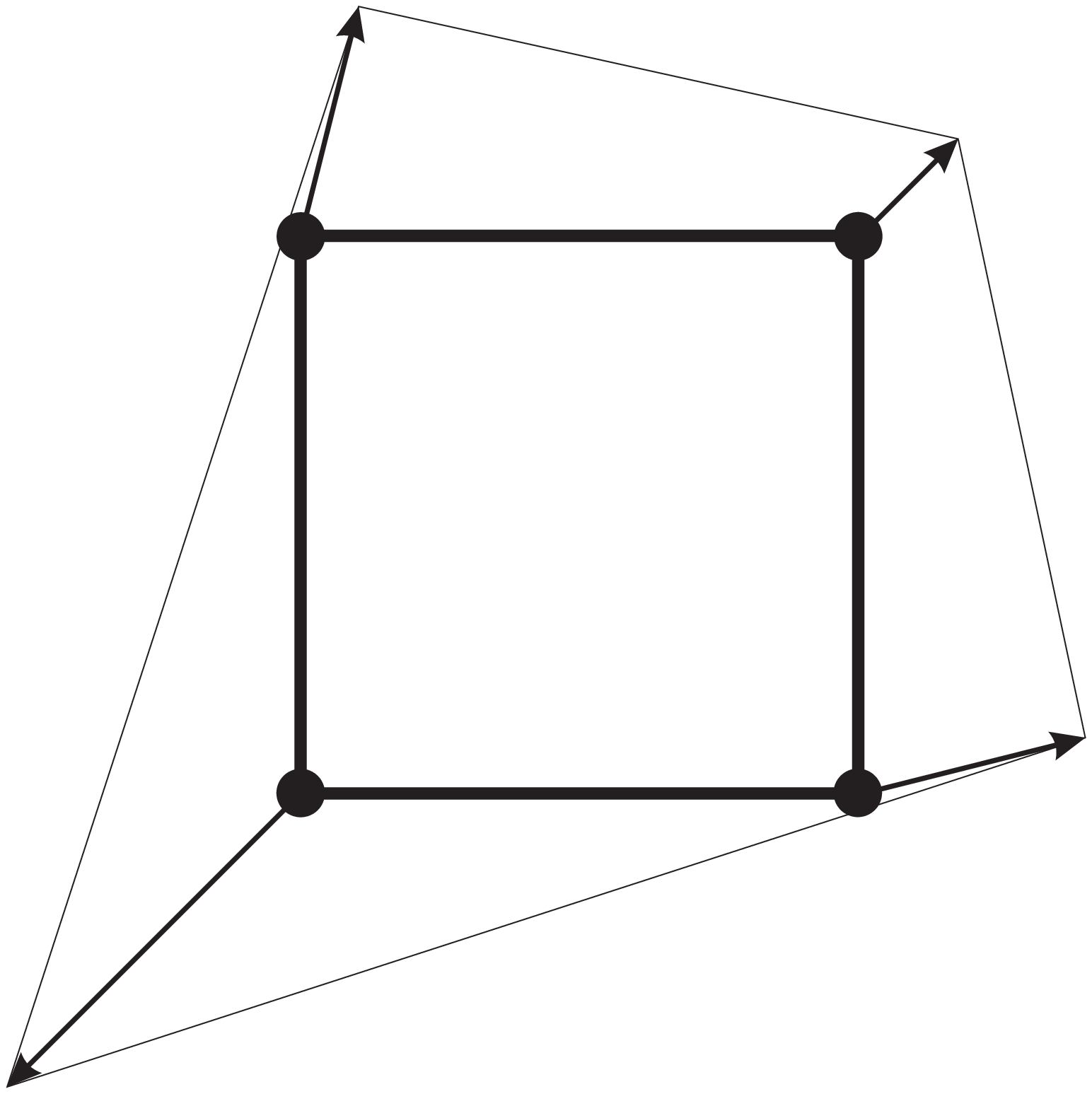,width=0.3\textwidth} &
                  & &\psfig{file=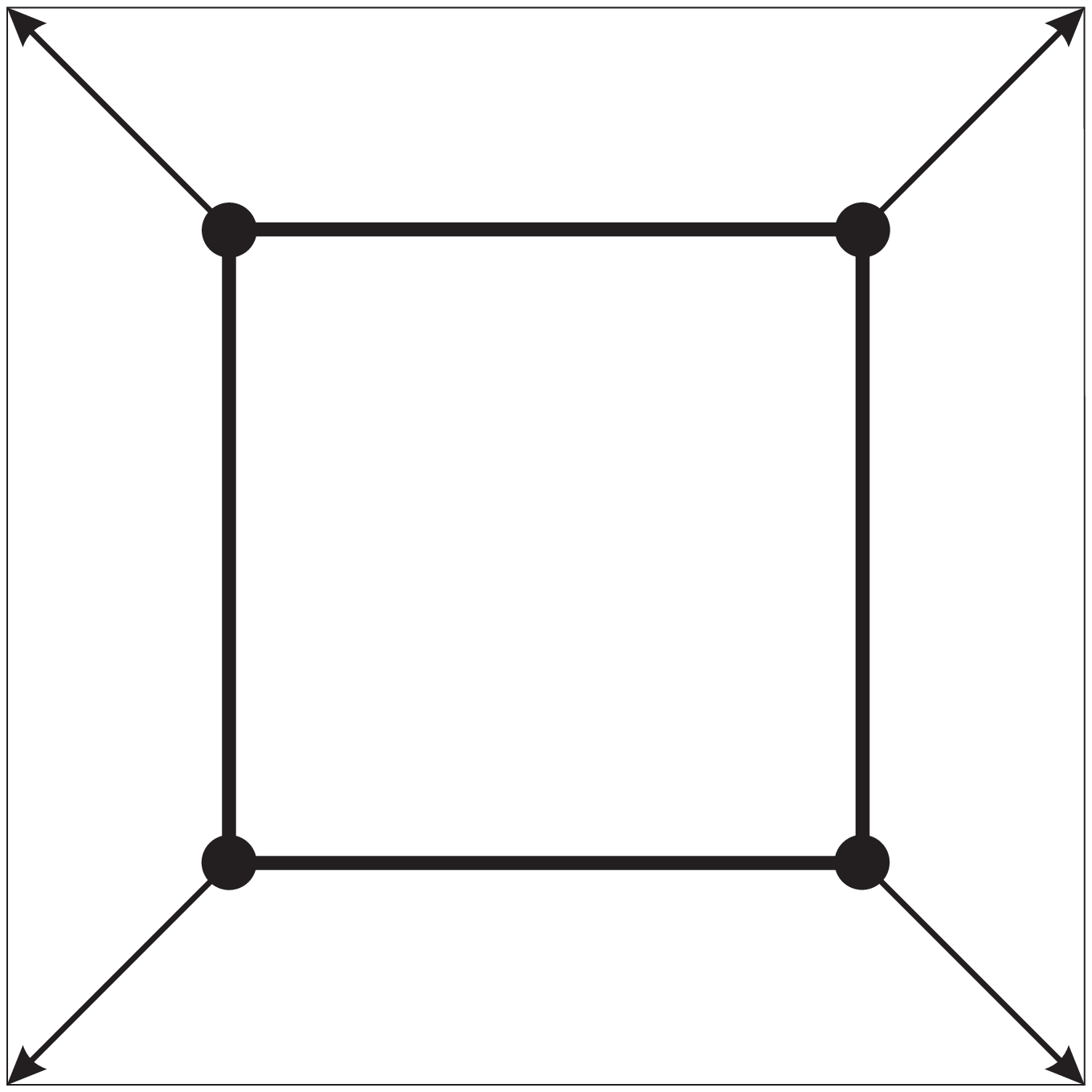,width=0.3\textwidth} \\
  \vspace*{2cm}  4) & $C_s^d$ &  & 8) & $C_{4v}$ & \\
  \hline
\end{tabular}
\caption{Different vibrational regimes (bushes of NNMs) of the
square molecule.\label{Fig2}}
\end{figure}

Indeed, the first configuration in Fig.~\ref{Fig2} with atoms
displaced \emph{arbitrarily} corresponds to the symmetry group
$G_1=C_1$. In this figure, we depict by arrows the atomic
displacements and by thin lines the resulting instantaneous
configurations of the molecule.

For our analysis, it is essential that the symmetry of the
instantaneous configuration of the mechanical system be preserved
during the vibrational motions. More precisely, if an atomic
pattern at time $t_0$ has a symmetry element $g$, this element
cannot disappear spontaneously at any $t>t_0$. This proposition,
which we call the "symmetry preservation theorem", can be proved
rigorously by examining the Hamiltonian equations of motion.

Let us note, however, that spontaneous lowering of the symmetry
can occur when the considered dynamical regime \textit{loses} its
\textit{stability} leading most frequently to the appearance of
another bush of higher dimension. This important phenomenon, which
may be regarded as the dynamical analogue of a phase transition,
will be discussed in Section III.

Of course, at the times when all atoms pass through their
equilibrium positions, the molecule configuration is again a
square. Nevertheless, it can be proved~\cite{DAI} that these
isolated instances do not affect such general dynamical properties
as the selection rules for excitation transfer between modes of
different symmetry.

Thus, all possible vibrational regimes of the square molecule can
be classified according to 8 subgroups of the group $G=C_{4v}$.
The different configurations of the vibrating molecule, therefore,
as depicted in Fig.~\ref{Fig2}, are the following: a pulsating
square ($G_8=C_{4v}$), a rotating and pulsating square
($G_5=C_4$), a rectangle ($G_6=C_{2v}^c$), a rhombus
($G_7=C_{2v}^d$), a trapezoid ($G_3=C_s^c$), a deltoid
($G_4=C_s^d$), a parallelogram ($G_2=C_2$) or an arbitrary
quadrangle ($G_1=C_1$). All these configurations vary in size as
time progresses, but the \textit{type} of the corresponding
quadrangle does not change.

Note that the different types of the above vibrational regimes of
the molecule are described by \textit{different numbers of degrees
of freedom}. Let us discuss this important fact more carefully:

Each of the eight types of vibrational regimes in Fig.~\ref{Fig2}
corresponds to a certain \textit{bush of NNMs}. For example, the
dynamical regime representing a pulsating square ($G=C_{4v}$) can
be characterized by only \emph{one} degree of freedom. The edge of
the square or the displacement of a certain atom from its
equilibrium position along the corresponding diagonal can be used
to represent this degree of freedom. Such a vibrational regime is
described by a \textit{one-dimensional} bush consisting only of
the so-called ``breathing'' nonlinear normal mode.

The rhombus-type vibration ($G_7=C_{2v}^d$) and rectangle-type
vibration ($G_6=C_{2v}^c$) are characterized by \emph{two} degrees
of freedom. The length of the diagonals in the former case and the
length of the adjacent edges in the latter can be chosen as these
degrees of freedom. Both of these vibrational regimes are
described by \emph{two-dimensional} bushes of modes. Similarly,
one can check that a trapezoid-type vibration ($G=C_s^c$)
corresponds to a \emph{three-dimensional} bush, the deltoid-type
vibration to a \emph{four-dimensional} bush and the vibration with
arbitrary quadrangle is represented by a \emph{five-dimensional}
bush of vibrational modes. Note that vibrations with $G_5=C_4$
group are described by a two-dimensional bush consisting of one
rotating and one pulsating mode.

Let us also emphasize  that the above-discussed classification of
vibrational regimes does not depend on the interparticle
interactions characterizing the mechanical system. On the other
hand, the time-evolution of the degrees of freedom corresponding
to a certain bush \textit{does} depend on these interactions in an
essential way, as we discuss in more detail below.

\subsection{Bushes of NNMs for a simple octahedral structure}

\begin{figure}[h]
\centering
\psfig{file=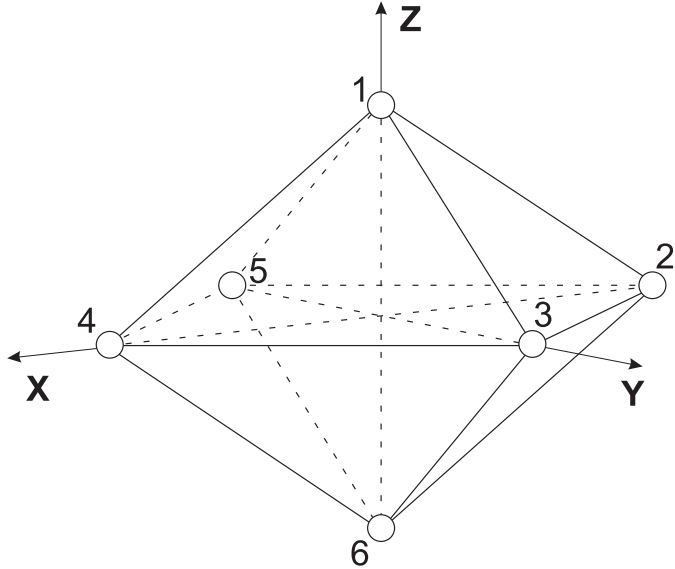,width=0.5\textwidth}
\caption{Octahedral mechanical system.\label{FigOcta}}
\end{figure}

Finally, let us describe the occurrence of bushes in a
3--dimensional mechanical system consisting of six mass points
(called particles or atoms) whose interactions are described by a
pair isotropic potential $u(r)$, $r$ being the distance between
two particles. We suppose that, at equilibrium, these particles
form a regular octahedron with edge $a_0$ as depicted in
Fig.~\ref{FigOcta}. Let us introduce a Cartesian coordinate
system. Four particles of the above octahedron lie in the $x,y$
plane and form a square with edge $a_0$. Two other particles lie
on the $z$-axis and are called ``top particle'' and ``bottom
particle'' with respect to the direction of the $z$-axis. The
distance between these particles is also equal to $a_0$.

At equilibrium, this system possesses the point symmetry group
$O_h$ and all its possible bushes of vibrational modes are listed
in~\cite{Gnezdilov}. We may suppose that $u(r)$ is the well-known
Lennard-Jones potential:
\begin{equation}
\label{eq99} u(r)=\frac{A}{r^{12}}-\frac{B}{r^6},
\end{equation}
as we have done in~\cite{Gnezdilov}, where we studied the
stability of the bushes in the octahedral molecule.  In the
present section, however, we shall consider $u(r)$ as an
\textit{arbitrary} pair isotropic potential.

The potential energy of our system in its vibrational state is
written in the form
\begin{equation}
\label{eq11} U(\vec X)=\sum_{{i,j}\atop{(i<j)}}u(r_{ij}),
\end{equation}
where $r_{ij}$ is the distance between the $i$th and $j$th
particles. The $N$-dimensional vector $\vec
X=\{x_1(t),x_2(t),\ldots,x_N(t)\}$ in Eq.~(\ref{eq11}) determines
the displacements of all particles at an arbitrary chosen instant
$t$. Here $N=18$ is the total number of degrees of freedom of the
octahedral mechanical system.

The configuration vector $\vec
X(t)=\{x_1(t),x_2(t),\ldots,x_{18}(t)\}$ determines all the
dynamical variables $x_i(t)$ as follows: The first three
components of this vector correspond to the $x$-, $y$- and
$z$-displacements of particle 1, the next three components
correspond to $x$-, $y$-, $z$-displacements of particle 2, etc.,
as numbered in Fig. \ref{FigOcta}. The dynamics of this system is
described by Newton's equations
\[
\ddot x_i=-\frac{\partial U}{\partial x_i},\quad i=1,2,\ldots,18
\]
with the masses of all particles equal to unity.

Each bush of modes corresponds to a certain subgroup $G_j$ of the
symmetry group $G_0=O_h$ of the mechanical system at equilibrium.
This means that the vibrational state described by the bush
possesses the symmetry group $G_j\subset G_0$ and, therefore,
there exist certain \textit{restrictions} on the dynamical
variables $x_i(t)$. As a result, the number $m$ of degrees of
freedom corresponding to the given bush (i.e.\ its dimension), is
smaller than the total number of dynamical variables of the
system.

Indeed, at any time $t$, the atom configuration in a vibrational
state represents a polyhedron characterized by the symmetry group
$G_j\subset O_h$. Thus, instead of the old variables $x_i(t)$, we
introduce $m$ new variables $y_j(t)$ $(j=1,\ldots,m)$, which
completely determine this polyhedron. Next, we employ the
well-known Lagrange method (see, for example,~\cite{LandauLiv})
for obtaining the dynamical equations in terms of the new
variables $y_j(t)$.

For this purpose, we introduce the Lagrange function $L=T-U$,
where $T$ is the kinetic and $U$ is the potential energy expressed
as functions of new variables $y_j$ and new momenta $\dot y_j$
$(j=1,\ldots,m)$. Then Euler--Lagrange equations in the new
variables read
\begin{equation}
\label{eq27} \frac{d}{dt}\left(\frac{\partial L}{\partial\dot
y_j}\right)-\frac{\partial L}{\partial y_j}=0,\quad j=1,\ldots,m,
\end{equation}
where $m$ is the dimension of the considered bush.

We will not present here the tedious derivation of such equations
for the octahedral molecule (see e.g.~\cite{Gnezdilov}, where
dynamical equations for bushes are obtained by means of an
appropriate \texttt{MAPLE}-program). Rather, we will give the
final results for the bushes B$[O_h]$, B$[D_{4h}]$ and
B$[C_{4v}]$. Here, in the square brackets next to the bush symbol
B, we give the symmetry group of the corresponding bush of
vibrational modes. The symmetry groups of the above bushes satisfy
the following group-subgroup relations:
\begin{equation}
\label{eq30} C_{4v}\subset D_{4h}\subset O_h.
\end{equation}

The bushes B$[O_h]$, B$[D_{4h}]$, and B$[C_{4v}]$ have one, two
and three dimensions, respectively, while their geometrical
features can be immediately seen from the symmetry groups that
correspond to them. Indeed, the one-dimensional bush B$[O_h]$
consists of only one (``breathing'') mode: The appropriate
nonlinear dynamical regime describes the evolution of a regular
octahedron whose edge $a=a(t)$ periodically changes in time.

The two-dimensional bush B$[D_{4h}]$ describes a dynamical regime
with two degrees of freedom. The symmetry group $G=D_{4h}$ of this
bush contains the four-fold axis symmetry coinciding with the $z$
coordinate axis and the mirror plane coinciding with the $x,y$
plane. This symmetry group restricts essentially the shape of the
polyhedron describing our mechanical system in the vibrational
state. Indeed, the presence of the four-fold axis demands that the
quadrangle in the $x,y$ plane be a square. For the same reason,
the four edges connecting the particles in the $x,y$ plane
(vertices of the above square) with the top particle lying on the
$z$-axis must have the same length, which we denote by $b(t)$.

\begin{figure}
\centering
\psfig{file=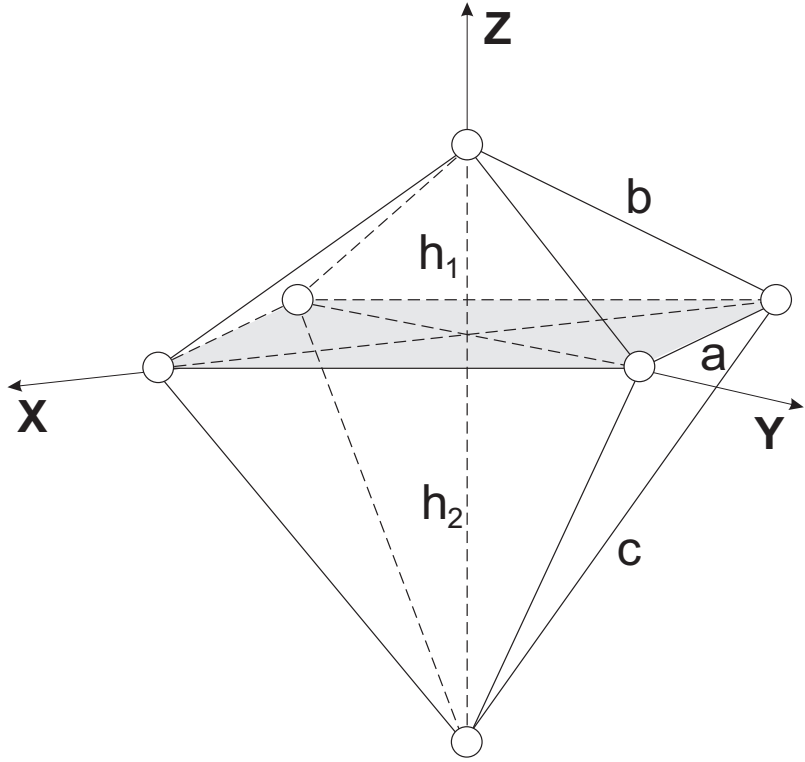,width=0.6\textwidth}
\caption{The distorted octahedron illustrating positions of
particles at fixed instant for the bush
B$[C_{4v}]$.\label{Fig3Octa}}
\end{figure}

Similarly, let the length of the edges connecting the bottom
particle on the $z$-axis with any of the four particles in the
$x,y$ plane be denoted by $c(t)$. In the case of the bush
B$[D_{4h}]$, $b(t)=c(t)$ for any $t$, because of the presence of
the horizontal mirror plane in the group $G=D_{4h}$. However, for
the three-dimensional bush B$[C_{4v}]$, this mirror plane is
absent and, therefore, $b(t)\ne c(t)$. We illustrate the
instantaneous configuration of our mechanical system vibrating
according to the bush B$[C_{4v}]$ in Fig.~\ref{Fig3Octa}.

Let us also introduce the heights, $h_1(t)$ and $h_2(t)$ of the
perpendicular distances, respectively, from the top and bottom
vertices of our polyhedron onto the $x,y$ plane. We can now write
the dynamical equations of the above bushes in terms of the purely
geometrical variables $a(t),b(t),c(t),h_1(t)$ and $h_2(t)$.

Choosing $a(t)$ and $h(t)\equiv h_1(t)\equiv h_2(t)$ as suitable
variables for describing the two-dimensional bush B$[D_{4h}]$ and
$a(t),h_1(t)$ and $h_2(t)$ as appropriate for describing the
three-dimensional bush B$[C_{4v}]$, we can write the potential
energy for our bushes of vibrational modes as follows:
\[
\begin{array}{lcl}
B[O_h]&:&U(a)=12u(a)+3u(\sqrt{2}a),\\
B[D_{4h}]&:&U(a,h)=4u(a)+2u(\sqrt{2}a)+8u\left(\sqrt{h^2+\frac{a^2}{2}}\right)+u(2h),\\
B[C_{4v}]&:&U(a,h_1,h_2)=4u(a)+2u(\sqrt{2}a)+4u(b)+4u(c)+u(h_1+h_2),
\end{array}
\]
where
$b=\sqrt{\frac{a^2}{2}+\left(\frac{5}{4}h_1-\frac{1}{4}h_2\right)^2}$,
$c=\sqrt{\frac{a^2}{2}+\left(\frac{5}{4}h_2-\frac{1}{4}h_1\right)^2}$.

Then, with the aid of Lagrange's equations, we obtain the
following dynamical equations for the above bushes of vibrational
modes:

\begin{equation}
\label{ur}
\begin{array}{ll}
B[O_h]:&\\
\qquad\ddot a=&-4u'(a)-\sqrt{2}u'(\sqrt{2}a);\\
B[D_{4h}]:&\\
\qquad\ddot a=&-2u'(a)-\sqrt{2}u'(\sqrt{2}a)-2u'(b)\frac{a}{b},\\
\qquad\ddot h=&-4u'(b)\frac{h}{b}-u'(2h);\\
B[C_{4v}]:&\\
\qquad\ddot a=&-2u'(a)-\sqrt{2}u'(\sqrt{2}a)-u'(b)\frac{a}{b}-u'(c)\frac{a}{c},\\
\qquad\ddot h_1=&-u'(b)\frac{5h_1-h_2}{b}-u'(h_1+h_2),\\
\qquad\ddot h_2=&-u'(c)\frac{5h_2-h_1}{c}-u'(h_1+h_2).
\end{array}
\end{equation}

Thus, we derive dynamical equations for our bushes of vibrational
modes in terms of variables having an explicit geometrical nature.
Each bush describes a certain nonlinear dynamical regime
corresponding to a vibrational state of the considered mechanical
system, so that at any fixed time the configuration of this system
is represented by a definite polyhedron characterized by the
symmetry group $G$ of the given bush.

The above dynamical equations for the bushes B$[O_h]$, B$[D_{4h}]$
and B$[C_{4v}]$ are \textit{exact}, but rather complicated to
solve. One may consider instead some \textit{approximate}
equations which can be obtained by expanding the potential energy
in Taylor series (near the equilibrium state) keeping only the
lowest order terms in the expansion.

Thus, using the first few leading terms in such decompositions, it
turns out that bushes belonging to systems of different physical
nature, with different symmetry groups and different structures
can produce \textit{equivalent dynamical equations}. More
precisely, the equations for many bushes can be written in
\textit{the same form}, and this fact leads to the idea of the
``classes of dynamical universality'' mentioned in \cite{DAI,
DAI2, Chechin-Sakhnenko}.

Returning to the dynamical equations~(\ref{ur}) for the bushes
B$[O_h]$, B$[D_{4h}]$, B$[C_{4v}]$, we see that every bush can be
considered as a \textit{reduced} dynamical system whose dimension
($m$) is less than the full dimension of the considered mechanical
system. If the original system is Hamiltonian, the reduced system
can also be proved to be Hamiltonian.

\section{Bushes of NNMs in modal space and stability analysis}\label{s2}

Vectors $\vec{X}(t)$, corresponding to bushes of NNMs, are
determined in the configuration space $\Re^N$. If we introduce in
this space a basis set of vectors  $\{\vec \varphi_1, \vec
\varphi_2, ..., \vec \varphi_N\}$, the dynamical regime of our
mechanical system can be represented by a linear combination of
the vectors $\vec \varphi_j$ with time dependent coefficients
$\nu_j(t)$ as:

\begin{equation}\label{eq60}
    \vec{X}(t)=\sum_{j=1}^N \nu_j(t)\vec \varphi_j \,
\end{equation}
where the functions $\nu_j(t)$ entering this decomposition
represent \textit{new dynamical variables}.

Every term in the sum (\ref{eq60}) has the form of a NNM, whose
basis vector $\vec \varphi_j$ determines a displacement pattern,
while the functions $\nu_j(t)$ determine the time evolution of the
atomic displacements. Because of this interpretation, we can
consider a given dynamical regime $\vec X(t)$ as a \textit{bush}
of NNMs. In fact, as will be shown later one often speaks about
\textit{root} modes and \textit{secondary} modes of a given bush.

Note that each NNM $\nu_j(t)\vec \varphi_j$ from (\ref{eq60}) is
not, in general, a solution of the dynamical equations of the
considered mechanical system, while a specific linear combination
of a number of these modes can represent such a solution and, thus
describe an exact dynamical regime. Sometimes, for brevity, we
will use the word "mode" not only for the whole term $\nu_j(t)
\vec \varphi_j$, but also for the time dependent function
$\nu_j(t)$.

\subsection{Normal coordinates and normal modes}

Normal coordinates and normal modes are commonly introduced in the
theory of \emph{small vibrations} of multiparticle systems (see,
for example,~\cite{LandauLiv}) in the framework of the
\textit{harmonic approximation}, where they represent
\textit{exact} solutions of the corresponding dynamical equations.

In this approximation, the potential energy $U(\vec X)$ is
decomposed into a Taylor series with respect to the atomic
displacements $x_i(t)$ from their equilibrium positions and all
terms whose orders are higher than $2$ are neglected (because the
displacements are supposed to be sufficiently small). As a result,
the dynamical equations
\begin{equation}
\label{eq45} m_i\ddot x_i=-\frac{\partial U}{\partial
x_i},\quad(i=1,\ldots,N)
\end{equation}
turn out to be \textit{linear differential equations}, with
constant coefficients.

Each normal mode is a particular solution to Eq.~(\ref{eq45}) of
the form
\begin{equation}
\label{eq47} \vec X(t)=\vec c\cos(\omega t+\varphi_0),
\end{equation}
where the $N$-dimensional constant vector $\vec
c=\left\{c_1,c_2,\ldots,c_N\right\}$ and the constant phase
$\phi_0$ determine the initial displacements of all particles from
their equilibrium state and $\omega$ is the frequency of
vibration.

Since $U(\vec X)$ has a quadratic form, substituting~(\ref{eq47})
into Eq.~(\ref{eq45})  and dividing the resulting equations by
$\cos(\omega t+\varphi_0)$, we reduce the problem of finding the
normal modes to the algebraic problem of evaluating the
eigenvalues and eigenvectors of the matrix $\mathcal K=\|k_{ij}\|$
with coefficients
\begin{equation}\label{e58}
k_{ij}=\left.\frac{\partial^2U}{\partial x_i\partial
x_j}\right|_{\vec X=0} \ \ \ \ \ \ \ \ \ i, j=1..N.
\end{equation}

Each eigenvalue of this matrix is the square of the frequency,
$\omega_j^2$, of the $j$th normal mode, while its eigenvector
$\vec c_j$ represents the pattern of this normal mode and is
called the \textit{normal coordinate}.

Being real and symmetric, the $N\times N$ matrix $\mathcal K$
possesses $N$ eigenvectors $\vec c_j$ ($j=1,\ldots,N$) and $N$
eigenvalues $\omega_j^2$. The complete collection of these
vectors, i.e. the $N$ normal coordinates, can be used as the
\textit{basis} of the configuration space, hence we may write
\begin{equation}
\label{eq50} \vec X(t)=\sum_{i=1}^N{\mu_j(t)\vec c_j},
\end{equation}
where $\vec X(t)=\left\{x_1(t),x_2(t),\ldots,x_N(t)\right\}$,
while $\mu_j(t)$ are new dynamical variables which we introduce
instead of the old variables $x_i(t)$ ($i=1,\ldots,N$).

If the transformation to normal coordinates is used in the absence
of degeneracies, the corresponding system of linear ODEs leads to
a set of uncoupled harmonic oscillators
\[
\ddot\mu_j(t)+\omega_j^2\mu_j(t)=0,\quad j=1,\ldots,N
\]
with the well-known solution
\begin{equation}
\label{equat101}
\mu_j(t)=a_j\cos\left(\omega_jt+\varphi_{0j}\right),
\end{equation}
where $a_j$ and $\varphi_{0j}$ are arbitrary constants.

Note the distinction we make between a \textit{normal coordinate},
represented by the eigenvector $\vec c_j$, and a normal mode,
referring to the product of the vector $\vec c_j$ and the
time-periodic function
$\mu_j(t)=\cos\left(\omega_jt+\varphi_{0j}\right)$.

\subsection{Symmetry--adapted coordinates as basis of the configuration space}

In general, the normal coordinates (normal modes) depend on the
interparticle interactions in the mechanical system. Indeed, the
concrete form of the potential energy is needed for their
construction according to the above-described procedure. Because
of this fact, it is more convenient to use
\textit{symmetry-adapted} coordinates, instead of normal
coordinates. The former are the basis vectors of the
\textit{irreducible representations} of a certain parent group.
More importantly, the symmetry-adapted coordinates do not depend
on the interparticle interactions and can be obtained by the
appropriate group theoretical methods.

Note that, for simple mechanical systems, the normal and the
symmetry-adapted coordinates may \textit{coincide} with each
other. This coincidence is a consequence of the fact that each
irreducible representation of the parent group is contained
\textit{only once} in the decomposition of the reducible
vibrational representation into its irreducible parts.

We complete this part of our study by recalling in Appendix A some
notions from the theory of matrix representations of symmetry
groups. We also discuss in Appendix A the theory of irreducible
representations and describe how they can prove especially helpful
in analyzing important problems of crystal vibrations in Solid
State Physics.

This concludes our brief outline of the group theoretical concepts
needed to find bushes of vibrational modes for arbitrary
mechanical systems. The interested reader will find more details
in~\cite{DAI,Chechin-Sakhnenko,IJNLM00, ComputMath}. Now,it is
time to discover how this approach of discrete symmetries can help
us analyze the stability properties of the periodic and
quasiperiodic solutions belonging to these bushes.

\subsection{Linear stability analysis of bushes of NNMs}

Let us begin by considering individual NNMs, representing
one--dimensional bushes of the FPU-$\beta$ Hamiltonian.

\vspace{.5cm} \textbf{1. \emph{Example 1}: Stability of the
$\pi$-mode in the FPU-$\beta$ chain }

The question of \textit{local stability} near NNMs of the form
$\hat{x}_{j}$ may be answered by studying the linearized equations
about them, setting $x_{j}=\hat{x}_{j}+y_{j}$ and keeping up to
linear terms in $y_{j}$. For example, in the case of the NNM
(\ref{eq5}) these equations become ($\beta=1$):

\begin{equation}\label{FPUNNM}
\ddot{y}_{j}=(1+48\hat{x}^{2})(y_{j-1}-2y_{j}+y_{j+1}),\;j=1,\ldots,N
\end{equation}
where $\hat{x}$ is the solution of (\ref{eq6}), see also equations
(\ref{eq7})-(\ref{eq10}), and $y_{1}=y_{N+1}$, $y_{0}=y_{N}$, due
to p.b.c..

According to Floquet theory, linear stability analysis of periodic
solutions is performed by studying the eigenvalues of the
monodromy matrix $M(T)$, whose $N$ columns are the fundamental
solutions of Eq.(\ref{FPUNNM}) evaluated at $t=T$ (the period of
the NNM), with $M(0)$ equal to the $N\times N$ identity matrix.
However, in the case of such simple modes as (\ref{eq5}), it is
much easier to diagonalize first the variational equations into
$N$ uncoupled Lam\'{e} equations
\begin{equation}\label{FPULame1}
\ddot{z}_{j}(t)+4(1+48\hat{x}^{2}){\sin}^{2}\biggl(\frac{\pi
j}{2N}\biggr)z_{j}(t)=0,\;j=1,\ldots,N
\end{equation}
where the $z_{j}$ variations are simple linear combinations of the
$y_{j}$'s. Changing variables to $u=\lambda t$, where
$\lambda^2=4/(1-2k^2)$ and using (\ref{eq7})-(\ref{eq10}) this
equation takes the form

\begin{equation}\label{FPULame2}
{z_{j}^{\prime\prime}}(u)+\bigl(1+22{k}^2-24{k}^2sn^{2}(u,{k}^{2})\bigr){\sin}^{2}\bigl(\frac{\pi
j}{2N}\bigr)z_{j}(u)=0,\;j=1,\ldots,N
\end{equation}
where primes denote differentiation with respect to $u$. Equation
(\ref{FPULame2}) is an example of Hill's equation
\begin{equation}\label{FPU_Floquet_equation}
{z^{\prime\prime}}(u)+Q(u)z(u)=0
\end{equation}
where $Q(u)$ is a $T$--periodic function  ($Q(u)=Q(u+T)$).
According to Floquet theory, its solutions are bounded or
unbounded depending on whether the eigenvalues of its monodromy
matrix lie on the unit circle or not. The first variation
$z_{j}(u)$ to become unbounded as $k\geq k_c$ determines the total
energy of the lattice
\begin{equation}
E_c=\frac{N{k_c}^2+2N{k_c}^4}{(1-2{k_c}^2)^2}
\end{equation}
at which this particular NNM becomes unstable for $0\leq k\leq
\frac{1}{\surd{2}}$ \vspace{.5cm}

\textbf{2. \emph{Example 2}: Stability of the $\pi$-mode in the
FPU-$\alpha$ chain}

Let us consider the stability of the one-dimensional bush
$B[\hat{a}^2, \hat{i}]$, representing the $\pi$-mode, in the
FPU-$\alpha$ chain with $N=4$ particles under p.b.c.

The equations for this nonlinear chain read
\begin{equation}
\label{eq9zh}
\begin{array}{rcl}
\ddot x_1 & = & f(x_2-x_1)-f(x_1-x_4),\\
\ddot x_2 & = & f(x_3-x_2)-f(x_2-x_1),\\
\ddot x_3 & = & f(x_4-x_3)-f(x_3-x_2),\\
\ddot x_4 & = & f(x_1-x_4)-f(x_4-x_3),
\end{array}
\end{equation}
where $f(x)=x+\alpha x^2$. Clearly, the $\pi$-mode represents here
the exact \textit{periodic} vibrational regime
$\vec{X}=\{A(t),-A(t),A(t),-A(t)\}$, with $A(t)=C_0\cos(2t)$.
According to the conventional prescription, we should linearize
the dynamical system~(\ref{eq9zh}) in the neighborhood of the
given bush and study the resulting system of ODEs. To achieve
this, we write
\begin{equation}
\vec X(t)=\vec C(t)+\vec\delta(t), \label{eq40zh}
\end{equation}
where $\vec C=\{A(t),-A(t),A(t),-A(t)\}$ is our NNM, while
$\vec\delta(t)=\{\delta_1(t),\delta_2(t),\delta_3(t),\delta_4(t)\}$
is an infinitesimal vector. Substituting~(\ref{eq40zh}) into
Eqs.~(\ref{eq9zh}) with $f(x)=x+x^2$ (i.e. $\alpha=1$) and
neglecting all terms nonlinear in $\delta_j(t)$, we obtain  for
this FPU-$\alpha$ model the system of linearized equations:
\begin{equation}
\ddot{\vec\delta}=\mat J(t)\vec\delta, \label{eq42zh}
\end{equation}
where $\mat J(t)$ is the \textit{Jacobi} matrix of (\ref{eq9zh})
for the FPU-$\alpha$ chain calculated by the substitution of the
vector $\vec X=\vec C=\{A(t),-A(t),A(t),-A(t)\}$. This matrix can
be expressed as follows:
\begin{equation}
\mat J(t)=\mat L+4A(t)\mat M, \label{eq43zh}
\end{equation}
where
\begin{equation}
\mat L=\left(
\begin{array}{rrrr}
  -2& 1& 0& 1\\
   1&-2& 1& 0\\
   0& 1&-2& 1\\
   1& 0& 1&-2
\end{array}\right)
,\qquad \mat M=\left(
\begin{array}{rrrrrr}
   0&-1& 0& 1\\
  -1& 0& 1& 0\\
   0& 1& 0&-1\\
   1& 0&-1& 0
\end{array}\right)
\label{eq44zh}
\end{equation}
are time-independent symmetric matrices.

It easy to verify that $\mat L$ and $\mat M$ commute with each
other; therefore, there exists a time-independent orthogonal
matrix $\mat S$ transforming both $\mat L$ and $\mat M$ into
diagonal form: $\widetilde{\mat S}\mat L{\mat S}=\mat L_{dia},\,$
$\widetilde{\mat S}\mat M{\mat S}=\mat M_{dia}$ (where
$\widetilde{\mat S}$ is the transpose of $\mat S$). In turn, this
means that the Jacobi matrix $\mat J(t)$ can be diagonalized at
\textit{any time} $t$ by one and the same
\textit{time-independent} matrix $\mat S$. Therefore, our
linearized system~(\ref{eq42zh}) for the considered bush B$[\hat
a^2,\hat\i]$ can be decomposed into four \textit{independent}
differential equations.

Observe that, due to similar reasons, we were able to reduce
(\ref{FPUNNM}) into a system of $N$ uncoupled ODEs, see
(\ref{FPULame1}), for the $\pi-$ mode of the FPU $\beta-$ chain in
Example 1.

The above matrix $\mat S$ can be obtained with the aid of the
theory of irreducible representations of the symmetry group $G$,
for the case $G=D_2$ considered here. The general method for
obtaining the matrix $\mat S$ uses the basis vectors of
irreducible representations of the group $G$, constructed in the
mechanical space of the considered dynamical system. In the simple
example of a monoatomic chain with $N=4$ particles, this method
leads to the following result
\begin{equation}
\mat S=\frac{1}{2} \left(
\begin{array}{rrrr}
  1& 1& 1& 1\\
  1& 1&-1&-1\\
  1&-1& 1&-1\\
  1&-1&-1& 1
\end{array}\right).
\label{eq501zh}
\end{equation}
The rows of the matrix $\mat S$ in~(\ref{eq501zh}) are simply the
characters of four one-dimensional irreducible representations
(irreps)~-- $\Gamma_1$, $\Gamma_2$, $\Gamma_3$, $\Gamma_4$~-- of
the Abelian group $D_2$, because each of these irreps is contained
once in the decomposition of the natural representation of the
group $G=D_2$ (see also Appendix A). Introducing new variables
$\vec Y=\{y_1,y_2,y_3,y_4\}$ by the transformation $\vec Y=\mat
S\vec\delta$, with $\mat S$ as in~(\ref{eq501zh}), we arrive at
the \emph{complete splitting} of the linearized
equations~(\ref{eq42zh}) for the FPU-$\alpha$ model:
\begin{subequations}
\label{eq511zh}
\begin{eqnarray}
&&\ddot y_1=0,\label{eq511zha}\\
&&\ddot y_2=-2[1+4A(t)]y_2,\label{eq511zhb}\\
&&\ddot y_3=-4y_3,\label{eq511zhc}\\
&&\ddot y_4=-2[1-4A(t)]y_4,\label{eq511zhd}
\end{eqnarray}
\end{subequations}
where $A(t)=C_0\cos(2t)$.

With the aid of Eqs.~(\ref{eq511zh}) one can now determine
precisely the threshold value for $C_0$, leading to the loss of
stability of the one-dimensional bush B$[\hat a^2,\hat\i]$.
Indeed, according to Eqs.~(\ref{eq511zh}), the variables $y_j(t)$
($j=1,2,3,4$) are independent from each other, and we can solve
for them separately: Eq.~(\ref{eq511zha}) for $y_1(t)$ describes
the uniform motion of the center of mass of our chain, since it
follows from the equations $\vec Y=\mat S\vec\delta$ that
$y_1(t)=(\delta_1(t)+\delta_2(t)+\delta_3(t)+\delta_4(t))/2$.
Therefore, considering \textit{vibrational} regimes only, we may
assume $y_1(t)\equiv0$.

If, in the solution of the system~(\ref{eq511zh}), $y_j(t)=0$ for
$j=1,2$ but not $j=3$, $\vec\delta=\widetilde{\mat S}\vec Y$ would
imply (since $\mat S$ is orthogonal and $\mat
S^{-1}=\widetilde{\mat S}$):
$\vec\delta(t)=\{y_3(t),-y_3(t),y_3(t),-y_3(t)\}$, where
$y_3(t)\sim\cos(2t)$. This leads to variations ``along'' the bush
$\vec X(t)=C_0\{\cos(2t),-\cos(2t),\cos(2t),-\cos(2t)\}$, which do
not affect the stability of the motion.

On the other hand, if $y_2(t)$ and $y_4(t)$ are nonzero, with
$A(t)=C_0\cos(2t)$, Eq.~(\ref{eq511zhb}) and (\ref{eq511zhd})
imply $\ddot y_j+[2+8C_0\cos(2t)]y_j=0, j=2,4$ and can thus be
transformed to the standard form of Mathieu's equation
\cite{AbramSteg}. Therefore, the stability threshold of the
considered bush B$[\hat a^2,\hat\i]$ for $N=4$ can be determined
directly from the well-known diagrams of stable and unstable
motion of the Mathieu equation. In this way we find that the
lowest critical value $C_c>0$ of the amplitude $C_0$ at which the
given bush loses its stability is $C_c \approx 0.303$.

Let us now consider the FPU-$\alpha$ chain with $N=6$ particles.
Linearizing the dynamical equations of the FPU-$\alpha$ chain with
$N=6$ in the vicinity of the bush B$[\hat a^2,\hat\i]$
($\pi$-mode), we also obtain a Jacobi matrix of the
form~(\ref{eq43zh}). However, unlike the case $N=4$, matrices
$\mat L$ and $\mat M$, \textit{do not} commute with each other. As
a consequence, \emph{we cannot diagonalize} both matrices $\mat L$
and $\mat M$, i.e.\ with one and the same orthogonal matrix $\mat
S$. Thus, it is impossible to diagonalize the Jacobi matrix $\mat
J(t)$ in equation $\ddot{\vec\delta}=\mat J(t)\vec\delta$ for all
time $t$. This means that there is no matrix $\mat S$ that
completely decouples the linearized system for the bush B$[\hat
a^2,\hat\i]$ for the chain with $N=6$ particles.

This difference between the cases $N=4$ and $N=6$ (generally, for
$N>4$) can be explained group-theoretically as follows: The group
$G=[\hat a^2,\hat\i]$ of the considered bush leads to
\textit{different} groups for the cases $N=4$ and $N=6$. Indeed,
for $N=4$ $\quad[\hat a^2,\hat\i]\equiv\{\hat e,\hat
a^2,\hat\i,\hat a^2\hat\i\}=D_2$, while for $N=6$ $\quad[\hat
a^2,\hat\i]\equiv\{\hat e,\hat a^2,\hat a^4,\hat\i,\hat
a^2\hat\i,\hat a^4\hat\i\}=D_3$. Group $D_3$, on the other hand,
is \emph{non-Abelian} ($\hat\i\hat a^4=a^2\hat\i$), unlike group
$D_2$ ($\hat\i\hat a^2=a^2\hat\i$) and, as a consequence,
possesses not only one-dimensional, but also two-dimensional
irreps. This is precisely why we are not permitted to completely
decouple the above linearized system\footnote{Actually, this fact
can be understood if one takes into account that the
two-dimensional irrep is contained two times in the decomposition
of the natural representation of the considered chain.}
(see~\cite{Chechin-Zhukov}).

In spite of this difficulty, we can still simplify the linearized
system $\ddot{\vec\delta}=\mat J(t)\vec\delta$ considerably using
group theoretical methods (see \cite{Chechin-Zhukov,
Columbus2007}). Let us present the final result of this splitting
for the case $N=6$:
\begin{subequations}
\label{eq850zh}
\begin{eqnarray}
 &&\ddot y_1=-4y_1,\label{eq850zha}\\
 &&\ddot y_2=0,\label{eq850zhb}\\ \nonumber\\
 &&\left\{
 \begin{array}{l}
  \ddot y_3+2y_3=P(t)y_5,\\
  \ddot y_5+2y_5=\bar P(t)y_3,
 \end{array}
 \right.\label{eq850zhc}\\ \nonumber\\
 &&\left\{
 \begin{array}{l}
  \ddot y_4+2y_4=P(t)y_6,\\
  \ddot y_6+2y_6=\bar P(t)y_4.
 \end{array}
 \right.\label{eq850zhd}
\end{eqnarray}
\end{subequations}
Here $P(t)=e^{\frac{i\pi}{3}}-4A(t)[1+e^{-\frac{i\pi}{3}}]$, while
$\bar P(t)$ is the complex conjugate of $P(t)$.

The stability of the $\pi$-mode (the bush B$[\hat a^2,\hat\i]$) has
already been extensively discussed in a number of
papers~\cite{Bud,Sand,Flach,FPU1,FPU2,Ruffo,Yoshimura,Shinohara,Shinohara2003,LeoLeo,AntiFPU}
by different methods and with emphasis on different aspects of
stability. In the papers~\cite{FPU1},~\cite{FPU2} the stability of
all one--dimensional and two--dimensional bushes of vibrational
modes, in both FPU-$\alpha$ and FPU$-\beta$ models was numerically
studied. In particular, this study revealed a remarkable property of
the FPU-$\alpha$ chain: the stability threshold of the $\pi$-mode
turns out to be \textit{one and the same} when it interacts with
\textit{any} other mode of the chain. By contrast, other nonlinear
normal modes of the FPU-$\alpha$ (or FPU-$\beta$) chain exhibit very
different stability thresholds when interacting with different
modes~\cite{FPU2}.

\subsection{Bushes of NNMs in monoatomic chains}

To study bushes of NNMs in monoatomic chains, we shall choose the
complete set of normal coordinates $\vec \varphi_k$ as the basis
of the configuration space.  Here, we use the normal coordinates
in the form presented in~\cite{Ruffo}:
\begin{equation}
\vec\varphi_k=\left\{\frac{1}{\sqrt{N}}\left.\left[\sin\left(\frac{2\pi
k}{N}n\right) +\cos\left(\frac{2\pi
k}{N}n\right)\right]\right|~n=1,\dotsc,N\right\}, \ \
k=0..N-1,\label{eq70fpu2}
\end{equation}
where the subscript $k$ refers to the mode and the subscript $n$
refers to the atom. The vectors $\vec\varphi_k$
($k=0,1,2,\dotsc,N-1$) form an orthonormal basis, in which we can
expand the set of atomic displacements $\vec X(t)$ corresponding
to a given bush as follows:
\begin{equation}
\vec X(t)=\sum_{k=0}^{N-1}\nu_k(t)\vec{\varphi}_k.
\label{eq71fpu2}
\end{equation}
For example, one obtains the following expressions for the bushes
B$[\hat a^4,\hat\i]$ and B$[\hat a^4,\hat a^2\hat\i]$ [see
Eqs.~(\ref{eq46a}), (\ref{eq46c})]:
\begin{eqnarray}
B[\hat a^4,\hat\i]: \ \ \ \vec X(t)&=&\{~x_1(t),x_2(t),-x_1(t),-x_2(t)~|~%
x_1(t),x_2(t),-x_1(t),-x_2(t)~|\cdot\cdot\cdot~\}\nonumber\\
&=&\mu(t)\vec\varphi_{N/2}+\nu(t)\vec\varphi_{3N/4},
\label{eq72fpu2}\\
B[\hat a^4,\hat a^2\hat\i]: \ \ \ \vec
X(t)&=&\widetilde\mu(t)\vec\varphi_{N/2}+\widetilde\nu(t)\vec\varphi_{N/4}.
\label{eq73fpu2}
\end{eqnarray}
From the complete basis~(\ref{eq70fpu2}) only the vectors
\begin{eqnarray}
\vec\varphi_{N/2}&=&\frac{1}{\sqrt{N}}(-1,1,-1,1,-1,1,-1,1,-1,1,-1,1,...),
\label{eq74fpu2}\\
\vec\varphi_{N/4}&=&\frac{1}{\sqrt{N}}(1,-1,-1,1,1,-1,-1,1,1,-1,-1,1,...),
\label{eq74afpu2}\\
\vec\varphi_{3N/4}&=&\frac{1}{\sqrt{N}}(-1,-1,1,1,-1,-1,1,1,-1,-1,1,1,...).
\label{eq74bfpu2}
\end{eqnarray}
contribute to the two-dimensional bushes (\ref{eq72fpu2}),
(\ref{eq73fpu2}), which are equivalent to each other and
constitute an examples of dynamical domains. For the bush B$[\hat
a^4,\hat\i]$, we can find the following relations between the old
dynamical variables $x_1(t)$, $x_2(t)$ (corresponding to the
configuration space) and the new dynamical variables $\mu(t)$,
$\nu(t)$ (corresponding to the modal space):
\begin{equation}
\begin{array}{l}
\mu(t)=-\frac{\sqrt{N}}{2}[x_1(t)-x_2(t)],\\
\nu(t)=-\frac{\sqrt{N}}{2}[x_1(t)+x_2(t)].
\end{array}
\label{eq75fpu2}
\end{equation}

Thus, each of the above bushes consists of two modes. One of these
modes is the \textit{root} mode ($\vec\varphi_{3N/4}$ for the bush
B$[\hat a^4,\hat\i]$ and $\vec\varphi_{N/4}$ for the bush B$[\hat
a^4,\hat a^2\hat\i]$), while the other mode $\vec\varphi_{N/2}$ is
the secondary mode. Indeed, according to \cite{DAI,
Chechin-Sakhnenko} the \emph{symmetry} of the \emph{secondary
modes} must be \emph{higher or equal} to the symmetry of the root
mode. In our case, as we deduce from Eq.~(\ref{eq74afpu2}), the
translational symmetry of the mode $\vec\varphi_{N/4}$ is $\hat
a^4$ (acting by this element on~(\ref{eq74afpu2}) produces the
same displacement pattern), while the translational symmetry of
the mode $\vec\varphi_{N/2}$ is $\hat a^2$, which has twice more
symmetry elements than that of $\vec\varphi_{N/4}$
[see~(\ref{eq74fpu2}),(\ref{eq74afpu2})]. Note that the
\emph{full} symmetry of the modes $\vec\varphi_{N/4}$ and
$\vec\varphi_{N/2}$ is represented by $[\hat a^4,\hat a^2\hat\i]$
and $[\hat a^2,\hat\i]$, respectively.

Let us now return to the Hamiltonian of the FPU system written as
follows:
\begin{equation}
H=T+V=\frac{1}{2}\sum_{n=1}^N\dot{x}_n^2+\frac{1}{2}\sum_{n=1}^N
(x_{n+1}-x_n)^2 +\frac{\gamma}{p}\sum_{n=1}^N(x_{n+1}-x_n)^p.
\label{eq80fpu2}
\end{equation}
Here $p=3$, $\gamma=\alpha$ for the FPU-$\alpha$ chain, and $p=4$,
$\gamma=\beta$ for the FPU-$\beta$ chain, while $T$ and $V$ are
the kinetic and potential energies, respectively. We assume again
p.b.c.

Let us consider the set of atomic displacements corresponding to
the two-dimensional bush B$[\hat a^4,\hat\i]$
\begin{equation}
\vec X(t)=\{~x,y,-y,-x~|~x,y,-y,-x~|~x,y,-y,-x~|~\cdots~\},
\label{eq81fpu2}
\end{equation}
where we rename $x_1(t)$ and $x_2(t)$ from Eq.~(\ref{eq72fpu2}) as
$x(t)$ and $y(t)$, respectively. Substituting the atomic
displacements from Eq.~(\ref{eq81fpu2}) into the
Hamiltonian~(\ref{eq80fpu2}) corresponding to the FPU-$\alpha$
chain we obtain:
\begin{equation}
T=\frac{N}{4}(\dot x^2+\dot y^2), \label{eq82fpu2}
\end{equation}
\begin{equation}
V=\frac{N}{4}(3x^2-2xy+3y^2)+\frac{N\alpha}{2}(x^3+x^2y-xy^2-y^3).
\label{eq83fpu2}
\end{equation}
These expressions are valid for an arbitrary FPU-$\alpha$ chain
with $N\mod 4=0$. The size of the extended primitive cell (EPC)
for the vibrational state~(\ref{eq81fpu2}) is equal to $4a$ and,
therefore, when calculating the energies $T$ and $V$, we may
restrict ourselves to summing over only one EPC. In the present
case, Lagrange's equations (\ref{eq27}) can be written as follows:
\begin{equation}
\left\{\begin{array}{lll}
\ddot{x}+(3x-y)+\alpha(3x^2+2xy-y^2)&=&0,\\
\ddot{y}+(3y-x)+\alpha(x^2-2xy-3y^2)&=&0.
\end{array}\right.
\label{eq85fpu2}
\end{equation}
Let us emphasize that these equations do not depend on the number
$N$ of the particles in the chain (only $N\mod 4=0$ must hold).

Eqs.~(\ref{eq85fpu2}) are written in terms of the atomic
displacements $x(t)$ and $y(t)$. From them, it is easy to obtain
the dynamical equations for the bush \emph{in terms of the normal
modes} $\mu(t)$ and $\nu(t)$. Using the relations~(\ref{eq75fpu2})
between the old and new variables, we find the following equations
for the bush B$[\hat a^4,\hat\i]$ in the modal space
\begin{eqnarray}
\ddot\mu+4\mu-\frac{4\alpha}{\sqrt{N}}\nu^2=0,
\label{eq86fpu2}\\
\ddot\nu+2\nu-\frac{8\alpha}{\sqrt{N}}\mu\nu=0. \label{eq86bfpu2}
\end{eqnarray}
The Hamiltonian for the bush B$[\hat a^4,\hat\i]$, considered as a
two-dimensional dynamical system, can be written in the modal
space as follows:
\begin{equation}
H[\hat a^4,\hat\i]=\frac{1}{2}(\dot\mu^2+\dot\nu^2)+(2\mu^2+\nu^2)-%
\frac{4\alpha}{\sqrt{N}}\mu\nu^2. \label{eq87fpu2}
\end{equation}

Note that dynamical equations for all one-dimensional and
two-dimensional bushes of vibrational modes for the FPU-chains
were presented in \cite{FPU2}.

\subsection{Stability of bushes of vibrational modes in nonlinear chains}

\subsubsection{Consideration in modal space}

Since the term ``stability'' is often used in the literature with
different meanings, let us explain in what sense we use it in the
present paper.

The stability of bushes of modes was discussed, in general,
in~\cite{DAI,Chechin-Sakhnenko}, while in the case of the FPU
chains it was considered in~\cite{FPU1,FPU2}. Following these
papers, we also discuss here the stability of a given bush of
normal modes with respect to its interactions \footnote{There is
an essential difference between the interactions of the modes that
belong and those that do not belong to a given bush: we speak
about ``force interaction'' in the former case and about
``parametric interaction'' in the last case
(see~\cite{Chechin-Sakhnenko}).} with the modes which \textit{do
not belong} to this bush. Let us illustrate this idea with an
example.

As was shown above, the two-dimensional bush B$[\hat a^4,\hat\i]$
for the FPU-$\alpha$ chain is described by
Eqs.~(\ref{eq86fpu2}),(\ref{eq86bfpu2}). These equations admit a
special solution of the form
\begin{equation}
\mu(t)\neq 0,\quad \nu(t)\equiv 0. \label{eq90fpu2}
\end{equation}
which can be excited by imposing the initial conditions:
$\mu(t_0)=\mu_0\neq 0$, $\dot\mu(t_0)=0$, $\nu(t_0)=0$,
$\dot\nu(t_0)=0$). Substitution of ~(\ref{eq90fpu2})
into~(\ref{eq86fpu2}) produces the dynamical equation of the
one-dimensional bush B$[\hat a^2,\hat\i]$ (see \cite{FPU2})
consisting of only one mode $\mu(t)$
\begin{equation}
\ddot\mu+4\mu=0 \label{eq92fpu2}
\end{equation}
with the simple solution
\begin{equation}
\mu(t)=\mu_0\cos(2t). \label{eq93fpu2}
\end{equation}
taking (with no loss of generality) the initial phase to be equal
to zero. Substituting~(\ref{eq93fpu2}) into Eq.~(\ref{eq86bfpu2}),
we obtain
\begin{equation}
\ddot\nu+\left[2-\frac{8\alpha\mu_0}{\sqrt{N}}\cos(2t)\right]\nu=0.
\label{eq94fpu2}
\end{equation}
This equation can be easily transformed into the standard form of
the Mathieu equation \cite{AbramSteg}
\begin{equation}
\ddot\nu+[a-2q\cos(2t)]\nu=0. \label{eq95fpu2}
\end{equation}

As is well--known, there exist domains of stable and unstable
motion of the Mathieu equation~(\ref{eq95fpu2}) in the $a-q$ plane
of its parameters \cite{AbramSteg}. The one-dimensional bush
B$[\hat a^2,\hat\i]$ is stable for sufficiently small amplitudes
$\mu_0$ of the mode $\mu(t)$, but becomes unstable when $\mu_0>0$
is increased . This phenomenon, similar to the well-known
parametric resonance, takes place at $\mu_0$ values which lie
within the domains of unstable motion of Mathieu's
equation~(\ref{eq94fpu2}).

The loss of stability of the dynamical regime~(\ref{eq90fpu2}),
representing the bush B$[\hat a^2,\hat\i]$), manifests itself in
the exponential growth of the mode $\nu(t)$, which was identically
zero for the vibrational state~(\ref{eq90fpu2}) and oscillatory
for the stable regime of (\ref{eq95fpu2}). As a result of
$\nu(t)\neq 0$, the \textit{dimension} of the original
one-dimensional bush B$[\hat a^2,\hat\i]$ increases and the bush
is transformed into the two-dimensional bush B$[\hat a^4,\hat\i]$.
This is accompanied by the \emph{breaking of the symmetry} of the
vibrational state (the symmetry of the bush B$[\hat a^2,\hat\i]$
is twice as high as that of the bush B$[\hat a^4,\hat\i]$).

In general, we may view a given bush as a stable dynamical object
if the \emph{complete collection of its modes} (and, therefore,
its dimension) \emph{does not change in time}. All other modes of
the system--according to the definition of a bush as a full
collection of active modes--possess zero amplitudes and are
therefore called ``sleeping'' modes. If we increase the intensity
of bush vibrations, some sleeping modes (because of parametric
interactions with the active modes \cite{Chechin-Sakhnenko}) can
lose their stability and become excited. In this situation, we
speak of the \textit{loss of stability} of the \textit{original
bush}, since the dimension of the vibrational state (the number of
active modes) becomes larger, while its symmetry becomes lower.
Thus, as a consequence of stability loss, the original bush
transforms into another bush of higher dimension. In Section IV we
will discuss in more detail the problem of stability of invariant
tori from a more general perspective, when we study quasiperiodic
orbits in which all modes may be active.

Note, however, that what we described above is the loss of
stability of the (one--dimensional) bush B$[\hat a^2,\hat\i]$ with
respect to its transformation into the (two--dimensional) bush
B$[\hat a^4,\hat\i]$, using Floquet theory and analyzing a simple
Mathieu's equation. In the case of a more general perturbation of
the $N$ particles though, one must examine the stability of the
bush B$[\hat a^4,\hat\i]$ with respect to \textit{all other}
modes, as well. This cannot be done using Floquet theory as the
variations that need to be studied are \textit{not} about a
periodic but a \textit{quasiperiodic} orbit with two rationally
independent frequencies. The stability analysis of a
multidimensional bush is described theoretically in the next
subsection and is performed efficiently and accurately employing
the GALI numerical method outlined in section IV.

Indeed, as we shall show in Section IV, it is possible that tori
may continue to exist at energies \textit{higher} than the
thresholds (obtained by Floquet theory) where their ``parent''
modes become linearly unstable. This ensures stability over a
wider domain than just an infinitesimal neighborhood of the simple
periodic orbits (one-dimensional bushes) representing the NNM's of
the FPU chain.

\subsection{Stability of bushes of NNMs in the general case}

\subsubsection{Some general comments}

Some aspects of bush stability analysis have already been
discussed in the previous sections. In this last subsection of
Section III, we would like to complete the discussion of bush
stability using the apparatus of the irreducible representations
of symmetry groups.

In~\cite{Chechin-Zhukov}, a general group theoretical procedure
was developed for simplifying the linear stability analysis of
periodic and quasipereodic nonlinear regimes in $N$-particle
mechanical systems with arbitrary groups of discrete symmetry.

This procedure allows us to split the \textit{linearized}
equations (near a given dynamical regime) into a number of
\textit{independent} subsystems whose dimensions can be much
smaller than that of the full system. The basis vectors of the
irreducible representations of the appropriate groups are needed
to perform this splitting explicitly. On the other hand, we can
obtain more easily very useful ``splitting schemes'', which
determine how many subsystems of different dimensions can appear
as a result of the above splitting with the aid of the
group-representation \textit{characters} only. Let us discuss
these problems in more detail.

In general, the study of stability of periodic and, especially,
quasiperiodic dynamical regimes of mechanical systems with many
degrees of freedom presents considerable difficulties. Indeed, for
this purpose, we often need to integrate large linearized systems
of differential equations with time-dependent coefficients. In the
case of a periodic regime, one can use the Floquet approach, which
requires integration over only one time-period to construct the
monodromy matrix. However, for quasiperiodic regimes this method
is not applicable, and one needs to solve a system of several
differential equations for very long times in order to detect
stability (especially, near an instability threshold).

In such a situation, a decomposition (splitting) of the full
linearized system into a number of independent subsystems of small
dimensions proves to be very useful. Moreover, this decomposition
can provide valuable information regarding the specific degrees of
freedom, which are responsible for the first destabilization of a
dynamical regime as the energy (or a parameter value) is
increased. We remark that the number of such ``critical'' degrees
of freedom can frequently be rather small.

\subsubsection{Main theorem of bush stability}

Let us consider an $N$-degrees-of-freedom mechanical system
described by the set of autonomous differential equations
\begin{equation}
\ddot{\vec X}=\vec F(\vec X), \label{eq100zh}
\end{equation}
where the configuration vector $\vec
X=\{x_1(t),x_2(t),\dots,x_N(t)\}$ determines the deviation from
the equilibrium state $\vec X=\{0,0,\dots,0\}$, while the
vector-function $\vec F(\vec X)=\{f_1(\vec X),f_2(\vec
X),\dots,f_N(\vec X)\}$ provides the right-hand-side of the
dynamical equations.

We assume that Eq.~(\ref{eq100zh}) is invariant under the action
of a discrete symmetry parent group $G_0$. This means that for all
$g\in G_0$ Eq.~(\ref{eq100zh}) is invariant under the
transformation of variables
\begin{equation}
\widetilde{\vec X}=\hat g\vec X, \label{eq101zh}
\end{equation}
where $\hat{g}$ is the operator associated with the symmetry
element $g$ of the group $G_0$ by the conventional definition.

Using~(\ref{eq100zh}) and~(\ref{eq101zh}), we may write $\vec
X=\hat g^{-1}\widetilde{\vec X}$, $\hat
g^{-1}\ddot{\widetilde{\vec X}}=\vec F(\hat g^{-1}\widetilde{\vec
X})$, and finally
\begin{equation}
\ddot{\widetilde{\vec X}}=\hat g\vec F(\hat g^{-1}\widetilde{\vec
X}). \label{eq102zh}
\end{equation}
On the other hand, renaming $\vec X$ from Eq.~(\ref{eq100zh}) as
$\widetilde{\vec X}$, one can write $\ddot{\widetilde{\vec
X}}=\vec F(\widetilde{\vec X})$. Comparing this equation with
Eq.~(\ref{eq102zh}), yields $\vec F(\widetilde{\vec X})=\hat g\vec
F(\hat g^{-1}\widetilde{\vec X})$, or
\begin{equation}
\vec F(\hat g\vec X)=\hat g\vec F(\vec X). \label{eq1500zh}
\end{equation}
This is the invariance condition of the dynamical
equations~(\ref{eq100zh}) under the action of the operator $\hat
g$. Note that it must hold for all $g\in G_0$, hence it is
sufficient to consider such equivalence only for the
\textit{generators} of the group $G_0$.

Let $\vec X(t)=\vec C(t)$ be an $m$-dimensional specific dynamical
regime in the considered mechanical system that corresponds to the
bush B$[G]$ ($G\subseteq G_0$). This means that there exist
functional relations between the individual displacements $x_i(t)$
($i=1,2,\dots,N$), which reduce system~(\ref{eq100zh}) to $m$ ODEs
in terms of independent functions denoted by $a(t)$, $b(t)$,
$c(t)$, etc. in the previous section.

The vector $\vec C(t)$ is, therefore, a general solution of the
equation [see, Eq.~(\ref{eq2993}) in Appendix A]
\[
\hat G\vec X=\vec X,
\]
where $G$ is the symmetry group of the given bush B$[G]$
($G\subseteq G_0$).

Suppose now we wish to study the \textit{stability} of the
dynamical regime $\vec C(t)$, corresponding to the bush B$[G]$. To
this end, we must linearize the dynamical
equations~(\ref{eq100zh}) in a vicinity of the given bush, or more
precisely, in a vicinity of the vector $\vec C(t)$. Thus, let
\begin{equation}
\vec X=\vec C(t)+\vec\delta(t), \label{eq104zh}
\end{equation}
where $\vec\delta(t)=\{\delta_1(t),\dots,\delta_N(t)\}$ is an
infinitesimal $N$-dimensional vector. Substituting $\vec X(t)$
from~(\ref{eq104zh}) into~(\ref{eq100zh}) and linearizing these
equations with respect to $\vec\delta(t)$, we obtain
\begin{equation}
\ddot{\vec\delta}=\mat J[\vec C(t)]\vec\delta, \label{eq105zh}
\end{equation}
where $\mat J[\vec C(t)]$ is the Jacobi matrix of the
system~(\ref{eq100zh}):
\[
\mat J[\vec C(t)]=\left\|\left.\frac{\partial f_i}{\partial
x_j}\right|_{\vec X=\vec C(t)}\right\|.
\]

In~\cite{Chechin-Zhukov}, we have proved the following theorem:
\begin{theorem}
The matrix $\mat J[\vec C(t)]$ of the linearized dynamical
equations near a given bush B$[G]$, determined by the
configuration vector $\vec C(t)$, commutes with all matrices $\mat
M(g)$ ($g\in G$) of the natural representation of the symmetry
group $G$ of the considered bush:
\[
\mat M(g)\mat J[\vec C(t)]=\mat J[\vec C(t)]\mat M(g).
\]
\label{theor1zh}
\end{theorem}
We introduce, hereafter, a simpler notation for the Jacobi matrix:
\begin{equation}\label{eq405zh}
\mat J\left[\vec{C}(t)\right]\equiv\mat J(t).
\end{equation}
\noindent\textit{Remark}. Theorem 1 suggests that if we take a
symmetry element $g\in G_0$ that is not contained in $G$ ($g\in
G_0\setminus G$), the matrix $\mat M_g$ corresponding to $g$
\textit{may not} commute with $\mat J(t)$.

Based on the above, we can now apply the well-known Wigner theorem
(see, for example, \cite{ElliottDawber}) to split the linearized
system $\ddot{\vec\delta}=\mat J(t)\vec\delta$ into a number of
independent subsystems. In fact, it is useful to formulate this
theorem in a way that is convenient for our present purposes.

Consider a reducible representation $\Gamma$ of the group $G$
which can be decomposed into a direct sum of the irreducible
representations (irreps) $\Gamma_j$ of this group:
\begin{equation}
\label{eq48zh} \Gamma=\sum_j{^{^{\bigoplus}}m_j\Gamma_j}.
\end{equation}

Here $m_j$ is the number of times that $\Gamma_j$ enters into this
decomposition (subduction frequency). We denote the dimension of
the irrep $\Gamma_j$ by $n_j$. Then the Wigner theorem asserts the
following:

\begin{theorem}
Any matrix $\mat H$ commuting with all the matrices of a
representation $\Gamma$ of the group $G$ can be reduced to the
block-diagonal form
\begin{equation}
\label{eq49zh} \mat H=\sum{^{^{\bigoplus}}\mat D_j}
\end{equation}
so that: (a) the dimension of the each block $\mat D_j$ is equal
to $m_j n_j$ and (b) the block $\mat D_j$ consists of sub--blocks
representing matrices proportional to the identity matrix $\mat
I_n$ of dimension $n_j$ which are repeated $m_j$ times along the
rows and columns of the block $\mat D_j$.
\end{theorem}

The structure of one such block $\mat D_j=\mat D$ characterized by
the numbers $n_j=n$, $m_j=m$ is as follows:
\begin{equation}\label{eq50zh}
\mat D=\left(\begin{array}{cccc}
\mu_{11}\mat I_n & \mu_{12}\mat I_n & \ldots & \mu_{1m}\mat I_n\\
\mu_{21}\mat I_n & \mu_{22}\mat I_n & \ldots & \mu_{2m}\mat I_n\\
\ldots & \ldots & \ldots & \ldots\\
\mu_{m1}\mat I_n & \mu_{m2}\mat I_n & \ldots & \mu_{mm}\mat I_n\\
\end{array}\right),
\end{equation}

Let us apply the Wigner theorem for splitting the linearized
system $\ddot{\vec\delta}=\mat J(t)\vec\delta$ near the dynamical
regime (a bush of modes) with symmetry group $G$. To this end, we
will assume that $\mat H$ and $\Gamma$ of the theorem are
respectively the Jacobi matrix $\mat J(t)$ and the natural
representation of the group $G$.

To implement this splitting explicitly we must pass from the old
basis $\vec\Phi_{old}=\{\vec e_1,\vec e_2,\ldots,\vec e_N\}$ of
the natural space to the new basis
$\vec\Phi_{new}=\{\vec\varphi_1,\vec\varphi_2,\ldots,\vec\varphi_N\}$
formed by the complete set of the basis vectors $\vec\varphi_k$
($k=1,2,\ldots,N$) of all the irreps of the group $G$. If
$\vec\Phi_{new}=\mat S\vec\Phi_{old}$, then the unitary
transformation
\begin{equation}
\mat J_{new}(t)=\mat S^\dag\mat J_{old}\mat S \label{eq51zh}
\end{equation}
produces the above-discussed block-diagonal matrix $\mat
J_{new}(t)$ of the linearized system $\ddot{\vec\delta}_{new}=\mat
J_{new}(t)\vec\delta_{new}$ (here $\vec\delta_{old}=\mat
S\vec\delta_{new}$).

\subsubsection{Splitting schemes}

The above-discussed implicit decomposition of the Jacobi matrix
$\mat J(t)$ is a cumbersome procedure and, therefore, it is
interesting to know beforehand to what extent it can be useful in
specific cases. This can be easily determined by means of the
theory of \textit{characters} of group representations.

Let us consider such an approach by determining the
\textit{splitting scheme} of the linearized system
$\ddot{\vec\delta}=\mat J(t)\vec\delta$. In other words, we would
like to find out how many subsystems of different dimensions can
be obtained as a result of the decomposition of the Jacobi matrix.

Each block $\mat D_j$ in the decomposition~(\ref{eq49zh}) of the
matrix $\mat H\equiv \mat J(t)$ generates an independent subsystem
of $n_j m_j$ equations of the linearized system. However, each
subsystem automatically splits into $n_j$ new subsystems
consisting of $m_j$ differential equations, as a consequence of
the specific structure of these blocks [see Eq.~(\ref{eq50zh})].
For example, if a certain block for the matrix $\mat J(t)$
possesses the form ($n_j=3$, $m_j=2$)
\[
\mat J_j(t)=\left(\begin{array}{cc}
\mu_{11}\mat I_3 & \mu_{12}\mat I_3 \\
\mu_{21}\mat I_3 & \mu_{22}\mat I_3 \\
\end{array}\right),
\]
it is easy to check that one can obtain from the system
$\ddot{\vec\delta}=\mat J(t)\vec\delta$ three independent $2\times
2$ subsystems with one and the same matrix
\[
\mat M=\left(\begin{array}{cc}
\mu_{11} & \mu_{12} \\
\mu_{21} & \mu_{22}
\end{array}\right).
\]

Thus, to obtain the splitting scheme of the linearized system
$\ddot{\vec\delta}=\mat J\vec\delta$, for each irrep $\Gamma_j$ of
the bush symmetry group $G$, one must find two numbers---$n_j$
(the dimension of $\Gamma_j$) and $m_j$ (the subduction frequency
from Eq.~(\ref{eq48zh})).

The multiplicity numbers $m_j$ can be found by means of
Eq.~(\ref{eq113}) of Appendix A in terms of the characters
$\chi_j(g)$ of the irreps $\Gamma_j$, which are well--known for
many groups of discrete symmetry (in particular the point symmetry
groups) and $\chi_\Gamma(g)$, which represents  the character of
the reducible representation $\Gamma$. These characters can be
determined without the explicit construction of the matrices of
the natural representation $\Gamma$.

A quick and simple method for obtaining $\chi_\Gamma(g)$ can be
found in many textbooks (see, for example,~\cite{LandauLivKv}) in
connection with studying the small vibrations of multiatomic
molecules. The main point of the method is that, for a fixed $g\in
G$, the only atoms that contribute to $\chi_\Gamma(g)$ are those
whose position does not change under the action of $g$ on the
molecule. Moreover, this contribution is determined by the trace
of the three-dimensional matrix associated with the symmetry
element $g$. For example, the contribution to $\chi_\Gamma(g)$
from any atom invariant under the action of a rotation by an angle
$\varphi$ is $1+2\cos(\varphi)$, while that for a mirror rotation
is $-1+2\cos(\varphi)$.

Let us illustrate this in the case of an $N$-particle monoatomic
chain for $N\gg 1$. We denote by B$[\hat a^m,\dots]$ the bush
B$[G]$, with the symmetry group $G$ containing the translational
subgroup $[\hat a^m]$ and dots standing for other generators of
the group $G$. In~\cite{Chechin-Zhukov}, the following theorem was
proved:

\begin{theorem}
The linear stability analysis of any bush B$[\hat a^m,\dots]$ in
an $N$-degrees-of-freedom monoatomic chain can be reduced to the
stability analysis of isolated subsystems of the second order
differential equations with time-dependent coefficients whose
dimension does not exceed the integer number $m$.
\end{theorem}

\noindent\textit{Corollary}. If the bush dimension is $d$, one can
perform the above analysis for subsystems of \textit{autonomous}
differential equations with dimensions not exceeding $(m+d)$.

For example, the scheme for the linearized system about a
one-dimensional bush $B[\hat{a}^3, \hat{i}]$ of an arbitrary
nonlinear chain with $N=12$ particles allows us to split the
$12$~linearized equations into two one-dimensional, two
two-dimensional and two three-dimensional subsystems, which are
all independent from each other.

\subsubsection{Stability diagrams for NNMs in the FPU chains}

\begin{figure}[ht]
\centering
\psfig{file=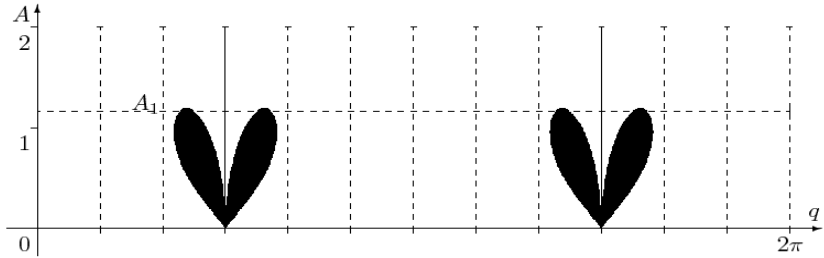,width=120mm,height=36mm}
\caption{\label{fig1zh} Regions of stability (white color) of
different modes of the FPU-$\beta$ chain, interacting
parametrically with the one-dimensional bush B$[a^4,iu]$.}
\end{figure}

The above discussed decomposition of the full linearized system
$\ddot{\vec\delta}=\mat J(t)\vec\delta$ into independent
subsystems of small dimension permits one to analyze efficiently
the stability of a given bush in monoatomic chains with
arbitrarily large number of particles. Using this idea, the
stability diagrams for all the one-dimensional bushes (NNMs) in
both FPU-$\alpha$ and FPU-$\beta$ chains were obtained
in~\cite{FPU2}. As an example, we reproduce here in
Fig.~\ref{fig1zh} the stability diagram for the bush B$[\hat
a^4,\hat\i\hat{u}]$ of the FPU-$\beta$ chain.

%\begin{figure}
%\centering \setlength{\unitlength}{1mm}
%\begin{picture}(130,45)(0,0)
%\put(5,5){\psfig{file=bB4aiu_new.eps,width=120mm,height=36mm}}
%\drawgridB
%\put(35,5){\line(0,1){36}}% 3
%\put(95,5){\line(0,1){36}}% 9
%\put( 5,26.0){\dashbox{1}(120,0){}}% A1
%\put(20,26.2){$A_1$}
%\end{picture}
%\caption{\label{fig1zh} Regions of stability (white color) of
%different modes of the FPU-$\beta$ chain, interacting
%parametrically with the one-dimensional bush B$[a^4,iu]$.}
%\end{figure}

In this diagram, each point $(A,q)$ provides a value of the bush
mode amplitude $A$ and the wave number $q=\frac{2\pi j}{N}$
associated with the index $j$ of a particular mode. The black
points $(A,q)$ correspond to the case where the mode
$j=q\frac{N}{2\pi}$ becomes excited because of its parametric
interaction with the mode of the bush B$[\hat a^4,\hat\i\hat{u}]$.
The white color denotes the opposite case: the corresponding mode
$j$ is zero initially and continues to be zero in spite of its
interaction with the considered bush. Such diagrams allow one to
study stability of one-dimensional bushes not only for finite $N$,
but also for the case $N\rightarrow\infty$ (for more details
see~\cite{FPU2}).

In the particular diagram depicted in Fig.~\ref{fig1zh}, we denote
the permissible values of the wave number $q$ for ${N=12}$ by the
dotted vertical lines (${q=2\pi j/12, j=1,2,...,12}$. The black
color corresponds to the regions of unstable motion in the plane
(${q-A}$). From Fig.~\ref{fig1zh} one can see that, for an
FPU-$\beta$ chain with ${N=12}$ particles, the NNM B$[a^4,iu]$
with ${\vec c=\{1,-1,-1,1~|~1,-1,-1,1~|~1,-1,-1,1\}}$ is
\textit{stable} for \textit{every} amplitude $A>0$ (at least up to
$A=20$), since the vertical dotted lines do not cross the rabbit--
ear shaped black (unstable) regions. In fact, the mode B$[a^4,iu]$
is also stable for chains with ${N=4}$ and ${N=8}$ particles,
since the corresponding vertical dotted lines are \textit{even
more distant} from each other than for ${N=12}$.

In the case ${N=16}$, however, there exist dotted vertical lines
which are \textit{tangent} to the rabbit-ear unstable regions,
while for $N>16$ (note that ${N\mod 4=0}$ must hold!) these lines
begin to \textit{cross} the unstable regions. Therefore, we can
conclude that the considered nonlinear normal mode in FPU-$\beta$
chains with ${N>16}$ becomes unstable for amplitude values, which
lie within the black regions in Fig.~\ref{fig1zh}. The case
${N=16}$ represents the ``boundary'' between stable and unstable
behavior of the B$[a^4,iu]$ NNM.

Using Fig.~\ref{fig1zh}, one can verify that the \textit{critical}
amplitude $A_c$ of the nonlinear normal mode B$[a^4,iu]$, at which
stability is lost, decreases with increasing $N$. Thus, one can
evaluate numerically the corresponding scaling law of the function
$A_c(N)$ in the thermodynamic limit, where $N\rightarrow\infty$
and $E\rightarrow\infty$ with $E/N$ fixed
\cite{ChechinRyabov2010}. It is also interesting to note that if
the amplitude  of the NNM $B[\hat{a}^4, \hat{i} \hat{u}$] exceeds
a certain critical value $A^{*}\approx 1.\!19$ (see
Fig.~\ref{fig1zh}), this NNM becomes stable again! Indeed, for
$A>A^{*}$, the horizontal lines of $A=$constant do not intersect
any unstable regions.

Let us summarize the results reviewed of this section concerning
the stability of bushes of NNMs:

(a) If the amplitude of a given NNM (one-dimensional bush) exceeds
a certain limit, this mode can lose its stability because of
specific interactions with some other NNMs of the considered
system. As a result, the symmetry of the vibrational state becomes
lower and the system passes to the dynamical regime of a higher
dimensional bush whose dimension $n>1$. This means that the
\emph{periodic motion} of the considered NNM becomes
\emph{quasiperiodic} with $n>1$ basic frequencies, which are
rationally independent.

(b) Of course, one cannot use Floquet theory to study stability of
bushes whose dimension $n>1$, but the group theoretical method
presented in subsection III.F  can simplify significantly the
analysis. Indeed, the methods we have described allow us to split
the linearized equations of motion near a bush into a number of
independent subsystems whose dimension can be much smaller than
that of the full system. Thus, when a given bush loses its
stability and transforms into another one of higher dimension, one
may ask: What new modes turn out to be involved into the
corresponding vibrational process?

The following questions also arise: How can one determine the
dimension $n$ of these quasiperiodic orbits and the frequencies of
the corresponding tori? What are the properties of the dynamics in
their vicinity? How can the stability (and hence the existence) of
these tori be determined accurately and efficiently so as to gain
a more global picture of the stability of motion of the system? To
answer these questions, we need to examine the phenomena of energy
localization and NNM interaction on low--dimensional tori,
employing analytical methods as well as the numerical approach of
the Generalized Alignment Indices (GALI), as we describe in
Section IV below.

\section{Energy localization on q--tori and stability of motion in FPU chains}

In sections I--III of this paper, we discussed extensively a group
theoretical method by which one can construct exact solutions of
mechanical systems described by $N$--degree of freedom
Hamiltonians. In particular, using as an example the FPU
$N$--particle chains, under periodic boundary conditions, we
exploited symmetry groups of their equations of motion to
determine all continuations of linear normal modes (termed here
nonlinear normal modes or NNMs), which constitute the so--called
one--dimensional bushes. Next, we considered linear combinations
of these NNMs, in ways which do not excite other modes and were
able to construct higher ($s>1$)--dimensional bushes, which at low
enough energies are \textit{quasiperiodic}, possessing $n$
generally incommensurate frequencies.

In this section, we complete our review by describing two more
recent developments in the study of low-dimensional tori of the
FPU system, in which one of us and his co-workers have played a
crucial role~\cite{SBAEPJ,SBAPhysD,BoMaChr,qtoriCBE}: The first
one starts, as in the earlier sections, with the construction of
$s(>1)$--dimensional bushes by linear combinations of NNMs of
\textit{low wave numbers} $q=1,2,3...$ and establishes the
important property of \textit{exponential localization} of the
energies of the Fourier modes belonging to the so--called
$q$--tori, through which the well--known paradox of FPU
recurrences can be better understood~\cite{BoMaChr,qtoriCBE}. It
is important to emphasize that this approach does not depend on
the symmetries of the system or the type of boundary conditions.

The second development concerns a more \textit{comprehensive
method} for studying the stability of quasiperiodic orbits,
without resorting to Floquet analysis, but using instead the
recently introduced spectrum of GALI
indicators~\cite{SBAEPJ,SBAPhysD}. Indeed, the GALI's asymptotic
behavior as $t\rightarrow\infty$ (typically up to $t=10^6 -
10^8$): (i) determines the \textit{dimension} $s$ of the torus, if
the GALI$_k$ indices tend to a constant for $k=2,3,...,s$ (and
decay by algebraic power laws for $s<k\leq 2N$) and (b) predicts
the \textit{instability} of a torus beyond an energy threshold
$E_{th}$ if, for $E>E_{th}$, \textit{all} GALI$_k$ start, at some
time t, to \textit{decay exponentially}, indicating that the orbit
we are following is \textit{not quasiperiodic} but diffuses
chaotically away from a torus that at energies $E\geq E_{th}$ no
longer exists.

\subsection{Energy localization on $q$--tori and FPU recurrences}

We shall use as our main example the FPU Hamiltonian for a chain
of $N$ particles, (\ref{fpuham}), which we rewrite here for
convenience:
\begin{equation}\label{fpuham}
H= {1\over 2}\sum_{k=1}^N p_k^2 + {  1   \over
2}\sum_{k=0}^N(x_{k+1}-x_k)^2 + {\alpha \over 3}
\sum_{k=0}^N(x_{k+1}-x_k)^3+{\beta \over 4}
\sum_{k=0}^N(x_{k+1}-x_k)^4
\end{equation}
We shall discuss primarily f.b.c., setting $x_0=x_{N+1}=0$, since
the case of p.b.c. does not present any major differences .

For the FPU $\beta$--model ($\alpha =0$) of this mechanical system
the normal mode canonical variables $(Q_q,P_q)$ are introduced by
the canonical transformations
\begin{equation}\label{lintra}
x_k=\sqrt{2\over N+1}\sum_{q=1}^N Q_q\sin\left({qk\pi\over
N+1}\right),~~~ p_k=\sqrt{2\over N+1}\sum_{q=1}^N
P_q\sin\left({qk\pi\over N+1}\right)
\end{equation}
Substitution of (\ref{lintra}) into (\ref{fpuham}) brings the
Hamiltonian (\ref{fpuham}) in the form $H=H_2+H_4$, where the
quadratic part represents $N$ uncoupled harmonic oscillators
\begin{equation}\label{fpuham2}
H_2=\sum_{q=1}^N {P_q^2+\Omega_q^2Q_q^2\over 2}
\end{equation}
with normal mode frequencies
\begin{equation}\label{fpuspec}
\Omega_q=2\sin\left({q\pi\over 2(N+1)}\right),~~~1\leq q\leq N~~~.
\end{equation}
On the other hand, the quartic part of the Hamiltonian becomes
\begin{equation}\label{fpuham4}
H_4={\beta\over 2(N+1)}\sum_{q,l,m,n=1}^N C_{q,l,m,n}
\Omega_q\Omega_l\Omega_m\Omega_n Q_qQ_lQ_mQ_n
\end{equation}
where the coefficients $C_{q,l,m,n}$ are non-zero only for
particular combinations of the indices $q,l,m,n$, i.e.
\begin{equation}\label{ccoef}
C_{q,l,m,n}= \left\{
\begin{array}{rl}
 1 & \mbox{if}~q \pm l \pm m \pm n = 0\\
-1 & \mbox{if}~q \pm l \pm m \pm n = \pm 2(N+1)
\end{array}
\right.
\end{equation}
with all possible choices of the $\pm$ signs taken into account.
Thus, in the new variables, the equations of motion are expressed
as follows:
\begin{equation}\label{eqmo}
\ddot{Q}_q+\Omega_q^2Q_q=-{\beta\over 2(N+1)}\sum_{l,m,n=1}^N
C_{q,l,m,n} \Omega_q\Omega_l\Omega_m\Omega_n Q_lQ_mQ_n~~~.
\end{equation}
The question we wish to pursue here is what happens to the
(linear) normal modes
\begin{equation}\label{LNM}
E_q=(P_q^2+\Omega_q^2Q_q^2)/2 ,~~~ q=1,2,....,N
\end{equation}
when $\beta\neq 0$? In other words, if we choose initial
conditions $x_k=x_k(0)$, $p_k=p_k(0)$, $k=1,2,...,N$ in
(\ref{lintra}) such that a small number (one, two, or more) modes
(\ref{LNM}) are excited, with energies $E_q$, $q=1,2,..$, will
these modes retain the total energy $E$ (and only exchange it
among themselves), or will they share it equally with all other
modes, as statistical mechanics of ergodic systems would require?

This was precisely the question asked by Fermi, Pasta and Ulam in
their famous experiments of the early 1950's \cite{Fermi1955}. As
is well--known, they obtained the surprising result that, at small
$E$ values, energy was \textit{not shared} among all modes, but
kept returning to the initially excited ones, giving rise to the
so--called FPU recurrences. There have been hundreds of papers to
date analyzing this phenomenon (for a recent review see
\cite{Berman}) and we shall also provide our own explanation here,
using the concept of \textit{q--tori} \cite{qtoriCBE}. The
difference between $q$--tori and the multidimensional bushes
discussed in earlier sections is that the set of NNMs excited in a
bush does not change in time, while in $q$--tori all modes
(\ref{LNM}) are active in general, albeit with exponentially
decreasing energies.

We now recall the theorem by Lyapunov \cite{Lyapunov} mentioned in
the Introduction, which states that if the frequencies $\Omega_q$
of the \textit{linear} normal modes (\ref{fpuspec})of an
$N$--degree--of--freedom Hamiltonian system are incommensurate,
these modes survive for $\beta\neq 0$. This condition of
non--commensurability is indeed satisfied in the f.b.c. case
(\ref{fpuspec}) for many $N$, e.g. a power of two minus 1, or a
prime number minus 1, among others \cite{Hemmer}. It fails,
however, in the p.b.c. case, for which the frequency spectrum
given by
\begin{equation}\label{fpuspecpbc}
\Omega_q=2\sin\left({q\pi\over N}\right),~~~1\leq q\leq N~~~.
\end{equation}
is commensurate \textit{for every $N$}, since
$sin(q\pi/N)=sin((N-q)\pi/N)$, for $q=1,2,...,[N/2]$. Thus, the
arguments and methods based on discrete symmetry groups, which we
presented in section II, turn out to be extremely useful for
proving the existence of NNMs in the p.b.c. case.

The group theoretical approach considers orbits as dynamical
states, see Eq. (\ref{eq4}), consisting of patterns (extended
primitive cells) which are periodically repeated, see e.g.
(\ref{eq13a})--(\ref{eqq20}), (\ref{eq11fpu2}). By construction,
therefore, these states are built by modes that correspond to high
wavenumbers $q=N/2,3N/4,...$, etc., while their symmetries may be
expressed by $N$--dimensional vectors of the type shown in
(\ref{eq70fpu2}). Indeed, these vectors provide an orthonormal
basis, by which one can construct multi--dimensional bushes using
linear combinations of such NNMs, as described in subsection
III.D.

There is, however, a more direct way to construct quasiperiodic
orbits, which neither employs symmetries nor uses special
$q$--values to achieve this purpose: It is the well--known
Poincar\'{e}--Linstedt method, which starts with any number of
modes with incommensurate frequencies $\omega_{q_i}$, for
$i=1,2,...,n$, and approximates the orbits through the solutions
of the linear problem, $x_i(t)=A_i cos(\omega_{q_i}t)$. Then,
through interactions provided by the nonlinear terms of the
equations of motion, new $q$--modes are born, which, interestingly
enough, have exponentially decaying energies in the $E_q$ vs. $q$
space. These so--called $q$--tori are seen to be $n$--dimensional
(with $n$ as low as 2 or 3), depending on how many modes are
initially excited and are \textit{stable} in the sense that the
tangent space of the corresponding quasiperiodic orbits is spanned
by $n$ linearly independent vectors, at least for very long times
and sufficiently low energies.

Let us see how all this is done explicitly on an example of the
FPU--$\beta$ model (\ref{fpuham}) with f.b.c. and $N=8$: We shall
construct a 2--dimensional torus representing the continuation of
the $q_1=1$ and $q_2=2$ linear normal modes,
$Q_1(t)=A_1\cos\Omega_1 t$, $Q_2=A_2\cos\Omega_2 t$, for
$\beta\neq 0$ and special values of $A_1, A_2$. Following the
Poincar\'{e} - Lindstedt approach, we seek solutions $Q_q(t)$,
$q=1,\ldots,8$ expanded as series in the parameter
$\sigma=\beta/2(N+1)$, namely
\begin{equation}\label{qser}
Q_q(t)=Q_q^{(0)}(t)+\sigma Q_q^{(1)}(t) + \sigma^2
Q_q^{(2)}(t)+\ldots,~~~q=1,\ldots 8~~~.
\end{equation}
For the motion to be quasi-periodic on a two--torus, the functions
$Q_q^{(r)}(t)$ must, at any order $r$, be trigonometric
polynomials involving only two frequencies $\omega_1$ and
$\omega_2$ (and their multiples). Moreover, these must arise as
small corrections of the linear normal mode frequencies
$\Omega_1$, $\Omega_2$, given by series expansions in powers of
$\sigma$, as:
\begin{equation}\label{omeser}
\omega_q=\Omega_q +
\sigma\omega_q^{(1)}+\sigma^2\omega_q^{(2)}+\ldots~~~q=1,2~~.
\end{equation}
As is well--known, the $\omega_q^{(i)}$, $i=1,2,...$ in
(\ref{qser}) are determined by the requirement that no secular
terms (of the type $t\sin\omega_qt$ etc.) arise in the equations
of motion.

For example, the equation for the first mode, contains on its
right hand side (r.h.s) terms of the form:
\begin{equation}\label{eqq1}
\ddot{Q}_1+\Omega_1^2Q_1=-\sigma(3\Omega_1^4Q_1^3
+6\Omega_1^2\Omega_2^2Q_1Q_2^2 +3\Omega_1^3\Omega_3Q_1^2Q_3+...)
\end{equation}
Substituting $\Omega_1$ on the l.h.s. of Eq.(\ref{eqq1}) by the
expression obtained by squaring (\ref{omeser}) and solving for
$\Omega_1^2$, inserting the expansions (\ref{qser}) into
(\ref{eqq1}) and grouping together terms of the same order, we
find that at zeroth order
$\ddot{Q}_1^{(0)}+\omega_1^2Q_1^{(0)}=0$, while at first order the
following equation is satisfied
\begin{eqnarray}\label{eqq11}
\ddot{Q}_1^{(1)}+\omega_1^2Q_1^{(1)}&=&2\Omega_1\omega_1^{(1)}Q_1^{(0)}
-3\Omega_1^4(Q_1^{(0)})^3-
6\Omega_1^2\Omega_2^2Q_1^{(0)}(Q_2^{(0)})^2\nonumber\\
&-&3\Omega_1^3\Omega_3(Q_1^{(0)})^2Q_3^{(0)}+...
\end{eqnarray}
Repeating the above procedure for modes 2 and 3, we find that
their zeroth order equations also take the harmonic oscillator
form, i.e. we have:
\begin{equation}\label{eqq2030}
\ddot{Q}_i^{(0)}+\omega_i^2Q_i^{(0)}=0,~~i=1,2~~~~~
\ddot{Q}_3^{(0)}+\Omega_3^2Q_3^{(0)}=0~~~~.
\end{equation}
Note that the {\it corrected} frequencies $\omega_1$, $\omega_2$
appear in the zeroth order equations for the modes 1 and 2, while
the {\it uncorrected} frequency $\Omega_3$ appears in the zeroth
order equation of the mode 3 (similarly,
$\Omega_4,\ldots,\Omega_8$ appear in the zeroth order equations of
the modes $q=4,...,8$). We now proceed with the solutions of
(\ref{eqq2030}) (with zero velocities at $t=0$) which read:
$$
Q_1^{(0)}(t)=A_1\cos\omega_1t,~~,Q_2^{(0)}(t)=A_2\cos\omega_2t,~~Q_3^{(0)}(t)=A_3\cos\Omega_3t
$$
where the amplitudes $A_1,A_2,A_3$ are arbitrary. If the orbit is
to lie on a two-torus with frequencies $\omega_1$ and $\omega_2$
only, we must set $A_3=0$. In the same way, the zeroth order
equations $\ddot{Q}_q^{(0)}+\Omega_q^2Q_q^{(0)}=0$, for the
remaining modes $q=4,...,8$, would yield solutions with new
frequencies and hence we need to set $A_4=A_5=...=A_8=0$ also.
Thus, we are left with only two non-zero free amplitudes to
determine: $A_1$, $A_2$.

As we explain in detail in \cite{qtoriCBE}, equation (\ref{eqq11})
can be simplified dramatically and upon substitution of
$Q_1^{(0)}(t)=A_1\cos\omega_1t$, $Q_2^{(0)}(t)=A_2\cos\omega_2t$
reduces to:
\begin{eqnarray}\label{eqq11n}
\ddot{Q}_1^{(1)}+\omega_1^2Q_1^{(1)}&=&2\Omega_1\omega_1^{(1)}A_1\cos\omega_1t
-3\Omega_1^4A_1^3\cos^3\omega_1t\nonumber\\
&-&6\Omega_1^2\Omega_2^2A_1A_2^2\cos\omega_1t\cos^2\omega_2t~~~.
\end{eqnarray}
We now fix $\omega_1^{(1)}$ so that no secular terms appear in the
solution, yielding
$$
\omega_1^{(1)}={9\over 8}A_1^2\Omega_1^3 +{3\over
2}A_2^2\Omega_1\Omega_2^2
$$
while we also determine $Q_1^{(1)}$ as:
\begin{eqnarray}\label{q11sol}
Q_1^{(1)}(t)&=&{3A_1^3\Omega_1^4\cos 3\omega_1t\over
32\omega_1^2}+{3A_1A_2^2\Omega_1^2\Omega_2^2\cos
(\omega_1+2\omega_2)t\over
2[(\omega_1+2\omega_2)^2-\omega_1^2]}\nonumber\\
&+&{3A_1A_2^2\Omega_1^2\Omega_2^2\cos (\omega_1-2\omega_2)t\over
2[(\omega_1-2\omega_2)^2-\omega_1^2]}~~.
\end{eqnarray}
Repeating the above analysis for the second mode, $Q_2(t)$, we
obtain the frequency correction of the second mode:
$$
\omega_2^{(1)}={9\over 8}A_2^2\Omega_2^3 +{3\over
2}A_1^2\Omega_1^2\Omega_2
$$
and the first order solution
\begin{eqnarray}\label{q21sol}
Q_2^{(1)}(t)&=&{3A_2^3\Omega_2^4\cos 3\omega_2t\over
32\omega_2^2}+{3A_1^2A_2\Omega_1^2\Omega_2^2\cos
(2\omega_1+\omega_2)t\over
2[(2\omega_1+\omega_2)^2-\omega_2^2]}\nonumber\\
&+&{3A_1^2A_2\Omega_1^2\Omega_2^2\cos (2\omega_1-\omega_2)t\over
2[(2\omega_1-\omega_2)^2-\omega_2^2]}~~
\end{eqnarray}
which has a similar structure as the first order solution of the
first mode. For the third order term there is no frequency
correction, and we find
\begin{eqnarray}\label{q31sol}
&~&Q_3^{(1)}(t)= {A_1^3\Omega_1^3\Omega_3\over 4}\left({3\cos
\omega_1t\over\omega_1^2-\Omega_3^2} + {\cos 3\omega_1t\over
9\omega_1^2-\Omega_3^2}\right) \nonumber\\
&+& {3A_1A_2^2\Omega_1\Omega_2^2\Omega_3\over 4}\left({\cos
(\omega_1-2\omega_2)t\over (\omega_1-2\omega_2)^2-\Omega_3^2} +
{\cos (\omega_1+2\omega_2)t\over
(\omega_1+2\omega_2)^2-\Omega_3^2}+{2\cos \omega_1t\over
\omega_1^2-\Omega_3^2}\right)~~.
\end{eqnarray}
We thus proceed up to the sixth mode and find that all the
functions $Q_3^{(0)},\ldots,Q_6^{(0)}$ are equal to zero, while
the functions $Q_3^{(1)},\ldots,Q_6^{(1)}$ are non zero. However,
a new situation appears when we arrive at the seventh and eighth
modes. A careful inspection of the equation for $Q_7^{(1)}(t)$
shows that there can be {\it no term} on the r.h.s. which does not
involve some of the functions $Q_3^{(0)},\ldots,Q_8^{(0)}$,
according to the selection rules for the coefficients of
Eq.(\ref{ccoef}). Since all these functions are zero, we must set
$Q_7^{(1)}(t)=0$, so as not to introduce a \textit{third}
frequency in the solutions, whence the series expansion
(\ref{qser}) for $Q_7(t)$ necessarily starts with terms of order
at least  $O(\sigma^2)$. The same holds true for the equation
determining $Q_8^{(1)}(t)$.

In \cite{qtoriCBE} it is shown that the Poincar\'{e}--Linstedt
scheme is \textit{consistent}, in the sense that the series
constructed above will \textit{never} encounter terms with zero
denominator, as long as our parameters lie in the complement of a
Cantor set in the $\omega_1,\omega_2$ (or $A_1,A_2$) space. We
then discuss the question of \textit{convergence} of the series,
which is in general a very difficult issue. However, as we also
explain below, the fact that the amplitudes squared
(or,equivalently  the energies) of the higher order terms of our
series are seen to decrease \textit{exponentially} with $q$
suggests that our series may indeed be absolutely convergent,
after all.

%__________________________________________________________
\begin{figure}
\centering
\psfig{file=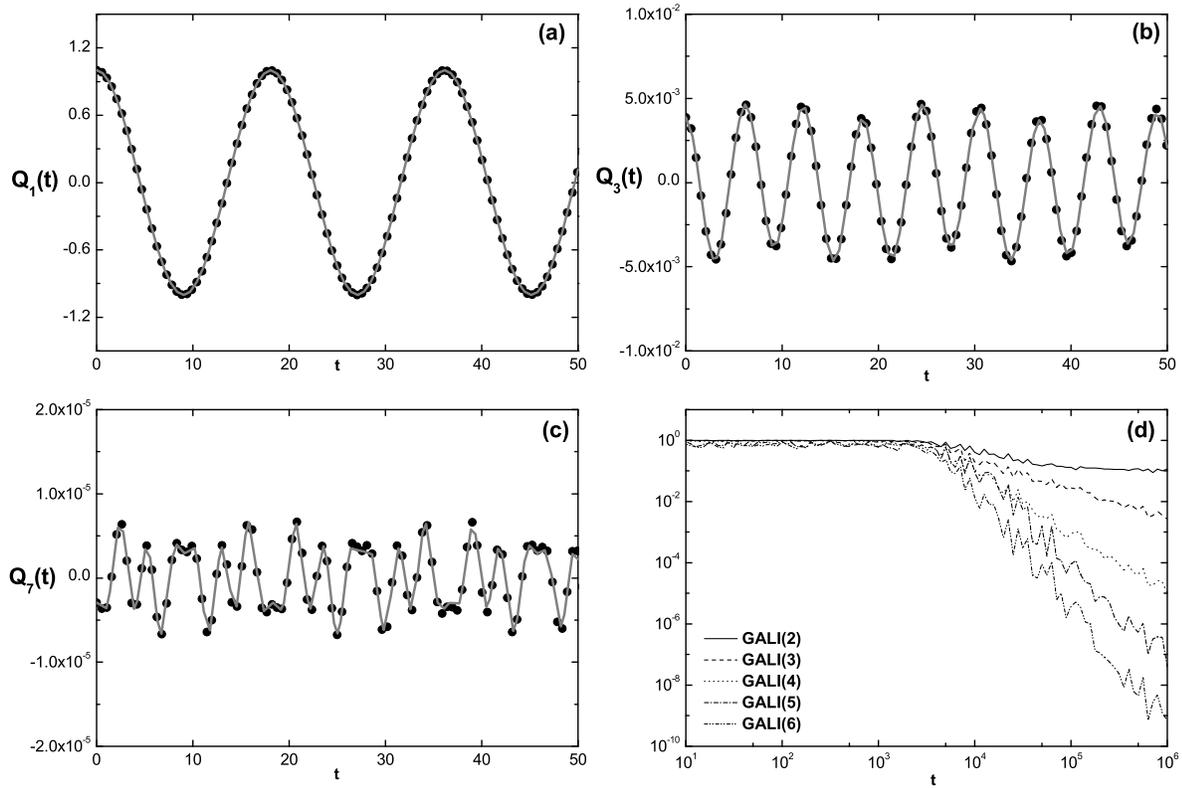,width=0.9\textwidth}
\caption{Comparison of numerical (points) versus analytical (solid
line) solutions, using the Poincar\'{e} - Lindstedt series up to
order $O(\sigma^2)$, for the temporal evolution of the modes (a)
$q=1$, (b) $q=3$, (c) $q=7$, when $A_1=1$, $A_2=0.5$, and $N=8$,
$\beta=0.1$ and (d) the time evolution of the indices GALI$_2$ to
GALI$_6$ up to a time $t=10^6$ show that the motion lies an
2--dimensional torus.} \label{qtor8}
\end{figure}
%_____________________________________________________________

To verify this numerically, let us use the analytical solutions
(\ref{qser}) at $t=0$ as initial conditions to integrate the
equations of motion and compare the computed $Q_i(t)$, $i=1,3,7$
with our analytical formulas. We also apply the GALI criterion
(see subsection IV.B) to show that the solutions lie indeed on
two-dimensional tori. As shown in Figures \ref{qtor8}a-c, the
numerically evaluated $Q_1(t),Q_3(t)$ and $Q_7(t)$ are in
excellent agreement with the analytical solution truncated at
second order with respect to $\sigma$, when $N=8$, $\beta=0.1$,
and $A_1=1$, $A_2=0.5$.

Concerning Fig. \ref{qtor8}d, let us note that, if an orbit lies
on a {\it stable} s-dimensional torus, the indicators
GALI$_2$,\ldots,GALI$_s$, discussed in the next subsection, are
nearly \textit{constant}, oscillating about {\it non-zero} values,
while the GALI$_{s+j}$, $j=1,2,...,2N-s$ decay asymptotically by
power laws, as $t^{-j}$ \cite{SBAPhysD,SBAEPJ,BoMaChr}.

This is precisely what we observe in Fig. \ref{qtor8}d. Namely,
the GALI$_2$ index stabilizes at a constant value GALI$_2\simeq
0.1$, while all subsequent indices, starting from GALI$_3$ decay
following a power law as predicted by the theory. Thus, we
conclude that the motion lies on a 2-torus, exactly as suggested
by the Poincar\'{e}-Lindstedt construction, despite the fact that
some excitation was provided initially to all modes. In fact, we
have carried out many such experiments, starting initially with
$s>2$ modes and, in every case, the analytical solutions were
found to be very accurate, while all the GALI indicators showed
that they lie on $s$--dimensional tori.

%_____________________________________________________________
\begin{figure}
\centering
\psfig{file=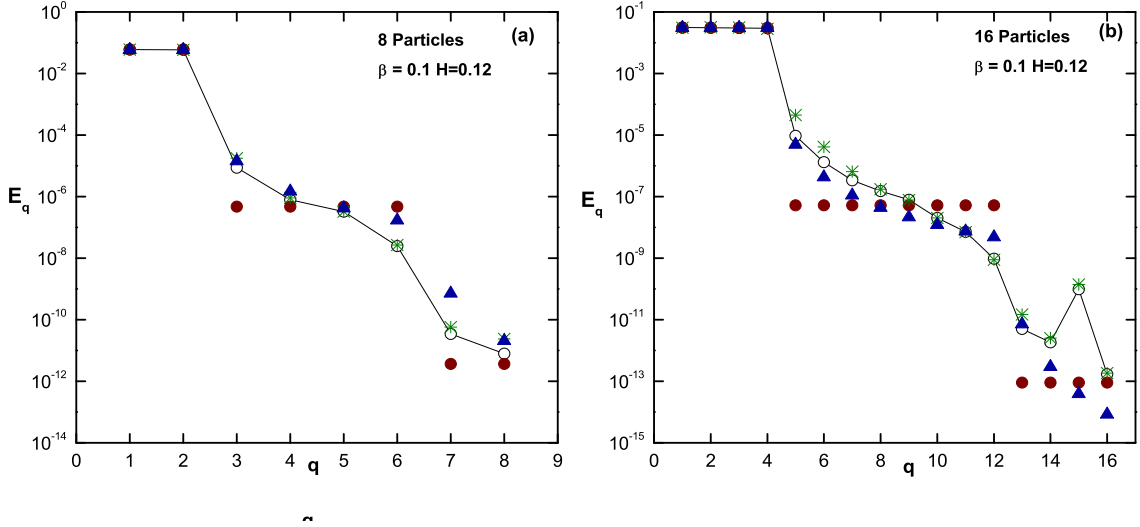,width=0.9\textwidth}
\caption{The average harmonic energy $E_q$ of the q-th mode as a
function of $q$, after a time $T=10^6$ for (a) the 2-torus of
Fig.\ref{qtor8}a and (b) a 4-torus solution (open circles). The
stars are $E_q$ values calculated via the analytical
representation of the solutions $Q_q(t)$ by the Poincar\'{e}
-Lindstedt series. The filled circles show a theoretical estimate
based on the average energy of suitably defined groups of modes
(see Eq.(\ref{eksp}) and relevant discussion in the text).}
\label{prof816}
\end{figure}
%_____________________________________________________________

Let us now evaluate, for such excitations, the average harmonic
energy $E_q$ of the $q$ mode and plot it as a function of the
wavenumber $q$, after a time interval $T=10^6$. As we observe in
Fig. \ref{prof816}a,b, the numerical results shown by the white
circles are very close to the analytical ones represented by the
stars, for an initial excitation with $s=2$ and one with $s=4$
modes respectively. Furthermore, what is clearly evident in these
semi-log plots, is that the magnitude of the $E_q$ falls
exponentially, though not with one and the same exponent. Let us
examine more closely this phenomenon of \textit{energy
localization}:

The average size of the oscillation amplitudes of each mode on an
s-dimensional q-torus follows from norm estimates of all terms of
the form $Q_{q^{(k)}}^{(k)}(t)$ appearing in equations such as
Eq.(\ref{q11sol}), where $q^{(k)}$ denotes the indices of all
modes excited at order $k$ of the perturbation. If we define by
$A^{(k)}$ the mean value of all the norms $||Q_{q^{(k)}}||$, we
obtain the estimate:
\begin{equation}\label{akfin}
A^{(k)} = {(Cs)^kA_0^{2k+1}\over 2k+1}
\end{equation}
where $A_0\equiv A^{(0)}$ is the mean amplitude of the
oscillations of all modes excited at the zeroth order of the
theory, and $C$ is a constant of order O(1). The proof of
(\ref{akfin}) is provided in \cite{qtoriCBE}, Appendix B, in which
the analytical estimate $C\simeq 3/2$ was given. In fact,
(\ref{akfin}) is a generalization of the estimate given by Flach
et al. \cite{flaetal2005,flaetal2006} for q-breathers, while the
two estimates become identical (except for the precise value of
$C$) if one sets $s=1$ in Eq.(\ref{akfin}), and $q_0=1$ in Flach's
q-breather formula.

Note, of course, that, with $|CsA_0^2|<1$, Eq. (\ref{akfin})
clearly implies that the amplitudes (and corresponding energies)
of the modes of these $s$--dimensional tori are decreasing
exponentially with $k$. This is reminiscent of the localization
properties of \textit{$q$--breathers}, studied by Flach et al.
\cite{flaetal2005,flaetal2006} in connection with the paradox of
FPU recurrences. We, therefore, decided to call all
$s$--dimensional tori whose modes are exponentially localized
\textit{$q$--tori}.

In fact, these $s$--dimensional $q$--tori, when excited at high
enough energies, give rise to the so--called ``natural packets" of
$m>s$ modes, which cause the breakdown of FPU recurrences by
becoming resonant, as suggested in a number of papers (see e.g.
\cite{beretal2004},\cite{deletal1999}) on the FPU $\beta$--chain.
Thus, $q$--tori offer a ``bridge" between the scenario described
in these references and the arguments based on energy localization
of q-breathers to provide a more complete interpretation of the
FPU recurrences.

Furthermore, using (\ref{akfin}), it is possible to obtain
`piecewise' estimates of the energy of each group, using a formula
for the average harmonic energies $E^{(k)}$ of the modes
$q^{(k)}$. To see how this is done, note that the total energy $E$
given to the system can be estimated as the sum of the energies of
the modes $1,\ldots,s$ (the remaining modes yield only small
corrections to the total energy), as
$$
E\sim s\omega_{q^{(0)}}^2 A_0^2\sim {\pi^2s^3A_0^2\over
(N+1)^2}~~.
$$
On the other hand, the energy of each mode $q^{(k)}$ can be
estimated from
$$
E^{(k)}\sim {1\over 2}\Omega_{q^{(k)}}^2 \left({\beta\over
2(N+1)}\right)^{2k}(A^{(k)})^2\sim
{\pi^2s^2(Cs\beta)^{2k}A_0^{4k+2}\over 2^{2k+1}(N+1)^{2k+2}}
$$
which, in terms of the total energy $E$, yields
\begin{equation}\label{ek}
E^{(k)}\sim {E\over s}\left({C^2\beta^2(N+1)^2E^2\over
\pi^4s^4}\right)^k~~.
\end{equation}
Eq.(\ref{ek}) is very similar to the corresponding equation for
q-breathers \cite{flaetal2006}
\begin{equation}\label{ekqb}
E_{(2k+1)q_0}\sim E_{q_0}\left({9\beta^2(N+1)^2E^2\over 64
\pi^4q_0^4}\right)^k
\end{equation}
where $q_0$ is the unique mode excited at zeroth order of the
perturbation theory and the integer $s$ plays in Eq.(\ref{ek}) a
role similar to that of $q_0$ in Eq.(\ref{ekqb}). This means that
the energy profile of a q-breather with $q_0=s$ obeys the same
exponential law as the energy profile of the s-dimensional
q-torus.

Still, the most important feature of the solutions on $q$--tori is
that their profile remains {\it invariant} as $N$ increases,
provided that: i) {\it a constant fraction} $M=s/N$ of the
spectrum is initially excited, (i.e. that $s$ increases
proportionally with $N$), and ii) the {\it specific energy
$\varepsilon=E/N$ remains constant}. Thus, in terms of the
specific energy $\varepsilon$, (\ref{ek}) takes the form
\begin{equation}\label{eksp}
E^{(k)}\sim {\varepsilon\over M}
\left({C^2\beta^2\varepsilon^2\over \pi^4M^4}\right)^k~~
\end{equation}
i.e. the profile becomes {\it independent} of $N$. A similar
behavior is recovered in the q-breather solutions provided that
the `seed' mode $q_0$ varies linearly with $N$, as was shown in
detail in refs.\cite{flaetal2007,kanetal2007}.

The `stepwise' profiles predicted by Eq.(\ref{ek}) in the case of
exact q-tori solutions are shown by filled circles in Fig.
\ref{prof816}, concerning examples of a 2--torus (see Fig.
\ref{qtor8}a) and a 4--torus solution (in all fittings we set
$C=1$ for simplicity). From these we see that the theoretical
`piecewise' profiles yield nearly the same ``plateaus'' and
average exponential slope of the profiles obtained numerically, or
analytically by the Poincar\'{e} - Lindstedt method.

Naturally, regarding the relevance of the q-tori solutions for the
interpretation of FPU recurrences, one must verify whether
Eqs.(\ref{ek}) or (\ref{eksp}) retain their predictive power in
the case of generic FPU trajectories which, by definition, are
started close to, but not exactly on a q-torus.

%_____________________________________________________________
\begin{figure}
\centering
\psfig{file=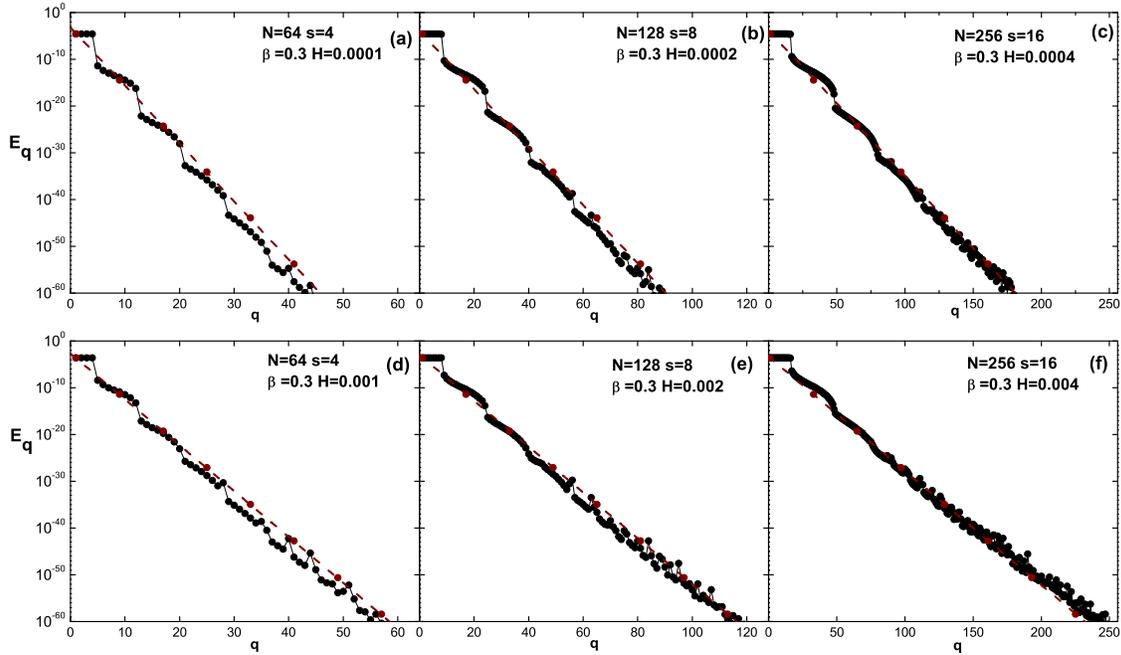,width=0.9\textwidth}
\caption{The average harmonic energy $E_q$ of the q-th mode over a
time span $T=10^6$ as a function of $q$ in various examples of
FPU-trajectories, for $\beta=0.3$, in which the $s(=N/16)$ first
modes are only excited initially via $Q_q(0)=A_q$,
$\dot{Q}_q(0)=0$, $q=1,\ldots,s$, with the $A_q$ selected so that
the total energy is equal to the value $E=H$ indicated in each
panel. We thus have (a) $N=64$, $E=10^{-4}$, (b) $N=128$,
$E=2\times 10^{-4}$, (c) $N=256$, $E=4\times 10^{-4}$, (d) $N=64$,
$E=10^{-3}$, (e) $N=128$, $E=2\times 10^{-3}$, (f) $N=256$,
$E=4\times 10^{-3}$. The specific energy is constant in each of
the two rows, i.e. $\varepsilon=1.5625\times 10^{-6}$ in the top
row and $\varepsilon=1.5625\times 10^{-5}$ in the bottom row. The
dashed lines represent the average exponential profile $E_q$
obtained theoretically by the hypothesis that the depicted FPU
trajectories lie close to q-tori governed by the profile
(\ref{eksp}).} \label{profpu1}
\end{figure}
\begin{figure}
\centering
\psfig{file=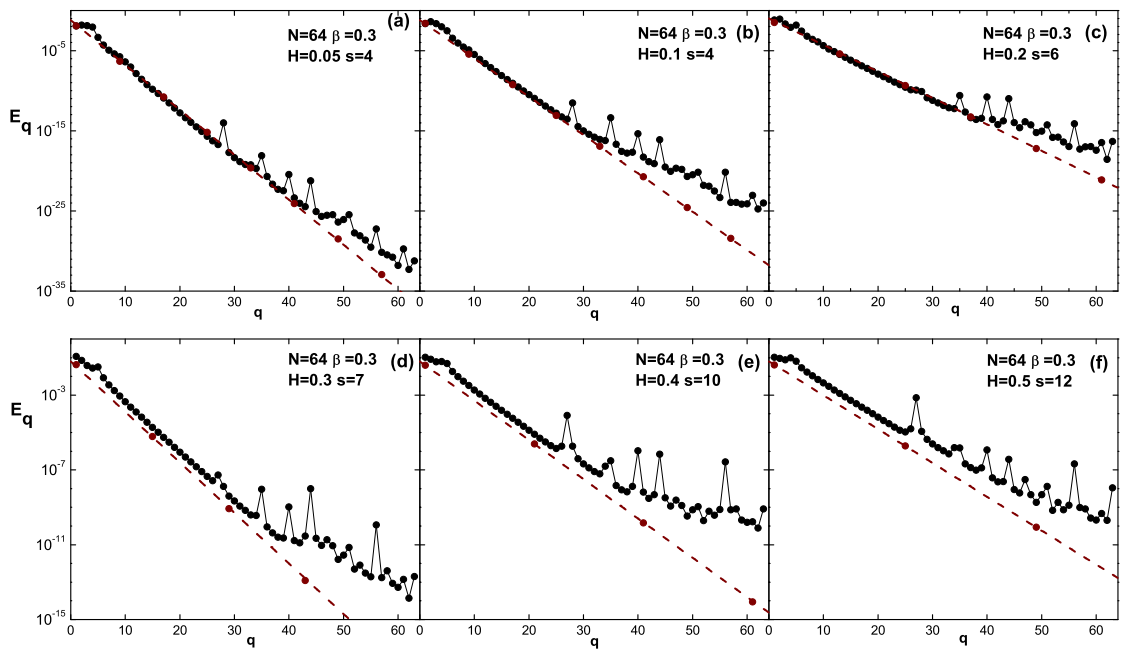,width=0.9\textwidth}
\caption{Same as in Fig.\ref{profpu1}a but for larger energies,
namely (a) $E=0.05$, (b) $E=0.1$, (c) $E=0.2$, (d) $E=0.3$, (e)
$E=0.4$, (f) $E=0.5$. Beyond the threshold $E\simeq 0.05$,
theoretical profiles of the form (\ref{ek}) yield the correct
exponential slope if $s$ is gradually increased from $s=4$  in (a)
and (b) to $s=6$ in (c), $s=7$ in (d), $s=10$ in (e), and $s=12$
in (f). } \label{profpu2}
\end{figure}

%_____________________________________________________________
To answer this question let us examine the results presented in
Figures \ref{profpu1} and \ref{profpu2}. Figure \ref{profpu1}
shows the energy localization profile in numerical experiments in
which $\beta$ is kept fixed ($\beta=0.3$), while $N$ takes the
values $N=64$, $N=128$ and $N=256$. In all six panels of Fig.
\ref{profpu1} the FPU-trajectories are computed starting with
initial conditions in which {\it only} the $s=4$ (for $N=64$),
$s=8$ (for $N=128$) and $s=16$ (for $N=256$) first modes are
excited at $t=0$, with the excitation amplitudes being compatible
with the values of the total energy $E$ indicated in each panel,
specific energy $\varepsilon=1.5625\times 10^{-6}$ in the top row
and $\varepsilon=1.5625\times 10^{-5}$ in the bottom row of Fig.
\ref{profpu1}. Thus, these FPU trajectories can be considered as
lying in the neighborhood of the q-torus solutions, at least
initially. The numerical evidence is that if $E$ is small, they
remain close to the q-tori even after relatively long times, e.g.
$t=10^6$.

This is evident in Fig. \ref{profpu1}, in which one sees that the
average energy profiles of the FPU-trajectories (at $t=10^6$)
exhibit the same behavior as predicted by Eq.(\ref{ek}), for an
exact q-torus solution with the same total energy as the FPU
trajectory in each panel. For example, based on the values of
their average harmonic energy, the modes in Fig. \ref{profpu1}a
(in which $s=4$) are clearly separated in groups, (1 to 4), (5 to
12) and (13 to 20),etc., as predicted theoretically for an exact
4-torus solution. The energies of the modes in each group have a
sigmoid variation around a level value characteristic of the
group, which is nearly the value predicted by Eq.(\ref{ek}). Also,
if we superpose the numerical data of the three top (or bottom)
panels we find that the average exponential slope is nearly
identical in all panels of each row, as implied by
Eq.(\ref{eksp}), according to which, for a given fraction $M$ of
initially excited modes, this slope depends on the specific energy
only, and is independent of $N$.

%_____________________________________________________________
%_____________________________________________________________

Of course, when the energy is increased, the FPU-trajectories
resulting from $s$ initially excited modes start to deviate from
their associated tori, when these become \textit{unstable}. As a
consequence, the energy profiles of the FPU-trajectories also
start to deviate from the energy profiles of the exact s-tori.
This is shown in Fig. \ref{profpu2}a by the fact that the profiles
of the FPU-trajectories become smoother and the groups of modes
less distinct, when the energy increases by a factor 50 (with
respect to Fig. \ref{profpu1}d), for the same values of $N$ and
$\beta$. We also observe in Fig. \ref{profpu2}a--f the formation
of the so-called `tail', i.e. an overall rise of the localization
profile at the high-frequency part of the spectrum. This is a
precursor of the evolution of the system towards equipartition,
which ultimately sets in at high enough energy.

\begin{figure}
\centering
\psfig{file=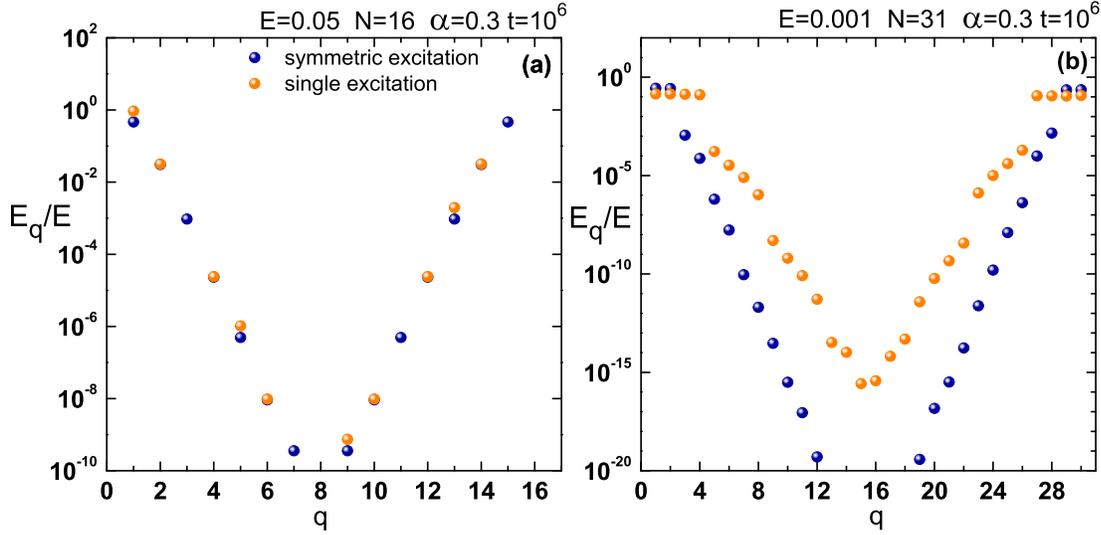,width=0.85\textwidth}
\caption{The average harmonic energy $E_q$ of the q-th mode as a
function of $q$ for the FPU--$\alpha$ model with $\alpha=0.3$ and
$\beta =0$ in (\ref{fpuham}) evaluated after a time $T=10^6$. (a)
A $q$--breather with $N=16$ is shown at energy $E=.05$ is (b) and
a 2-torus and 4-torus solution with $N=31$ are depicted at energy
$E=.001$ . Note that in (a), when only the $q=1$ mode is excited,
the profile is not as symmetric as it becomes when \textit{both}
$q=1$ and $q=N-1$ modes are excited equally.} \label{FPUalphapbc}
\end{figure}

This exponential localization of energy profiles also occurs in
FPU particle chains, when p.b.c. are imposed. In fact, the
profiles in this case become perfectly symmetric with respect to
the middle modes if one respects the symmetry of the (linear)
normal mode spectrum, (\ref{fpuspecpbc}), and excites equally the
$q$ mode \textit{and} $N-q$ mode, which have the same frequency,
see Fig. \ref{FPUalphapbc}. The existence and stability of the
corresponding $q$-tori at low energies is also evident here and is
expected to play an equally important role in the phenomenon of
FPU recurrences, as in the case of the fixed boundary conditions.

%_____________________________________________________________
%_____________________________________________________________

\subsection{Stability of the motion on q-tori and the GALI method}

As we have frequently emphasized, one of the most important
properties concerning the dynamics of conservative mechanical
systems, is whether the motion is ``regular'' or ``predictable''
for a given set of parameters and initial conditions. Since stable
periodic orbits are a set of measure zero, this typically implies
that the solutions we seek are quasiperiodic in time. In
$N$--degree--of --freedom Hamiltonian models, quasiperiodic orbits
are generically located around stable periodic orbits and lie on
$N$--dimensional tori.

In particular, when studying the phenomenon of FPU recurrences in
subsection IV.A, we realized that it is crucial to follow NNMs
with low wave numbers $q=1,2,3,...$. These are simple periodic
solutions called $q$--breathers (due to their exponential
localization in $q$--space), whose initial conditions are very
close to those of the recurrent FPU trajectories. Owing to this
proximity, one might expect that as long as $q$--breathers are
stable recurrences will continue for arbitrarily long times.

In fact, the situation is somewhat more complicated. The reason is
that exponential localization in $q$--space is also a property
shared by the tori surrounding these NNMs, which we called
$q$--tori in subsection IV.A. Actually it is \textit{their}
stability that matters and not that of the $q$--breathers, since,
as the energy increases, $q$--tori are seen to persist even after
the NNMs at their midst have turned unstable!

Of course, the stability of q-breathers can be studied by Floquet
theory, as shown e.g. in \cite{flaetal2006}, where it was
demonstrated that q-breathers with low $q=q_0=1,2,3,...$ are
linearly stable as long as
\begin{equation}\label{qbst}
E_{q_0}\leq \frac{\pi^2}{6\beta(N+1)}+O(1/N^2)
\end{equation}
This result is obtained by analyzing the eigenvalues of the
monodromy matrix of the linearized equations about a q-breather
constructed by the Poincar\'{e} - Lindstedt series. Furthermore,
examining also the associated eigenvectors one gains a complete
understanding of the tangent dynamics in the vicinity of these
simple periodic orbits.

In the case of q-tori, however, the above techniques are no longer
available. Stability is not decided by the eigenvalues of a
solution matrix and one must seek different ways of exploring the
variational equations linearized about quasiperiodic orbits. One
analytical approach for example, described in section III, is to
use discrete symmetries of the system to write the linearized
equations in the simplest form, expanding the solutions in terms
of linear combinations of NNMs, so that they ``uncouple'' as much
as possible.

Interesting as this approach may be, a more accurate stability
analysis should focus on the full set of variational equations and
examine the behavior of their solutions for \textit{arbitrary}
choices of initial conditions. This can be achieved by employing
the recently proposed method of the Generalized Alignment Indices
(GALI), which we outline below (see \cite{SBAEPJ}):

Consider an orbit $\gamma(t)=(\textbf{p}(t),\textbf{q}(t))$,
representing a solution of the Hamiltonian equations of motion
(\ref{eq:Hamiltonian_ODEs}) with initial conditions
$(\textbf{p}(0),\textbf{q}(0))$. The variational equations along
this orbit describe the dynamics on the tangent space of the
solutions of (\ref{eq:Hamiltonian_ODEs}) and are given by the
vector field
\begin{eqnarray}
\label{eqmetabolwn} \frac{d\textbf{w}}{dt}=J\cdot
M(\textbf{p}(t),\textbf{q}(t))\textbf{w}
\end{eqnarray}
where $J=\left(
\begin{array}{cc}
O & I_{N} \\
-I_{N} & O%
\end{array}%
\right) $, $I_{N}$ the $N\times N$ unit matrix and
$M(\textbf{p}(t),\textbf{q}(t))$ is the Hessian matrix of the
Hamiltonian function evaluated along $\gamma (t)$, i.e.
\begin{eqnarray} \label{Mmartix}
M(\gamma (t))=\frac{\partial^2 H }{
\partial \textbf{p}
\partial \textbf{q}} \mid _{\gamma (t)}~~~.
\end{eqnarray}

Let us now solve equations (\ref{eqmetabolwn}) for
$\textbf{w}_1(t)$, $\textbf{w}_2(t)$, ..., $\textbf{w}_k(t)$,
using as initial conditions $k$ randomly chosen orthocanonical
vectors and construct the unit vectors
\begin{eqnarray} \label{unitdev}
\hat{w}_i=\frac{\textbf{w}_i}{\parallel \textbf{w}_i \parallel
},~~i=1,...,k~~~.
\end{eqnarray}
The Generalized Alignment Index GALI$_k(t)$ is defined at time t
as the \textit{volume} of the parallelepiped produced by these $k$
unit deviation vectors $\hat{w}_i$, $i=1,...,k$ expressed, at time
t, by the wedge product
\begin{eqnarray} \label{GALIdef}
GALI_k(t)=\parallel \hat{w}_1 \wedge \hat{w}_2 \wedge ... \wedge
\hat{w}_k \parallel
\end{eqnarray}
As is well--known, this wedge product can be written with respect
to the basis $\hat{e}_{i_k}$ in the form
\begin{equation}
\hat{w}_1\wedge \hat{w}_2\wedge \cdots \wedge\hat{w}_k = \sum_{1
\leq i_1 <i_2 < \cdots < i_k \leq 2N} \left|
\begin{array}{cccc}
w_{1 i_1} & w_{1 i_2} & \cdots & w_{1 i_k} \\
w_{2 i_1} & w_{2 i_2} & \cdots & w_{2 i_k} \\
\vdots & \vdots &  & \vdots \\
w_{k i_1} & w_{k i_2} & \cdots & w_{k i_k} \end{array} \right|
\hat{e}_{i_1}\wedge \hat{e}_{i_2} \wedge \cdots \wedge
\hat{e}_{i_k} \label{eq:volume_2}
\end{equation}
From this it is easy to see that if at least two of the normalized
deviation vectors $\hat{w}_i$, $i=1,2,\ldots,k$ are linearly
dependent, all the $k \times k$ determinants will become zero and
the volume will vanish. Clearly, the norm of this quantity, $\|
\hat{w}_1(t)\wedge \hat{w}_2(t)\wedge \cdots \wedge\hat{w}_k(t)
\|$, is obtained by evaluating the sums \newline
\begin{equation}
GALI_{k}(t)=\left\{\sum_{1 \leq i_1 < i_2 < \cdots < i_k \leq 2N}
\left|
\begin{array}{cccc}
w_{1 i_1} & w_{1 i_2} & \cdots & w_{1 i_k} \\
w_{2 i_1} & w_{2 i_2} & \cdots & w_{2 i_k} \\
\vdots & \vdots &  & \vdots \\
w_{k i_1} & w_{k i_2} & \cdots & w_{k i_k} \end{array} \right|^2
\right\}^{1/2} \label{eq:norm}
\end{equation}
As was described in detail in a number of papers
\cite{SBAPhysD,SBAEPJ,BoChr}, it is possible to analyze the
\textit{asymptotic} behavior of these sums by estimating the
largest determinants as $t\rightarrow \infty$, both for chaotic as
well as quasiperiodic motion. Thus, the following asymptotic
estimates are derived:

I) For chaotic orbits:
\begin{equation}\label{eq:ch_GALI_decay}
\mbox{GALI}_k(t)  \propto e^{-\left[ (\sigma_1-\sigma_2) +
(\sigma_1-\sigma_3)+ \cdots+ (\sigma_1-\sigma_k)\right] t}\, .
\end{equation}
where
\begin{equation}
\sigma_1 \geq \sigma_2 \geq \cdots \geq \sigma_{N-1} \geq
\sigma_{N}=\sigma_{N+1}=0 \geq \sigma_{N+2} \geq \cdots \geq
\sigma_{2N}.
\end{equation}
are the Lyapunov exponents of the orbit \cite{SkoLyap}.

II) For quasiperiodic orbits lying on an $s$--dimensional torus:
\begin{equation}
\mbox{GALI}_k (t) \sim \left\{ \begin{array}{ll} \mbox{constant} &
\mbox{if $2\leq k \leq s$} \\
\frac{1}{t^{k-s}} & \mbox{if $s< k \leq 2N-s$} \\
\frac{1}{t^{2(k-N)}} & \mbox{if $2N-s< k \leq 2N$} \\
\end{array}\right. .
\label{eq:GALI_order_all}
\end{equation}
In the ``generic case'' $s=N$, (\ref{eq:GALI_order_all}) implies
that the GALI$_k$ remain constant for $2\leq k \leq N$ and
decrease to zero as $\sim 1/t^{2(k-N)}$ for $N< k \leq 2N$.
However, as we have seen in this review, the ``non--generic'' case
of low--dimensional tori, with $s<<N$, is also of great physical
interest and occurs quite frequently in mechanical systems of the
type considered here.

The accuracy of the above formulas, when compared against
numerical experiments, is remarkable. Indeed, as has been
demonstrated in many papers \cite{SBAPhysD,SBAEPJ,BoMaChr,BoChr}
the GALI$_k$ reach their asymptotic values after relatively short
times. In the chaotic case, the indices follow very closely the
theoretical values even when the Lyapunov exponents (and their
differences) are very small and have not as yet converged to their
limiting values. Indeed, it is important to note that what makes
the GALI method so efficient in distinguishing chaos from order is
the fundamental difference between exponential and power law
decay. In the case of chaotic orbits, \textit{all} indices
decrease exponentially and chaos is most quickly detected by
inspecting the ones with high k values, while for quasiperiodic
orbits all GALI$_k$ decrease by power laws, except for the first
few which are nearly constant and signify the dimension of the
torus.

Concerning the calculation of the GALI$_k$, it is clear that, for
large $N$, the computations become prohibitively time--consuming,
due to the number and size of determinants in (\ref{eq:norm}).
There is, however, a much faster way to proceed: Observe that the
normalized deviation vectors $\hat{w}_i$, $i=1,2,\ldots,2N$,
discussed above, when expressed in terms of the usual orthonormal
basis of the n--dimensional Euclidian space can be written as
\begin{equation}
\hat{w}_i=\sum_{j=1}^{2N} w_{ij} \hat{e}_j \,\,\, , \,\,\,
i=1,2,\ldots,k \label{eq:vec}
\end{equation}
Thus, we can write equations (\ref{eq:vec}) in matrix form as:
\begin{equation}
\left[\begin{array}{c} \hat{w}_1 \\ \hat{w}_2 \\ \vdots \\
\hat{w}_k
 \end{array} \right] = \left[
\begin{array}{cccc}
w_{11} & w_{12} & \cdots & w_{1\, 2N} \\ w_{21} & w_{22} & \cdots
& w_{2\, 2N} \\ \vdots & \vdots & & \vdots \\ w_{k1} & w_{k2} &
\cdots & w_{k\, 2N} \end{array} \right] \cdot
\left[\begin{array}{c} \hat{e}_1
\\ \hat{e}_2 \\ \vdots \\ \hat{e}_{2N} \end{array} \right] =
\textbf{A} \cdot \left[\begin{array}{c} \hat{e}_1 \\ \hat{e}_2 \\
\vdots \\ \hat{e}_{2N} \end{array} \right].\,\,\,
\label{eq:matrix_a}
\end{equation}
Recall that GALI$_k$ measures the volume of a $k$--parallelepiped
$P_k$ having as edges the $k$ unitary deviation vectors
$\hat{w}_i$, $i=1,\ldots, k $. This volume is given by
(\cite{HH}):
\begin{equation}
\mbox{vol}(P_k)=\sqrt{\det(\textbf{A} \cdot
\textbf{A}^{\mathrm{T}})}\,\,, \label{eq:vol_k}
\end{equation}
since $\det(\textbf{A} \cdot \textbf{A}^{\mathrm{T}})$ is equal to
the sum appearing in (\ref{eq:norm}) by virtue of Lagrange's
identity \cite{B58} (where $(^{\mathrm{T}})$ denotes transpose).
Thus we obtain a much more efficient way of computing GALI$_k$ by
the formula:
\begin{equation}
\mbox{GALI}_k = \sqrt{\det(\textbf{A} \cdot
\textbf{A}^{\mathrm{T}})},\label{eq:gali_det}
\end{equation}
where only the multiplication of two matrices and the square root
of one determinant appears.

Let us see how one can apply the above spectrum of indices to the
problems considered in this review. First of all, we have already
seen the usefulness of the GALI$_k$ in determining the
dimensionality of the $q$--tori described in subsection IV.A, see
Fig. \ref{qtor8}d and the discussion below it. Indeed, the near
constancy of the GALI$_k$, for $k=1,2,...,s$, at least for the
time intervals studied, is always seen to characterize stable tori
formed by exciting s linear normal modes of our FPU chains, at low
energies.

Quasiperiodic tori, however, can become \textit{unstable} by
losing their smoothness and eventually breaking down into sets of
points called cantori \cite{MacKayMeiss}. This means that the
variational equations about them possess a non--empty unstable
manifold through which small variations about the torus are
expected to grow exponentially. It follows, therefore, that beyond
the critical energy at which a $q$--torus becomes unstable, all
deviation vectors lying on the unstable manifold will tend to
align in the direction with the largest exponent, making all GALI
indices fall exponentially fast. Thus, the instability of the
torus can be verified, e.g. following the $k=2$ index, GALI$_2$,
which is the fastest to compute, even though the indices for
higher k values tend more quickly to zero.

%_____________________________________________________________
\begin{figure}
\centering
\psfig{file=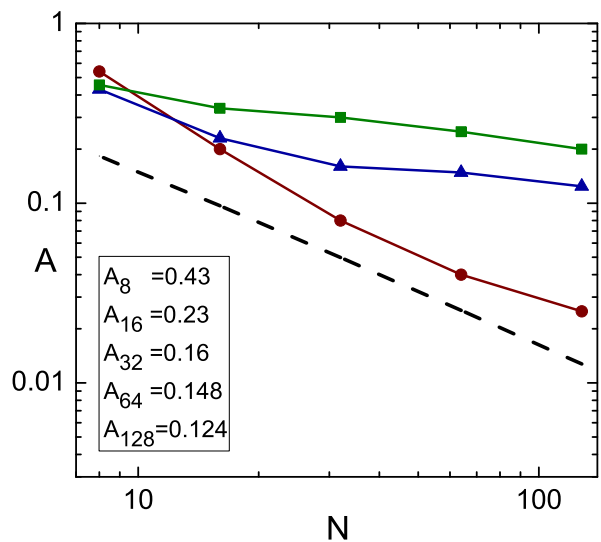,width=0.5\textwidth}
\caption{Fixing $\beta$ we plot in the upper curve (squares) the
dependence of the mean oscillation amplitude $A$ on $N$, for
critical energies $E_c$ determined by the destabilization of a
3-dimensional $q$--torus for FPU-trajectories identified as
quasiperiodic by teh GALI$_k$=const. for $k=1,2,3$. The middle
curve (triangles) corresponds to a similar calculation for an FPU
trajectory started by exciting initially a 2--dimensional
$q$--torus, while the lower one (circles) refers to a q-breather
solution ($q_0=1$). The dashed line represents the law of
Eq.(\ref{qbst}) and all calculations are carried out up to
$t=t_{max}=10^7$.} \label{linsta}
\end{figure}
%_____________________________________________________________

Using this criterion, it is easy to test the stability of q-tori
and determine approximately the value of the critical energy $E_c$
at which they become unstable. Fig. \ref{linsta} shows an example
of such a calculation using trajectories started close to a
3-torus (upper curve),  2-torus (middle curve), together with one
starting close to a $q$--breather (lower curve) for fixed $\beta$.
Observing the dependence of this instability threshold on N, we
discover that the $E_c$ values of at which the $q$--tori
destabilize are significantly {\it higher} than those
corresponding to the destabilization of $q$--breathers.

This is also inferred from the fact that the square and triangle
data in Fig. \ref{linsta} show a much weaker dependence of $A$ on
$N$ than the $N^{-1}$ law predicted by (\ref{qbst}) for
q-breathers. The relation $A\propto N^{-1}$ is seen to hold well
only for FPU-trajectories started close to a q-breather with $q=1$
(filled circles), whose numerical curve is close to the dashed
line predicted by Eq.(\ref{qbst}). These results strongly suggest
that the critical thresholds $E_c$ at which $q$--tori (obtained by
exciting the low modes) break down are significantly higher than
the energies at which the corresponding q--breathers become
unstable.

%_____________________________________________________________
\begin{figure}
\centering
\psfig{file=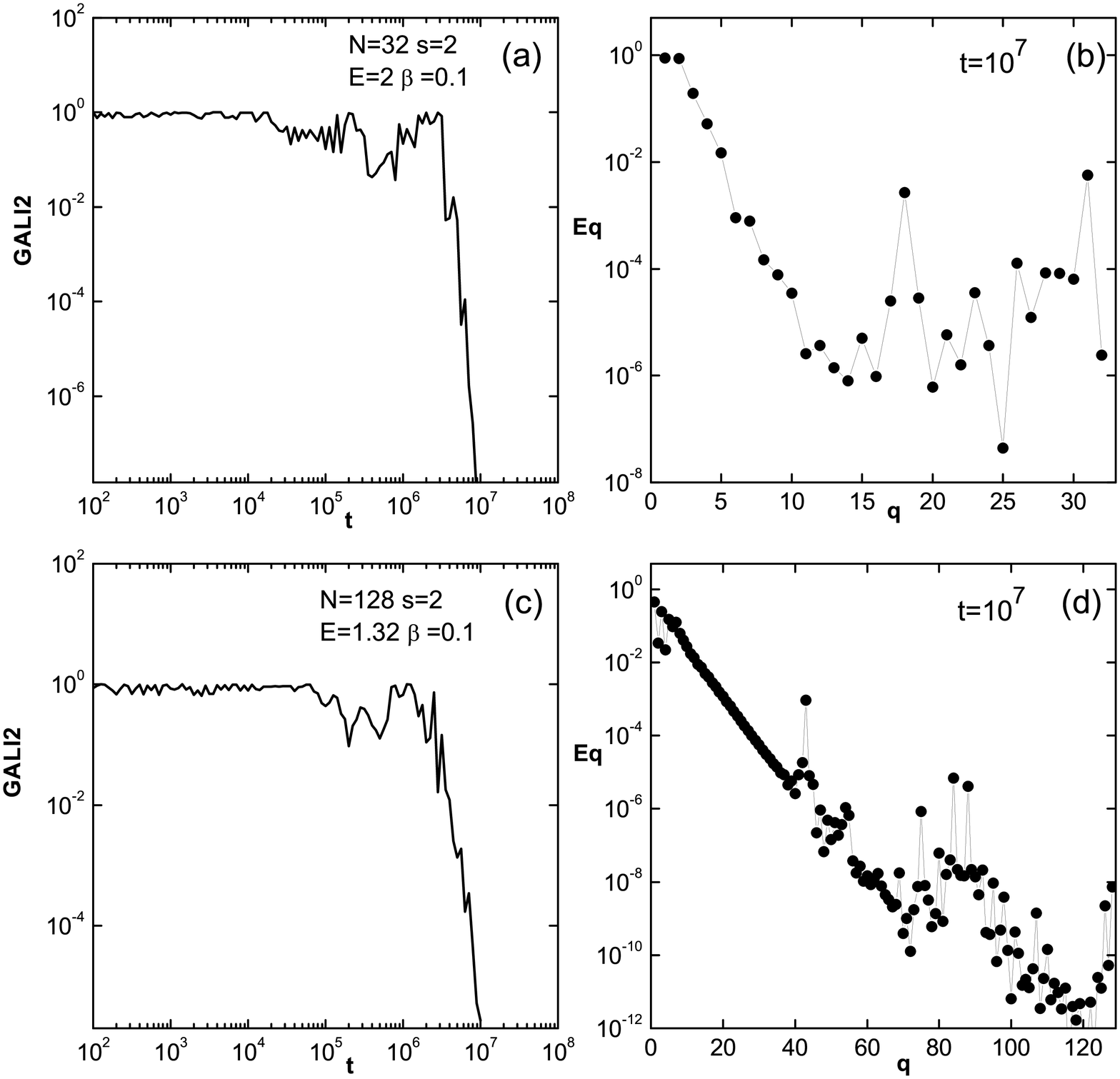,width=0.8\textwidth}
\caption{(a) Time evolution of the GALI$_2$ index up to $t=10^7$
for an FPU trajectory started by exciting the $q=1$ and $q=2$
modes in the system $N=32$, $\beta=0.1$, with total energy $E=2$,
i.e. higher than $E_c=1.6$. (b) {\it Instantaneous} localization
profile of the FPU trajectory of (a) at $t=10^7$. (c) Same as in
(a) but for $N=64$, $E=1.323$ (in this case the critical energy is
$E_c=1.24$). (d) Same as in (b) but for the trajectory of (c).}
\label{linsta2}
\end{figure}
%_____________________________________________________________

We, therefore, conclude this section by emphasizing that perhaps
the most important feature concerning the dynamics of orbits
forming q-breathers or q-tori is the exponential localization of
their energies, which persists even after these solutions have
become linearly unstable. This behavior is exemplified in Figure
\ref{linsta2}, where panels (a) and (c) show the time evolution of
the GALI$_2$ index for two FPU trajectories started in the
vicinity of 2-tori of FPU--$\beta$ systems with $N=32$ and
$N=128$, for $\beta=0.1$ and $t_{max}=10^7$. In both cases, the
energy satisfies $E>E_c$, as the exponential decay of the GALI$_2$
index evidently shows and the corresponding $q$--tori are
unstable. Still, as Figs. \ref{linsta2}b,d clearly demonstrate,
the exponential localization of the energies, at least for the
lower part of the spectrum, holds in this case also.

This localization property of the $q$--tori, therefore, provides,
in our view, a more complete interpretation of the phenomenon of
FPU recurrences. If there is still something ``paradoxical'' about
it, it is that it is not known to date if it holds to
$t\rightarrow \infty$, for small but non--zero energies. What is
certainly true (as Figs. \ref{linsta2}b,d show) is that the
destabilization of the $q$--tori brings about the
\textit{delocalization} of \textit{higher} $q$--modes and the
break down of recurrences, signifying the eventual equipartition
of energy, which is expected to spread to all modes at
$t>T_{eq}(\epsilon)$, where $\epsilon=E/N$ is the specific energy
and become practically observable when $\epsilon$ is large enough.

\appendix{}

\section{Elements of group representation theory}

\subsection{Natural representation}

We begin with an especially useful representation, called here
\textit{natural representation} of the parent group, using as an
example the square molecule model, depicted in Fig.~\ref{Fig1}. We
can describe the atomic displacement patterns of the vibrational
states of this molecule at any fixed time t by eight--dimensional
configuration vectors of the form
\begin{equation}\label{eq70}
    \vec{X} = \{\vec r_1 | \vec r_2 | \vec r_3 | \vec r_4\} = \{x_1, x_2|x_3, x_4|x_5, x_6|x_7, x_8\}.
\end{equation}
where $x_1$, $x_2$ are displacements of the first atom from the
equilibrium position (see Fig.~\ref{Fig1}) along the axes $X$ and
$Y$, while $\{x_3, x_4\}$, $\{x_5, x_6\}$, $\{x_7, x_8\}$ are
those of the atoms $2$, $3$ and $4$, respectively.

Let us act on an instantaneous  vibrational configuration of the
molecule by a certain symmetry element $g_i$, see
Eqs.(\ref{eq20}). Actually, we must act on the \textit{displaced}
atoms whose coordinates are determined by vectors $\vec R_j =\vec
R^0_j+\vec r_j $, where $\vec R_j^0 = (x_j^0, y_j^0)$ is the
equilibrium position of the $j$-th atom, while $\vec r_j = (x_j,
y_j)$ is the local vector which determines the displacement of
this atom from its equilibrium position.

Since $g_i \vec R_j=g_i \vec R_j^0+g_i \vec r_j$, one can
introduce \textit{operators} $\hat{g}_i$, which are
\textit{induced} by the symmetry elements \footnote{Note that
symmetry element $g$ acts on the vectors $\vec r$ of
three--dimensional Euclidean space, while operator $\hat{g}$ acts
on vectors of the eight-dimensional configuration space. In
general, the induced operator $\hat{g}$, acting on an arbitrary
function $f(\vec r)$, is determined by the equation $\hat{g}f(\vec
r) = f(g^{-1}\vec r)$, as shown in any textbook on the theory of
symmetry groups (see, for example, \cite{Hammermesh,
ElliottDawber}).} $g_i \in G_0$, acting on the eight dimensional
configuration vector (\ref{eq70}) by the following two step
process:

1) \textit{transpose} the local displacement vectors $\vec r_j =
(x_j, y_j)$ in (\ref{eq70}) according to the transposition of atom
equilibrium positions under the action of the symmetry element
$g_i^{-1}$;

2) \textit{act} on each local displacement vector $\vec r_j$ by
the symmetry element $g_i^{-1}$.

For example, we obtain for the symmetry element $g_2$:
\begin{equation}\label{eq72}
    \hat{g}_2\vec X  = \{g_4\vec r_2|g_4\vec r_3|g_4\vec r_4|g_4\vec r_1\} =
    \{y_2,-x_2|y_3,-x_3|y_4,-x_4|y_1,-x_1\},
\end{equation}
taking into account relations $g_2^{-1} = g_4$ and $g_4(x,y) = (y,
-x)$ ($g_2$ and $g_4$ are rotations about the Z axis through the
angles $90^\circ$ and $270^\circ$, respectively).

The above definition of the operators $\hat{g}_i$ induced by
symmetry elements $g_i$ allows one to construct an
eight-dimensional natural representation of the symmetry group
$G_0 = C_{4v}$ for the considered square molecule. To this end, we
may choose the ``natural'' basis
$\vec{\Phi}=\{\vec{e}_1,\vec{e}_2,\ldots,\vec{e}_N\}$ of
orthogonal unit vectors $\vec{e}_i$ (having 1 in their ith entry
and zero everywhere else) to describe all possible displacements
in the configuration space.

Next, acting by an operator $\hat g$ ($g\in G_0$) on the vector
$\vec{e}_j$, we can write $\hat{g}\vec{e}_j$ as a certain linear
combination of all basis vectors:
\begin{equation}
\hat g\vec{e}_j=\sum_{i=1}^{N}\mathcal M_{ij}(g)\vec{e}_i,\qquad
j=1,2,\dots,N. \label{eq6001zh}
\end{equation}
This equation associates the matrix $\mathcal M(g)= \{\mathcal
M_{ij}(g)\}$ with the operator $\hat g$ and, therefore, with the
symmetry element $g\in G_0$:
\begin{equation}
g\Rightarrow\hat g\Rightarrow\mathcal M(g).
 \label{eq6002zh}
\end{equation}
The set of matrices $\mathcal M(g)$ corresponding to all $g\in
G_0$ forms the natural representation $\Gamma$ for our mechanical
system.

In the case of the square molecule, $N = 8$, therefore, this
representation turns out to be eight-dimensional~--- and
represents a collection of $8\times8$ matrices $\mathcal M(g)$
which correspond to all the elements $g$ of the parent group
$G_0$.

Moreover, these matrices satisfy the same multiplication rule as
the symmetry elements corresponding to them: if $g_ig_j = g_k$,
then $M(g_i)\cdot M(g_j) = M(g_k)$. Among them there is the unit
matrix [$M(g_1) = I$], and for every matrix there also exists its
inverse [$M(g^{-1}) = M^{-1}(g)$]. Taking into account the
associative rule for matrix multiplication, we can show that all
different matrices belonging to the set $\{M(g)\,|\, \forall g\in
G_0\}$ generate a certain \textit{matrix group}
\textit{homeomorphic} to the parent group $G_0$ (In general, a
matrix representation can contain a number of identical matrices).

It is well--known that any finite group possesses an infinite
number of representations whose dimensions can be arbitrarily
large. However, there exists only a \emph{finite number} of
nonequivalent \emph{irreducible representations} (irreps). Let us
consider this issue in more detail.

\subsection{Irreducible representations}

Let us first use the basis $\Phi=\{\vec{e}_1,...,\vec{e}_N\}$ to
construct the natural representation of the parent group $G_0 =
C_{4v}$. Of course, one may use any other basis of the
configuration space for the same purpose. This leads to
considering a new form of matrices corresponding to symmetry
elements $g\in G_0$ as follows:
\begin{equation}\label{eq77}
    M_{new}(g) = S^{-1}M(g)S, \ \ \ \ \ \ \ \forall g\in G_0.
\end{equation}
where $S$ is a non--singular matrix. Based on this idea, we may
ask: "How can one choose the basis of our configuration space to
make the matrices $M(g)$ of the natural representation as simple
as possible?"

Indeed, let us recall that if a number of matrices commute with
each other they can be transformed to the simplest possible~---
\emph{diagonal}~--- form via a transformation of the type
(\ref{eq77}), using a suitably chosen matrix $S$. On the other
hand, if the matrices do not commute, they cannot be diagonalized
simultaneously by transformation of the form (\ref{eq77}), i.e.
such a matrix $S$ does not exist.

The natural representation is only one example of a matrix
representation of a given symmetry group. Continuing  our
discussion of arbitrary matrix representations of finite groups,
we now consider representations $\Gamma_j$ of the group $G$ by
\textit{complex} $n\times n$ matrices $\mathcal M(g), g\in G$.
These matrices act on the vectors of an $n$-dimensional complex
vector space which is called the \textit{carrier} space of
$\Gamma$.

In physics, we usually deal with \textit{unitary}
representations~\cite{Hammermesh, ElliottDawber}. All matrices
$\mathcal M(g)$ of such representations are unitary matrices,
satisfying $\mathcal M^\dag(g)=\mathcal M^{-1}(g)$, $\forall g\in
G$.

\begin{definition}
Two matrix representations $\Gamma=\{~\mathcal M(g)~|~g\in G~\}$
and $\Gamma'=\{~\mathcal M'(g)~|~g\in G~\}$ of the group $G$ are
called \textit{unitary equivalent}, if there exists a unitary
matrix $S$ (i.e. $\mathcal S^\dag=\mathcal S^{-1}$), such that all
their matrices obey the relation
\begin{equation}\label{eq503}
\mathcal M'(g)=\mathcal S^\dag\mathcal M(g)\mathcal S,
\end{equation}
\end{definition}

The transformation~(\ref{eq503}) appears when we pass from one
orthonormal basis of the carrier space of a given representation
to another. Let us now define the matrix representation $\Gamma$
associated with a certain basis $\vec\Phi=\{\vec\varphi_1(\vec
r),\vec\varphi_2(\vec r),\ldots,\vec\varphi_n(\vec r)\}$ of the
carrier space:
\begin{definition}
If
\begin{equation}\label{eq333}
\hat g\vec\Phi=\widetilde{\mathcal M}(g)\vec\Phi\quad(\text{for
all }g\in G),
\end{equation}
where operator $\hat g$ acts on the basis vectors
$\vec\varphi_j(\vec r)$ as $\hat g\vec\varphi_j(\vec
r)=\vec\varphi_j(g^{-1}\vec r)$, then
\begin{equation}\label{eq334}
\Gamma=\{~\mathcal M(g)~|~\forall g\in G~\}
\end{equation}
is a matrix representation of the group $G$. Note that
$\widetilde{\mathcal M}(g)$ denotes the transpose of the matrix
$\mathcal M(g)$.
\end{definition}

Let us also denote by the symbol $V$ the carrier space of the
matrix representation $\Gamma$ of the group $G$ and suppose we can
write $V$ as a \textit{direct} sum
\begin{equation}\label{eq80}
    V = \sum_{j=1}^K V_j
\end{equation}
of a number $K>1$ of \textit{invariant subspaces} $V_j$, which are
\textit{independent} of each other under the action of operators
$\hat{g}$ induced by all the elements of the group $G$. More
precisely the above invariance can be explained as follows: Each
operator $\hat{g}$ ($g\in G$) acting on any vector $\vec v$
belonging to $V_j$ transforms it into another vector of the
\textit{same subspace}, i.e. $\hat{g}\vec v$ cannot
\textit{coincide} with a vector of any other subspace in the sum
(\ref{eq80}) different from $V_j$.

If we choose an arbitrary basis in the subspace $V_j$, we obtain a
matrix representation $\Gamma_j$ of the group $G$ which is
constructed on this basis according to the general definition
(\ref{eq333},\ref{eq334}).

Since the basis of the full space $V$ can be chosen as the union
of all subspaces $V_j$ entering Eq. (\ref{eq80}), we find that the
representation $\Gamma$ can be written as a \emph{direct sum} of
all the representations $\Gamma_j$ constructed using the bases of
the individual subspaces $V_j$:
\begin{equation}\label{eq81}
    \Gamma = {\sum_{j=1}^N}^{\bigoplus}\Gamma_j.
\end{equation}

The dimension $n$ of the full carrier space $V$ is the sum of the
dimensions $n_j$ of all its invariant subspaces $V_j$:
\begin{equation}\label{eq82}
    n = \sum_{j=1}^N n_j.
\end{equation}

If a subspace $V_j$ \textit{cannot} be split into invariant
subspaces of \textit{smaller} dimensions, it is called
\textit{irreducible} and so is the representation $\Gamma_j$,
defined on it. On the other hand, the representation $\Gamma$,
defined on the full space $V$ is called "reducible", if it can be
decomposed into a finite number of $K>1$ irreducible
representations.

As a consequence of Eq. (\ref{eq81}), a reducible representation
$\Gamma$ possesses \textit{block diagonal} form, whose blocks turn
out to be the individual irreducible representation $\Gamma_j$.
For the given matrix $D_i(g)\in \Gamma_i$, each of these blocks,
$\mathcal D_{nj}^{(i)}(g)$, represents an $n\times n$ matrix
corresponding to $g\in G$, while $i$ numbers the blocks of one and
the same dimension ($n$). Observe that, when a certain irrep
$\Gamma_j$ is contained in the reducible representation $\Gamma$ a
number of times $m_j$ (the subduction frequency), some blocks may
be identical.

We may, therefore, rewrite equation~(\ref{eq81}) in the  form:
\begin{equation}\label{eq112}
\Gamma=\sum_j{^{^{\bigoplus}}m_j\Gamma_j},
\end{equation}
where $m_j$ indicates how many times the irrep $\Gamma_j$ is
contained in the decomposition of the reducible representation
$\Gamma$. Let us note that the basis vectors of different copies
of the same irrep $\Gamma_j$ are essentially \textit{different}.
One should also emphasize that matrices of all irreps $\Gamma_j$
must be known to obtain the explicit decomposition of the given
reducible representation, while for the \emph{schemes of the
decomposition}~(\ref{eq112}) one needs to know only the
\emph{characters} of these irreps.

\begin{definition}
A character $\vec\chi[\Gamma]$ of the $n$-dimensional
representation is the $n$-dimensional vector constructed by the
\textit{traces} of all the matrices $\mathcal M(g)$ of this
representation:
\[
\vec\chi[\Gamma]=\frac{1}{\sqrt{\|G\|}}\{\text{Tr}\mathcal
M(g)~|~g\in G\},
\]
where $\|G\|$ denotes the order of the group $G$.
\end{definition}

There exists a powerful group theoretical apparatus using the
irreps and their characters for solving a wide class of crystal
and molecular systems. The characters of the irreps of a given
group are orthonormal vectors. For calculating the subduction
frequency $m_j$ of the irrep $\Gamma_j$ from Eq.~(\ref{eq112}),
one can use the following formula~\cite{Hammermesh,
ElliottDawber}:
\begin{equation}\label{eq113}
m_j=\left(\vec\chi[\Gamma],\bar{\vec\chi}[\Gamma_j]\right)\equiv
\frac{1}{||G||}\sum_{g\in
G}\chi_{\footnotesize{\Gamma}}(g)\bar{\chi}_{j}(g).
\end{equation}
Here $\vec\chi[\Gamma]$ is the character of the reducible
representation $\Gamma$, while $\vec\chi[\Gamma_j]$ is the
character of the irreducible representation $\Gamma_j$
($\bar{\vec\chi}[\Gamma_j]$ is the complex conjugate value with
respect to $\vec\chi[\Gamma_j]$).

\subsection{Some properties of irreducible representations and their role in Physics}

Irreducible representations (irreps) possess many remarkable
properties which can be found in any textbook on group theory
(see, for example, \cite{Hammermesh, ElliottDawber}). In
particular, it is well--known that any group $G$ of finite
order$\|G\|$ possesses a \textit{finite number} of different
nonequivalent irreps, and this number is equal to the number of
classes of conjugate elements of the given group. The dimension
($n_j$) of each irrep ($\Gamma_j$) must be a divisor of the group
order $\|G\|$, i.e. $\|G\|$ mod $n_j = 0$. Dimensions of all the
irreps of the group $G$ satisfy the Burnside theorem:
\begin{equation}\label{eq85}
    \sum_{j=1}^N n_j^2 = \|G\|.
\end{equation}
Among irreps of any group $G$ there is the so-called
\textit{identity} representation which associates with every
element $g\in G$ one and the same one-by-one unit matrix.

Let us now return to the group $G = C_{4v}$ of the square
molecule. The order $\|G\|$ of this group is equal to $8$ (all its
elements are listed in Eq.(\ref{eq20})). According to
Eq.~(\ref{eq85}) there exist only three variants for the possible
set of dimensions of the irreps of the group $G = C_{4v}$:
\begin{equation}\label{eq86a}
    2^2 + 2^2 = 8,
\end{equation}
\begin{equation}\label{eq86b}
    1^2 + 1^2 + 1^2 + 1^2 + 1^2 + 1^2 + 1^2 + 1^2 = 8,
\end{equation}
\begin{equation}\label{eq86c}
    1^2 + 1^2 + 1^2 + 1^2 + 2^2 = 8.
\end{equation}

In variant (\ref{eq86a}), there are two 2-dimensional irreps ($n_1
= 2, n_2 = 2$). This is impossible because any group must possess
\textit{at least} one 1-dimensional irrep~: the identity irrep.
Variant (\ref{eq86b}) ($n_j = 1, j = 1...8$) is also impossible
because only for Abelian group \textit{all} the irreps can be
1-dimensional, while the group $G = C_{4v}$ is non-Abelian.
Therefore, only variant (\ref{eq86c}) turns out to be admissible:
the group $G = C_{4v}$ possesses \textit{five} irreps~: four
1-dimensional ($n_1 = n_2 = n_3 = n_4 = 1$) and one 2-dimensional
($n_5 = 2$), as presented in Table~1 below.

\begin{center}
Table 1. Symmetry elements of the group $G$ and their irreps
\begin{tabular}{|c|c|c|c|c|c|c|c|c|}
  \hline
   & $g_1$ & $g_2$ & $g_3$ & $g_4$ & $g_5$ & $g_6$ & $g_7$ & $g_8$ \\
  \hline
  $\Gamma_1$ & $1$ & $1$ & $1$ & $1$ & $1$ & $1$ & $1$ & $1$ \\
  \hline
  $\Gamma_2$ & $1$ & 1 & 1 & 1 & -1 & -1 & -1 & -1 \\
  \hline
  $\Gamma_3$ & $1$ & -1 & 1 & -1 & 1 & -1 & 1 & -1 \\
  \hline
  $\Gamma_4$ & $1$ & -1 &1 & -1 & -1 & 1 & -1 & 1 \\
  \hline
  $\Gamma_5$ & $\Biggl(\begin{array}{l}
1 \ \ 0\\
0 \ \ 1\\
\end{array}\Biggl)$
 &
 $\Biggl(\begin{array}{l}
\ 0 \ \ \ 1\\
-1 \ \ 0\\
\end{array}\Biggl)$
& $\Biggl(\begin{array}{l}
-1 \ \ \ \ 0\\
\ \ 0 \ \ -1\\
\end{array}\Biggl)$
& $\Biggl(\begin{array}{l}
0 \ \  -1\\
1 \ \  \ \ \ 0\\
\end{array}\Biggl)$
& $\Biggl(\begin{array}{l}
1 \ \ 0\\
0 \ \ 1\\
\end{array}\Biggl)$
& $\Biggl(\begin{array}{l}
0 \  \ 1\\
1 \  \ 0\\
\end{array}\Biggl)$
& $\Biggl(\begin{array}{l}
-1 \ \  0\\
 \  0 \ \ \ 1\\
\end{array}\Biggl)$
& $\Biggl(\begin{array}{l}
\ 0 \ \ -1\\
 -1 \ \ \ 0\\
\end{array}\Biggl)$\\
  \hline
\end{tabular}
\end{center}
\vspace{1cm}

Let us note that the construction of the irreps for a given group
$G$ is not a simple mathematical problem. However, the explicit
form of all the irreps of point and space groups can be found in
many textbooks (see, for example, \cite{kovalev}).

The importance of irreducible representations of symmetry groups
is illustrated by a famous theorem due to Wigner, according to
which eigenstates of any quantum system are classified by irreps
of the symmetry group $G$ of its Hamiltonian. Namely, every energy
level  of the quantum system is associated with a certain irrep
$\Gamma_j$ of the group $G$, and the \textit{degeneracy} of this
level is determined by the \textit{dimension} $n_j$ of the irrep
$\Gamma_j$.

The normal modes, representing exact vibrational regimes in the
harmonic approximation, can be classified in the same manner.
Indeed, every eigenstate of the force constant matrix $\mathcal K$
(see Eq. (\ref{e58})) represented by the natural frequency
$\omega_j$ and the corresponding eigenvector $\vec c_j$ can be
associated with a certain irrep $\Gamma_j$ of the parent symmetry
group of the considered mechanical system. In both cases, the
differential equations describing the physical systems (quantum or
classical) are \textit{linear}. In this regard, it is interesting
to investigate any possible symmetry~-- determined classification
of exact dynamical regimes in \textit{nonlinear} classical systems
with certain parent symmetry groups. This problem is fundamental
in the theory of bushes of nonlinear normal modes, as we
demonstrate below.

\subsection{Selection rules for excitation transfer between modes of different symmetry}

Let us return to the discussion of $N$-degree-of-freedom
mechanical systems with the parent symmetry group $G_0$ whose
vibrational state is described by the $N$-dimensional
configurational vector $\vec X(t)$.

As has already been discussed, all the vibrational states can be
classified by the subgroups of the parent group $G_0$. Let us
consider the vibrational state characterized by a certain subgroup
$G\subset G_0$. This means that the configuration vector $\vec
X(t)$ must be \textit{invariant} under the action of all the
operators $\hat g\in\hat G$:
\begin{equation}\label{eq2993}
\hat g\vec X(t)=\vec X(t), \ \ \ \ \quad\forall\hat g\in\hat G.
\end{equation}
The group $\hat{G}=\{~\hat g~|~\forall g\in G~\}$ consists of all
the operators $\hat g$ which are associated with the elements $g$
of the symmetry group $G$.

Let us now decompose the configuration space into the minimal
subspaces, which are invariant with respect to the parent group
$G_0$. Every one of these subspaces $V_j$ is the carrier space of
a certain irrep $\Gamma_j$ of the group $G_0$. As a result of this
transformation, the mechanical representation $\Gamma$ turns out
to be decomposed into a number of irreps:
\begin{equation}
\label{eq3103} \Gamma=\sum_j{^{^{\bigoplus}}\Gamma_j}.
\end{equation}
The set of all the basis vectors $\vec\varphi_j^{(i)}$ of the
irreps $\Gamma_j$ entering in the direct sum~(\ref{eq3103}) may be
chosen as a new basis of the configuration space. Therefore, we
can decompose the configuration vector $\vec X(t)$ in the above
mentioned basis as follows:
\begin{equation}
\label{eq3113} \vec
X(t)=\sum_{ji}{\mu_j^{(i)}(t)\vec\varphi_j^{(i)}}=\left(\vec\mu_j(t),\vec\Phi_j\right)=\sum_j{\vec\Delta_j(t)},
\end{equation}
where $\vec\Delta_j(t)$ is the contribution to $\vec X(t)$ from
the irrep $\Gamma_j$, while $i$ indicates different basis vectors
of the $n_j$-dimensional irreducible representation $\Gamma_j$.
Here we introduce the ``supervector''
$\vec\Phi_j=\{\vec\varphi_j^{(1)},\vec\varphi_j^{(2)},\ldots,\vec\varphi_j^{(n_j)}\}$
which determines the set of the basis vectors of the irrep
$\Gamma_j$, the vector
$\vec\mu_j(t)=\{\mu_j^{(1)}(t),\mu_j^{(2)}(t),\ldots,\mu_j^{(n_j)}(t)\}$,
which determines the time-dependence of the individual vibrational
modes, and their \textit{formal scalar product}
$\left(\vec\mu_j(t),\vec\Phi_j\right)$. From the condition of
invariance of the configuration vector $\vec X(t)$ with respect to
the group $\hat{G}$, (\ref{eq2993}), it can be deduced (see
\cite{DAI, Columbus2007}) that the following invariance conditions
also hold regarding the individual irreps $\Gamma_j$:
\begin{equation}
\label{eq3453} \left(\Gamma_j\downarrow
G\right)\vec\mu_j=\vec\mu_j.
\end{equation}
 Here we have introduced the \textit{restriction}
$\left(\Gamma_j\downarrow G\right)$ of the irrep $\Gamma_j$ of the
parent group $G_0$ onto the subgroup $G\subset G_0$. The
restriction $\left(\Gamma_0\downarrow G\right)$ is the set of all
the matrices of the representation $\Gamma_0$ of the group $G_0$
which are associated only with the elements of its subgroup
$G\subset G_0$.

Eq.~(\ref{eq3453}) provides the important \emph{selection rules
for excitation transfer} between modes of different symmetry. Let
us consider this point in more detail. This equation indeed holds
for every individual irrep $\Gamma_j$ of the parent group $G_0$.
Thus, from (\ref{eq3453}), we can see that $\vec \mu_j$ is
\textit{invariant} under all matrices $\mathcal
M_j(g)\in\left(\Gamma_j\downarrow G\right)$, i.e.\ $\vec\mu_j$ is
the \textit{common eigenvector} of all the matrices $\mathcal
M_j(g)$, $g\in G$ with \textit{unit} eigenvalue.

As the general solution of Eq.~(\ref{eq3453}), the vector
$\vec\mu_j$ depends on a number of arbitary constants denoted by
the letters $a,b,c,d$, etc., in the examples of section III. To
find all modes contributing to a given dynamical regime with
symmetry group $G$, expressed in Eq.~(\ref{eq3113}) by the vector
$\vec X(t)$, one must solve the linear algebraic
equations~(\ref{eq3453}) for each irrep $\Gamma_j$ of the group
$G_0$. As a result, the invariant vectors $\vec\mu_j$ for some
irreps $\Gamma_j$ may turn out to be equal to zero and thus
\textit{not contribute} to the considered dynamical regime. On the
other hand, some nonzero invariant vectors $\vec\mu_j$ for
multidimensional irreps can be very special due to certain
relations between their components (for example, some components
are equal to each other, or differ only by sign). Naturally, the
contributions $\vec \Delta_j(t)$ to $\vec X(t)$ from such irreps
$\Gamma_j$ also possess very special features.

Actually, Eq.~(\ref{eq3453}) can be considered as a source of
certain \emph{selection rules} for excitation transfer from the
``root'' mode to a number of other (secondary) modes. Indeed, if a
certain mode with the symmetry group $G$ is excited initially
(called the ``root'' mode), this group determines the symmetry of
the whole bush. The condition that the appropriate dynamical
regime $\vec X(t)$ must be invariant under the action of the above
group $G$ leads to Eq.~(\ref{eq2993}) and then to
Eq.~(\ref{eq3453}). If the vector $\vec\mu_j$ for a given irrep
$\Gamma_j$ is the zero vector, there are no modes, belonging to
this irrep, which contribute to $\vec X(t)$, i.e., the initial
excitation \textit{cannot be transferred} from the root mode to
the secondary modes associated with the irrep $\Gamma_j$.

Note that basis vectors associated with a given irrep $\Gamma_j$
in Eq.~(\ref{eq3113}) turn out to be equal to zero when this irrep
is not contained in the decomposition of the full natural irrep
$\Gamma$ into its irreducible parts $\Gamma_j$. This gives rise to
\textit{additional selection rules} which reduce the number of
possible vibrational modes in the considered bush. Trying every
irrep $\Gamma_j$ in Eq.~(\ref{eq3453}) and analyzing the above
mentioned decomposition of the natural representation $\Gamma$, we
obtain explicitly the whole bush of modes with the symmetry group
$G$.

\section{Acknowledgements}
One of us (T.B.) is grateful for the hospitality of the Centro
Brasileiro de Pesquisas Fisicas, at Rio de Janeiro, during March 1
- April 5 and of the Max Planck Institute for the Physics of
Complex Systems in Dresden, during April 5 - June 25, 2010, when
part of this paper was written. T. B. also acknowledges many
useful discussions on Hamiltonian systems with Professor C.
Tsallis, Dr. P. Tempesta and G. Ruiz at Rio de Janeiro and Dr. S.
Flach and Dr. Ch. Skokos at Dresden.

\end{document}